\documentclass[journal]{IEEEtran}
\ifCLASSINFOpdf
\fi

\usepackage[table,xcdraw]{xcolor}
\usepackage{threeparttable}
\usepackage{booktabs}
\usepackage{xcolor}
\usepackage{bbding}
\usepackage{pdfpages}
\usepackage{changes}
\usepackage[cmex10]{amsmath}
\usepackage{rotating}
\usepackage{threeparttable}
\usepackage{CJK}
\usepackage{amsmath}
\usepackage{amsthm,verbatim,amssymb,amsfonts,amscd}
\usepackage{array}
\usepackage{bm}
\usepackage{multicol}
\usepackage{multirow}
\usepackage{graphicx}
\usepackage{url}
\usepackage{subfigure}                     
\usepackage{subcaption}
\usepackage{stfloats}
\graphicspath{{figure/}}
\usepackage{epstopdf}
\usepackage{cite,amsfonts,amssymb,color, mathtools,setspace,comment,float,array}
\usepackage{stfloats}
\usepackage{bigstrut}
\usepackage[utf8]{inputenc}
\usepackage{fancyhdr}
\usepackage{tabularx}
\usepackage{makecell}
\usepackage{authblk}
\usepackage{hyperref}
\usepackage{longtable}
\usepackage{caption}

\def\BibTeX{{\rm B\kern-.05em{\sc i\kern-.025em b}\kern-.08em
		T\kern-.1667em\lower.7ex\hbox{E}\kern-.125emX}}
\newcolumntype{L}{>{\varwidth[c]{\linewidth}}l<{\endvarwidth}}
\newcolumntype{M}{>{$}l<{$}}

\begin{document}
\title{6G Channel Modeling: Requirement, Measurement, Methodology and Simulator}

\author{Jianhua~Zhang, Jiaxin~Lin, Pan~Tang\thanks{Pan Tang is corresponding author.}, Yuxiang~Zhang, Huixin~Xu, Tianyang~Gao, Haiyang~Miao, Huiwen~Gong, Changsheng~Zhao, Yameng~Liu, Yichen~Cai, Zhiqiang~Yuan, Lei~Tian, Shaoshi~Yang, Liang~Xia, Guangyi~Liu, and Ping~Zhang
	\IEEEcompsocitemizethanks{
 
\hyphenpenalty=5000
\tolerance=1000

		\par Jianhua~Zhang, Jiaxin~Lin, Pan~Tang, Yuxiang~Zhang, Huixin~Xu, Tianyang~Gao, Haiyang~Miao, Huiwen~Gong, Changsheng~Zhao, Yameng~Liu, Yichen~Cai, Zhiqiang~Yuan, Lei~Tian, Shaoshi~Yang, and Ping~Zhang are with State Key Lab of Networking and Switching Technology, Beijing University of Posts and Telecommunications, Beijing, 100876, P.R. China (email: jhzhang@bupt.edu.cn; linjx@bupt.edu.cn; tangpan27@bupt.edu.cn; zhangyx@bupt.edu.cn; xuhuixin@bupt.edu.cn; gtyss@bupt.edu.cn; hymiao@bupt.edu.cn;	birdsplan@bupt.edu.cn; zhaochangsheng@bupt.edu.cn;
        liuym@bupt.edu.cn; caiyichen@bupt.edu.cn; yuanzhiqiang@bupt.edu.cn; tianlbupt@bupt.edu.cn; shaoshi.yang@bupt.edu.cn; pzhang@bupt.edu.cn).
            \par Liang~Xia, and Guangyi~Liu are with the China Mobile Research Institute, Beijing 100053, P.R. China (e-mail: xialiang@chinamobile.com; liuguangyi@chinamobile.com).
		}
        
}

\maketitle
\begin{abstract}
Sixth-generation (6G) mobile communications have attracted substantial attention in the global research community of information and communication technologies (ICTs). 6G systems are expected to support not only extended 5G usage scenarios but also new usage scenarios, such as integrated sensing and communication (ISAC), integrated artificial intelligence (AI) and communication, and communication and ubiquitous connectivity. To achieve this goal, channel characteristics must be comprehensively studied and properly exploited to promote the design, standardization, and optimization of 6G systems. In this paper, we first summarize the requirements and challenges in 6G channel research. Our focus is on channels for six promising technologies enabling 6G, including ISAC, extremely large-scale MIMO (XL-MIMO), mid-band and terahertz (THz) technologies, reconfigurable intelligent surfaces (RISs), and space–air–ground integrated networks (SAGINs). A survey of the progress in 6G channel research regarding the above six promising technologies is presented in terms of the latest measurement campaigns, new characteristics, modeling methods, and research prospects. To support testing, optimization and evaluation, existing 6G channel simulators are summarized. Then, BUPTCMCCCMG-IMT2030 is introduced as an example of a simulator that was developed on the basis of the ITU/3GPP 3D geometry-based stochastic model (GBSM) methodology. We also address open issues covering standardization activities, AI-enabled methods, and system performance analysis in the context of 6G channel research. This paper offers in-depth, hands-on insights into the best practices of channel measurements, modeling, and simulations for the evaluation of 6G technologies, the development of 6G standards, and the implementation and optimization of 6G systems.
\end{abstract}


\IEEEpeerreviewmaketitle


\section{Introduction}\label{section1}
\subsection{Overview of 6G}
As fifth-generation (5G) mobile communication networks have been deployed worldwide, studies on sixth-generation (6G) mobile communication technologies, which are expected to be rolled out for 2030 and beyond, are being vibrantly conducted. Several 6G research projects and organizations have been established. Representative examples include the Finnish ``6Genesis Flagship Program" \cite{6Genesis}, the European ``Hexa-X" project \cite{HeraX}, the Chinese ``IMT-2030 (6G) Promotion Group" \cite{6GPromGroup}, and the North American ``Next G Alliance" \cite{NextG}. Simultaneously, international standardization organizations have launched several key initiatives. For example, in 2022, the International Telecommunication Union (ITU) identified future technology trends \cite{TRENDS} for International Mobile Telecommunications (IMT) systems toward 2030 and beyond. Very recently, the recommendation framework for IMT-2030, which is commonly known as the 6G vision, was finalized \cite{VISION}. This means that the global 6G standardization campaign has been formally initiated. With this recommendation framework, the ITU establishes the overall objectives, procedures and timelines for the development of 6G mobile communication standards. Following this framework, the 3rd Generation Partnership Project (3GPP) aims to complete standardization no later than 2030. As part of the technical preparation, the 3GPP has already initiated several study items in Release 20, including efforts on advanced channel modeling.

6G aims to establish a more inclusive information society than 5G does. Its goal is to intelligently connect humans, machines, and various other things \cite{VISION}. Some key drivers of 6G have been discussed in \cite{TRENDS}. These drivers can be understood from the perspective associated with continuously growing mobile traffic and emerging applications. Specifically, it is forecast that by 2030, global mobile traffic will reach 5016 exabytes (EB) per month \cite{ITU2370}, which is approximately 80 times greater than that in 2020. The potential emerging applications of 6G \cite{TRENDS} can be categorized into several main usage scenarios \cite{VISION}. One extends enhanced mobile broadband (eMBB), which was introduced in 5G, to target immersive applications such as holographic communications, tactile and haptic internet applications, and extended reality (XR). Similarly, hyper-reliable and low-latency communications and massive communications are extensions of the 5G usage scenarios of ultra-reliable and low latency communications (URLLC) and massive machine type communications (mMTC), respectively. In addition, three other usage scenarios have been identified, i.e., ubiquitous connectivity, integrated artificial intelligence (AI) and communication, and integrated sensing and communication (ISAC) \cite{VISION}. The applications of integrated AI/sensing and communication include perceptive mobile networks, digital twins, and the proliferation of networked AI. As a result, the capabilities of 6G are enhanced across multiple dimensions, including the data rate, spectrum efficiency, reliability, latency, mobility, and connection density \cite{UXzp,3-1,3-3,3-4,3-8,3-9,3-10,VISION}. Moreover, new capabilities, such as all-domain coverage, sensing-related and AI-related capabilities, sustainability and computing power, are being considered \cite{VISION}.

\subsection{Requirements and Challenges of 6G Channel Research}
\begin{figure*}[htbp]
 	\centering
 	\includegraphics[width=1\textwidth]{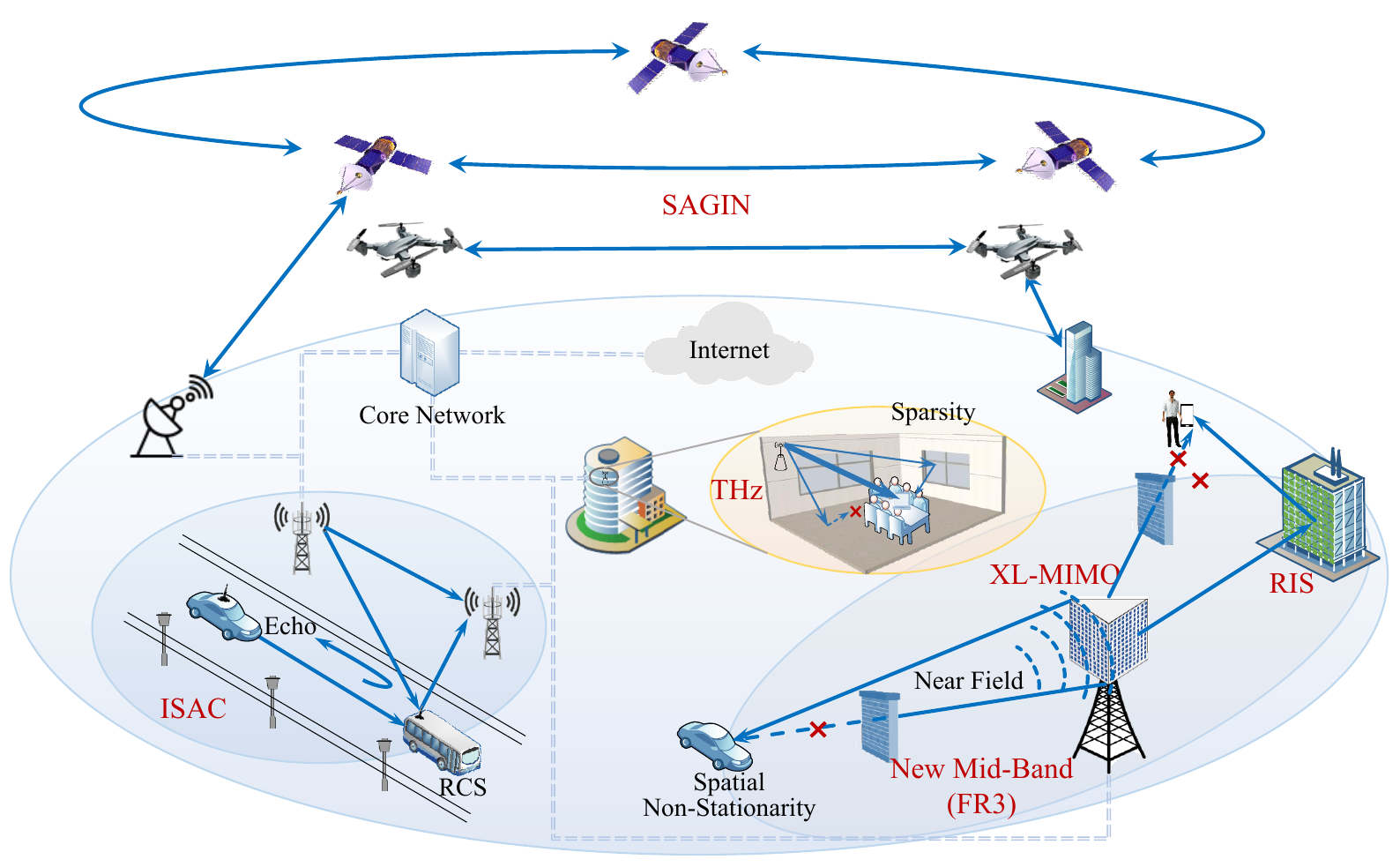}
\caption{Illustration of 6G channel research requirements.} \label{fig_1-2}
\end{figure*}

The radio channel is the transmission medium between a transmitter (Tx) and receiver (Rx). It plays a crucial role in determining the ultimate performance limit of mobile communications systems \cite{Molisch2012,3-2,THz_Ref_3}. As shown in Fig.~\ref{fig_1-2}, 6G communication systems will incorporate various novel technological components \cite{TRENDS,3-2,secIIall1,3-3,secIIall2,secIIall3,secIIall4,UXzp,secIIall5}, such as the integration of sensing into the communication system (i.e., ISAC) \cite{ISAC8,secIIISAC1,ISAC_sensing}, the extension of antenna arrays to extremely large-scale multiple-input multiple output (XL-MIMO) \cite{3-1,secIIall5,MIMO1-23, EMIMOch1_2021Decarli}, the extension of the frequency band to new mid-band (6--24 GHz) and terahertz (THz) ranges \cite{secIITHz1,2022THzSurveyHC,THz_Ref_61,MIMO1-1-0,THz_Ref_3,secIITHz2}, the introduction of a programmable wireless environment by using reconfigurable intelligent surfaces (RISs) \cite{secIIRIS1,secIIRIS2,RIS722}, and the extension of coverage to space-air-ground integrated networks (SAGINs) \cite{SAGIN2022Cui,SAGIN_1,SAGIN_2}. Each of these may result in distinct requirements and characteristics, as summarized below.

\subsubsection{ISAC, target scattering, and sharing feature}
ISAC is one of the key technologies for 6G\cite{ISAC4}. Compared with conventional systems with separate devices, ISAC technology can integrate functions into one system, enabling base stations (BS) or terminals to communicate while sensing the surrounding environment. In ISAC systems, sensing aims to obtain the state information, such as the speed and location, of the targets in a particular environment through echoes and the collaboration between BSs \cite{ISAC3}. Therefore, in the study of ISAC channels, it is essential to introduce the radar cross section (RCS), a target-related feature, and consider the impact of the environment on the sensing channel \cite{ISAC_wjl}. Moreover, owing to the simultaneous use of the propagation environment by the communication and sensing signals, some shared propagation scatterers may be involved in the ISAC system \cite{ISAC_wjl, ISAC18}. The clusters contributed by these shared scatterers may coexist and disappear in both communication and sensing channels. Hence, the communication channel component and the sensing channel component of an ISAC channel may be partially related to each other. These novel propagation characteristics facing ISAC technology cannot be adequately captured by existing communication channel models. Therefore, an ISAC channel model incorporating these characteristics on the basis of realistic channel measurements is highly important for the design and performance evaluation of ISAC systems.

\subsubsection{Larger-scale antenna array, near-field propagation, and spatial non-stationarity}
In the 1970s, researchers suggested introducing MIMO technologies into wireless communication systems. In 1995, Telatar theoretically derived MIMO channel capacity under the fading condition \cite{MIMO0-original}. It was demonstrated that the MIMO technology can substantially increase the capacity of wireless communication system. Then, the MIMO technology was introduced into the fourth generation (4G) mobile communication system, which can support eight array elements and highly improve the system spectrum efficiency \cite{MIMO0-original_1}. However, the channel model used in 4G systems only supports the azimuth dimension. With the proposal of massive MIMO, three-dimensional (3D) MIMO \cite{MIMO0-original_2} is a typical massive antenna realization. Therefore, the elevation dimension was introduced into the ITU/3GPP 5G standard channel model\cite{2412,38901}. The 5G channel model captures the elevation-azimuth-time-frequency channel characteristics, and meets the requirements of 3D MIMO and massive MIMO technologies. In the 6G era, the number of array elements is expected to increase significantly, reaching a range from a few hundreds to thousands. As a result, the XL-MIMO technology based on the much larger-scale antenna arrays is envisaged to be deployed. The XL-MIMO with 1024$\times$1024 array elements can offer abundant spatial degrees of freedom for supporting 100 times peak data rate improvement in 6G mobile communication systems \cite{MIMO0-original_3}. However, some new channel characteristics emerge in XL-MIMO systems. To elaborate a little further, when the propagation distance is shorter than the far-field Rayleigh distance, the plane wave assumption is invalid in the near-field region \cite{MIMO1-23,MIMO_yzq}. Therefore, the the near-field spherical wave and spatial non-stationary (SnS) characteristics \cite{MIMO1-1-0, MIMO1-8-1} cannot be ignored in XL-MIMO communication systems. As shown in Fig.~\ref{fig_1-2}, the XL-MIMO can be modeled using near-field spherical propagation in the near-field region. In addition, with the expansion of physical array aperture, different array elements see different scatterers, resulting in SnS on the array. However, these characteristics are not covered by the ITU/3GPP 5G channel model framework. Therefore, it is necessary to investigate the impact of the near-field spherical wave and the SnS characteristic through channel measurements and simulations for 6G channel modeling.

\subsubsection{Multi-frequency, propagation attenuation, and channel sparsity}

At present, the new mid-band (6--24 GHz) has attracted widespread attention and has the potential to be an important candidate frequency band for 6G. In June 2022, the 3GPP RAN Plenum adopted a standard modification proposal for the 5925--7125 MHz band, and this band was introduced into 5G-Advanced Rel-18. At the World Radiocommunication Conference 2023 (WRC-23), 6425--7125 MHz was identified for mobile use in every ITU region, and the agenda for WRC-27 was defined. The new WRC cycle will focus on the following bands for the IMT: 4400--4800 MHz, 7125--8400 MHz, and 14.8--15.35 GHz. In December 2023, the 3GPP TSG RAN Rel-19 discussed research work in the 7--24 GHz band. In \cite{THz_Sparsitylxm}, with the increase of frequency, channel sparsity becomes more prominent. However, the sparsity of the channel matrix derived from the 3GPP channel model is similar regardless of the frequency band, which leads to inaccurate prediction results for different frequency bands.

The deployment of 5G has already achieved highly efficient utilization of the sub-6 GHz frequency band and harnessed millimeter wave (mmWave) frequencies to deliver mobile devices with unparalleled multi-Gbps data rates \cite{THz_rappaport2017overview}. The THz frequency band (0.3-3 THz according to ITU radio regulations or 0.1-10 THz as a commonly used wider range) possesses an atmospheric transmission window with low molecular absorption losses, and abundant bandwidth was allocated to terrestrial mobile communication services at the WRC in 2019 \cite{THz_ITU_RR}. As the era of 6G begins, the demand for a peak data rate of 1 Tbps will require a comprehensive exploitation of the THz frequency band \cite{3-2}. Prototype demonstration and testing systems for THz communication have already been developed to achieve high-speed transmission. For example, high-capacity wireless transmission at 1.44 Tbps have been achieved within the THz frequency band of 135-170 GHz \cite{THz_NTT}. However, it is essential to recognize that the THz band has distinctive characteristics in contrast to the sub-6 GHz and mmWave frequency bands. The shorter wavelength ($\leq$ 3 mm) signals in the THz band reflect the roughness of the objects in the environment. Previously smooth surfaces, such as walls and tree leaves, appear rougher in the THz band. The specular reflection may lose its privileged position, allowing diffuse scattering to become the predominant reflection behavior. In general, THz waves experience severe transmission losses and molecular absorption, thereby causing changes in path loss characteristics and the emergence of sparsity. Sparsity refers to the condition where only a small number of multipath components (MPCs) carry the majority of the received signal power. Currently, the ITU/3GPP channel model \cite{2412,38901} only supports frequencies up to 100 GHz for 5G. Therefore, a comprehensive THz channel model, that is capable of covering typical scenarios (such as indoor hotspot (InH) and urban microcell (UMi)) and incorporating distinctive THz channel characteristics (such as high path loss and sparsity) is urgently needed.

\subsubsection{RIS, reflection coefficient characteristics, and concatenated channel}
RIS is considered as a potential technology that can overcome the challenges in traditional beam-space antenna array based beamforming \cite{VISION}. It is capable of significantly improving the channel gain by actively controlling the wireless propagation environment \cite{3-10}. For example, when the line-of-sight (LOS) path is obstructed, RIS can provide a high-power reflection path. Additionally, it promises low cost and high energy efficiency. However, since RIS can program the wireless propagation environment to some degree, the role of the RIS itself should be taken into account when studying the wireless channels that involve RIS. This makes the traditional wireless channel model no longer suitable for the concatenated channel of RIS, and the channel modeling theory for RIS is not yet mature. Furthermore, for RIS, both the decoupling of the channel model and the construction of a high-precision electromagnetic response model are urgent problems to be solved \cite{3-10}. In addition, the high path loss of the RIS concatenated channel and the flexibility of RIS deployment locations pose new challenges for RIS channel measurement.

\subsubsection{SAGIN, meteorological effects, and dynamic properties}
SAGIN is a heterogeneous architecture that integrates different networks like space networks, aerial networks, and terrestrial communication networks. It is expected to be realized in the future 6G era \cite{SAGINNature}. The terrestrial networks achieve normal coverage of various scenarios in urban and rural areas as discussed above. Space and aerial networks, in which devices are located on the satellite, high altitude platform stations (HAPS), and unmanned aerial vehicles (UAV), can realize ubiquitous coverage in remote areas, sea, and airspace. Due to the ability to achieve all-domain coverage, SAGIN has attracted much attention from both academia and the industry. In recent years, several organizations have started their projects to deploy satellite constellations in low earth orbit (LEO). These constellations contain hundreds or thousands of satellites and are deployed to provide worldwide coverage. Examples of these constellations include the SpaceX Starlink, which is announced to comprise over 10000 satellites, the Norwegian STEAM network, which is claimed to contain 4257 satellites, and the OneWeb constellation, which is a smaller constellation with 720 satellites \cite{SAGIN_2}. The ubiquitous coverage of SAGIN benefits a variety of new services and applications. It also leads to more diverse scenarios and more complex channel characteristics. Since SAGIN involves space, air, and ground, different types of communication segments should be considered, such as space-to-ground (SG), air-to-ground (AG), and others. As for the SG segment, the transmission signal may suffer from an extra loss caused by meteorological environmental factors like atmosphere, rain, clouds, and fog. When the signal travels through the ionosphere, fluctuations of amplitude and phase (scintillation), rotations of polarization angle (Faraday rotation), delay dispersion, and others are found due to the ionosphere effects. As for the AG segment, the high change rate of channel characteristics because of the UAV velocities, buildings, and ground surface composition is exhibited. Besides, characteristics of AG channels like Doppler spread should be considered. For terrestrial networks, channel characteristics are discussed above. For the channel of SAGIN, both each segment channel and the integration effect on channel characteristics require consideration and being well addressed.

\subsection{Paper Contributions and Organization}
Research on wireless channels plays a vital role in the design, standardization and optimization of 6G systems. At the time of this writing, a few surveys on some aspects of 6G channels \cite{3-1,3-2,2022THzSurveyHC,2022THzSurveySD,RISSurveyWCX} have been published. Nevertheless, there has not been a comprehensive survey of the entire portfolio of 6G channel research. We provide a comprehensive survey of the channel models of potential 6G enabling technologies, including ISAC, XL-MIMO, mid-band and THz technologies, RISs and SAGINs. 6G channel simulators supporting potential enabling technologies are summarized, and the Beijing University of Posts and Telecommunication and China Mobile Communications Corporation Channel Model Generator-IMT2030 (BUPTCMCCCMG-IMT2030) is introduced as an example, which supports various features of potential 6G technologies. Considering mainstream models, the simulator uses the 5G standard channel modeling framework ITU-R M.2412 \cite{2412} (principles described in Section \uppercase\expandafter{\romannumeral2}) as the basis for expansion to 6G. Our major contributions via this paper are as follows:

\begin{itemize}
\item{An overview of ISAC channel measurements is provided, including measurement platforms based on time-domain correlation and vector network analyzers (VNAs), as well as monostatic and bistatic sensing modes. Then, an in-depth analysis of the new channel characteristics introduced by ISAC is conducted, with a particular focus on target scattering, EO, Doppler shift, background effect, and sharing feature. A cluster-based ISAC channel model is introduced to characterize these characteristics accurately. Finally, the pros and cons of various ISAC channel modeling methods are discussed.}

\item{The XL-MIMO channel measurement platforms, including frequency domain and time domain platforms, are systematically introduced. Next, a near-field spherical wave model of XL-MIMO and the characteristics of SnS are discussed in detail. At the same time, the division of near-field range is analyzed. In addition, several channel modeling methods such as the statistical, deterministic and hybrid channel modeling are compared, then a realistic yet low-complexity SnS channel modeling framework is discussed. Finally, the future research directions of XL-MIMO are pointed out.}
\item{The representative multi-frequency channel measurement campaigns, covering both mid-band and THz frequencies, are reviewed and analyzed. Then, a detailed discussion on the multi-frequency channel characteristics is given, specifically focusing on reflection, high loss and sparsity. Finally, the pros and cons of stochastic, deterministic, and hybrid modeling approaches for multi-frequency channels are further discussed.}
\item{Some measurement campaigns related to the RIS channel are summarized, especially regarding the power gain brought by RIS. In addition, the characteristics of the RIS channel are introduced in detail, including the response characteristics of the important intermediate node RIS and the new loss characteristics presented due to the concatenation of two subchannels and RIS response. Then, to accurately describe the RIS channel, we provide various types of RIS channel modeling approaches, including statistical, deterministic, and hybrid approaches. Finally, future research directions are discussed.}
\item{Channel measurements in SAGIN are reviewed, including the SG, AG, and air-to-air (AA) channels, which is followed by a detailed description of the channel characteristics of SAGIN. In addition, channel models for simulating SAGIN are introduced. Finally, a brief summary of potential research directions for future work is presented.}
\item{6G channel simulators that support potential enabling technologies are summarized, and the BUPTCMCCCMG-IMT2030 simulator developed by our team is introduced, including its principles, structure, functions, and applications.} 

\end{itemize}
\begin{figure*}[htbp]
 	\centering
 	\includegraphics[width=0.75\textwidth]{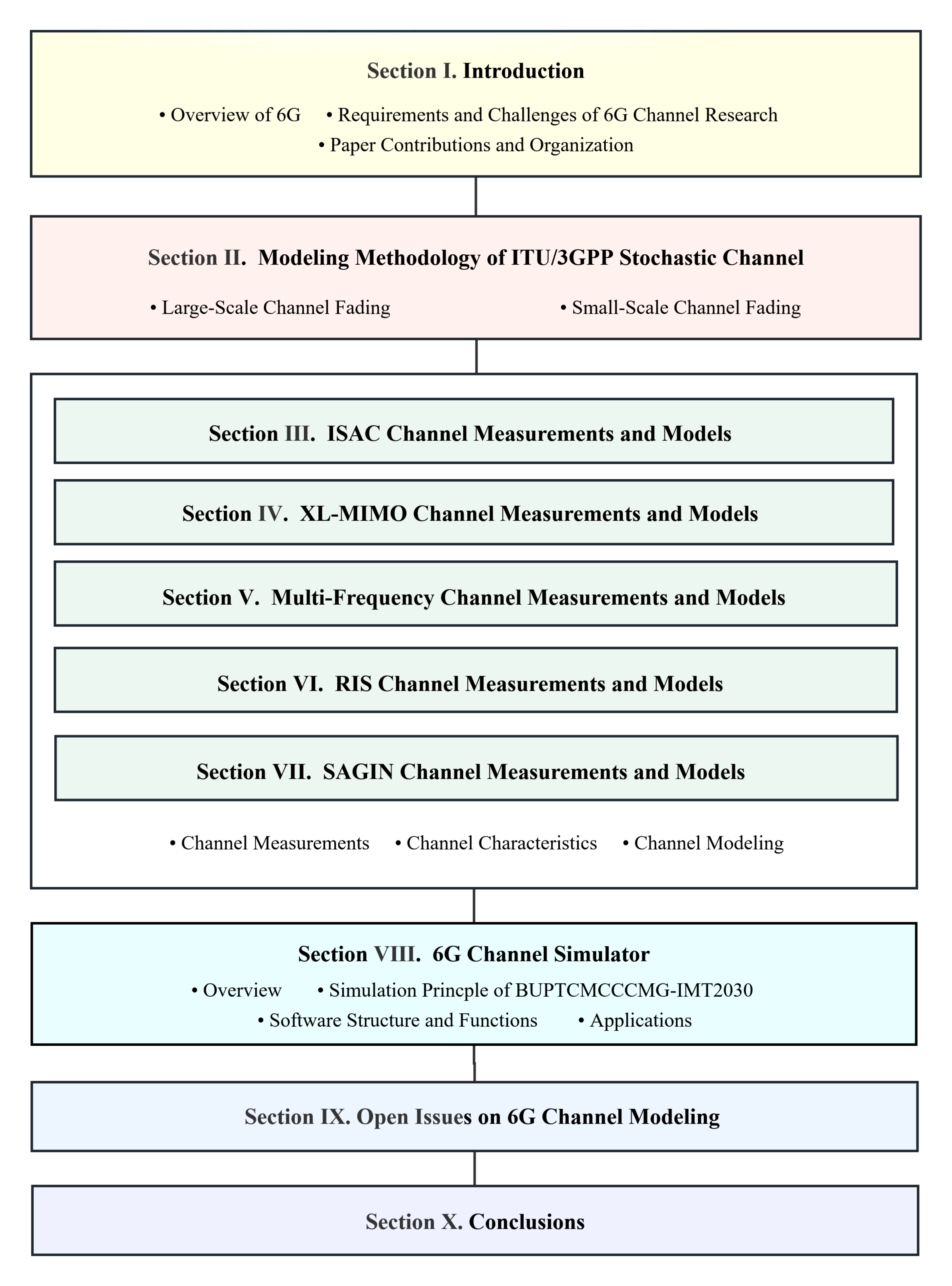}
\caption{Organization of the paper.} \label{fig_1Str}
\end{figure*}

The rest of this paper is organized as follows. In Section~\ref{SecISAC}--\ref{SecSAGIN}, we provide a survey of the state-of-the-art channel research on ISAC, XL-MIMO, mid-band and THz technologies (as a multi-frequency study), RISs and SAGINs in terms of measurements, characteristics and modeling and discuss the prospects. Section~\ref{SecSimu} summarizes 6G channel simulators that support potential enabling technologies and introduces BUPTCMCCCMG-IMT2030 as a standards-based implementation. Section~\ref{SecDis} presents open issues in 6G channel research, including recent 3GPP studies on channel modeling for 7--24 GHz bands and ISAC, as well as AI-enabled methods in channel research and system performance analysis for practical channel models. Finally, conclusions are drawn in Section~\ref{SecCon}. For clarity, the organization of the paper is shown in Fig.~\ref{fig_1Str}.

\section{Modeling Methodology of the ITU/3GPP Stochastic Channel}\label{Sec2}

This section introduces the standardized 5G channel models defined in 3GPP TR 38.901 \cite{38901} and ITU-R M.2412\cite{2412}. These models serve as the foundational framework upon which our proposed 6G channel extensions are built. Specifically, the new characteristics and modeling approaches discussed in Sections~\ref{SecISAC} to \ref{SecSAGIN}—including those for ISAC, XL-MIMO, and other key 6G technologies—are all based on extensions to this baseline model.

In wireless communications, transmitted signals are subject to propagation losses, which are broadly classified into large-scale and small-scale fading effects \cite{Tse2005}. In practice, both types of fading coexist \cite{Molisch2012} and can significantly impact various aspects of communication system design and performance \cite{3-2}. Large-scale fading affects signal strength over long Tx–Rx distances (typically hundreds or thousands of meters,) and is crucial for coverage prediction and interference analysis, thus playing a vital role in wireless system deployment.
On the other hand, small-scale fading refers to rapid signal fluctuations over short distances (on the order of a few wavelengths) or brief time intervals \cite{RappaportBook}. This type of fading is particularly important for the design of reliable and efficient physical-layer transmission techniques \cite{Tse2005}. The delay and angular characteristics of multipath components (MPCs), for instance, are critical for optimizing channel estimation, beamforming, and MIMO performance \cite{DDAWPL}. This section outlines the modeling principles of the 5G stochastic channel model from both large-scale and small-scale perspectives, forming the basis for subsequent 6G model development.

\subsection{Large-Scale Channel Fading}
Large-scale fading comes from path loss as a function of distance and shadowing by large objects such as buildings and hills \cite{Tse2005}. Path loss represents the reduction in radio wave power over a long distance \cite{RappaportBook}. Shadow fading is due mainly to multipath blockage, which can also cause a significant attenuation of signal strength. The 5G standard provides three types of path loss models, including the close-in (CI) model, floating-intercept (FI) model, and $\textrm{alpha–beta–gamma}$ (ABG) model \cite{Cheng2017,MacCartney2013,Sun2016a}. The FI model is expressed as
\begin{equation}\label{equadd2}
\textrm{PL}_{\textrm{FI}}(d)[\textrm{dB}]=\beta_{\textrm{FI}} +10\alpha_{\textrm{FI}}\textrm{lg}d+X_{\sigma}^{\textrm{FI}} ,
\end{equation}
where $\textrm{PL}_{\textrm{FI}}(d)$ denotes the path loss in dB, which is a function of the Tx–Rx separation distance $d$. $\beta_{\textrm{FI}} $ and $\alpha_{\textrm{FI}} $ are two fitted parameters. $X_{\sigma}^{\textrm{FI}}$ is a zero-mean Gaussian variable with a standard deviation of $\sigma$ in dB, representing shadow fading. Note that $\beta_{\textrm{FI}} $ is frequency dependent.

The CI path loss model can be provided by \eqref{equadd1}
\begin{equation}\label{equadd1}
\begin{split}
\textrm{PL}_{\textrm{CI}}(f,d)[\textrm{dB}]=&20\textrm{lg}\frac{4\pi d_0f}{\textrm{c}}+10\alpha_{\textrm{CI}}\ \textrm{lg}\frac{d}{d_0}+X_{\sigma}^{\textrm{CI}}\\ &\textrm{for}\ d\geq d_0,\ d_0=1\,\textrm{m},
\end{split}
\end{equation}
where $f$ is the center frequency, $d$ is the Tx–Rx separation distance greater than the reference distance $d_0$, ${\textrm{c}}$ denotes the speed of light, $\alpha_{\textrm{CI}}$ is the single model parameter known as the path loss exponent (PLE), and $X_{\sigma}^{\textrm{CI}}$ is the zero-mean Gaussian variable, which describes the large-scale signal fluctuations about the mean path loss over distance.

The third type of path loss model, the ABG model, is represented as a function of frequency and distance \cite{MacCartney2013,Maccartney2015}. It can describe the path loss at various frequencies and is expressed as
\begin{equation}\label{equ3}
\begin{split}
\textrm{PL}_{\textrm{ABG}}(f,d)[\textrm{dB}]=&10\alpha_{\textrm{ABG}}\  \textrm{lg}\frac{d}{d_0}+\\
&\beta_{\textrm{ABG}}+10\gamma_{\textrm{ABG}}\ \textrm{lg}\frac{f}{1\ \textrm{GHz}}+X_{\sigma}^{\textrm{ABG}}\\ &\textrm{where}\ \ d_0=1\,\textrm{m},
\end{split}
\end{equation}
where $d_0$ is the received power reference point, typically set to 1 m; here, $f$ is in GHz. $\alpha_{\textrm{ABG}}$, $\beta_{\textrm{ABG}}$ and $\gamma_{\textrm{ABG}}$ are fitting parameters, where $\alpha_{\textrm{ABG}}$ and $\gamma_{\textrm{ABG}}$ characterize the distance dependence and frequency dependence of path loss, respectively, and where $\beta_{\textrm{ABG}}$ is an offset value in dB. The typical value of $\gamma_{\textrm{ABG}}$ is 2, which corresponds to free space propagation. $X_{\sigma}^{\textrm{ABG}}$ is the zero-mean Gaussian variable. In the ABG model, path loss varies with frequency and distance. \cite{Sun2016} demonstrated that the CI and ABG models offer comparable modeling performance when field measurement data are used.

\subsection{Small-Scale Channel Fading}
Geometry-based statistical channel modeling is the mainstream approach adopted by both the ITU and 3GPP in 4G and 5G. According to the definition of the geometry-based stochastic model (GBSM) \cite{WINNER2}, the channels are generated by summing the contributions of rays on the basis of geometry principles and specific small-scale statistical channel parameters from measurements. Geometry-based modeling enables the channel to be independent of the antenna and allows the configuration of different antennas and different element patterns. Additionally, for individual channel snapshots, the channel parameters, such as delay, power, angle of arrival, and angle of departure, are stochastically determined on the basis of statistical distributions extracted from channel measurements. In the GBSM-based channel model of the 5G standard, a new dimension of elevation is introduced at both Tx and Rx, enabling support for 3D MIMO configurations.

\begin{figure*}[]
 	\centering
 	\includegraphics[width=0.75\textwidth]{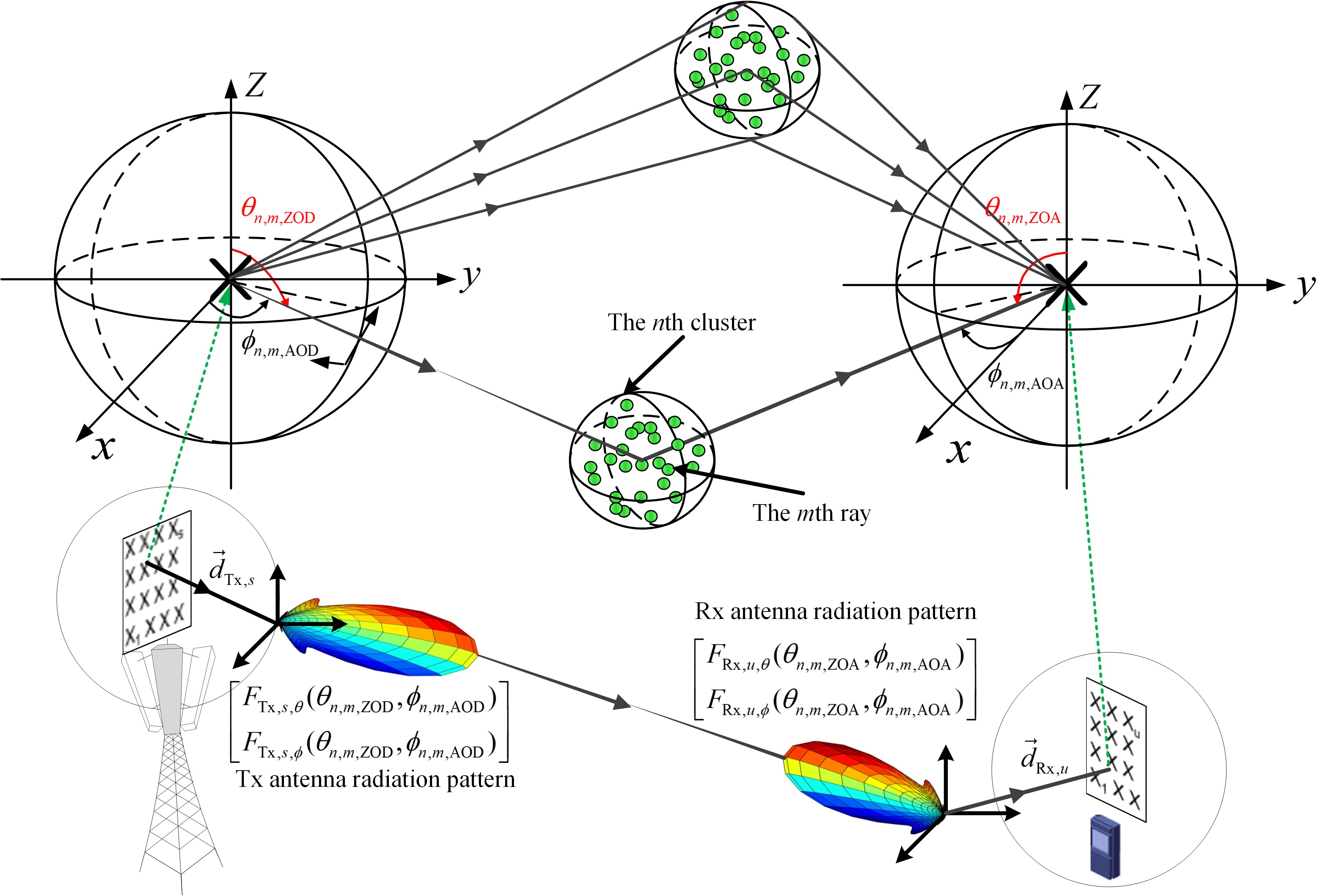}\\
\caption{Illustration of the 3D GBSM-based channel model \cite{Zhang3DMIMO}.}\label{3Dchannelmodel}
\end{figure*}

The 3D GBSM-based channel model is illustrated in Fig.~\ref{3Dchannelmodel}. The small sphere with several dots inside represents a scattering region that generates a cluster of propagation paths. Each cluster consists of $M$ rays, and $N$ clusters are assumed. The Tx and Rx are equipped with $S$ and $U$ antennas, respectively. The small-scale parameters (SSPs), such as the delay $\tau_{n, m}$, zenith angle of arrival $\theta_{n, m, \textrm{ZOA}}$, azimuth angle of arrival (AOA) $\phi_{n, m, \textrm{AOA}}$, azimuth angle of departure (AOD) $\phi_{n, m, \textrm{AOD}}$, and zenith angle of departure (ZOD) $\theta_{n, m, \textrm{ZOD}}$, are assumed to differ for the $m$th ray in the $n$th cluster. Radio waves propagate in 3D space, and MPCs exhibit not only horizontal-dimension characteristics but also vertical-dimension characteristics.

In the ITU-R 5G stochastic channel model, the channel impulse response (CIR) for the ray from the $s$th transmitting antenna to the $u$th receiving antenna of the $n$th cluster is a function of time $t$, delay $\tau$, angle of arrival $\bm{\Omega}$ and angle of departure $\bm{\Psi}$ given by

\begin{equation}\label{equ8}
\begin{split}
&h_{u,s,n}(t,\tau,\bm{\Omega},\bm{\Psi})=\sum_{m=1}^{M_n}a_{n,m}F_{\textrm{Rx},u}(\bm{\Omega}_{n,m})\\
&\textrm{exp}(j2\pi\lambda^{-1}(\hat{r}_{\textrm{Rx},n,m}\cdot\vec{d}_{\textrm{Rx},u}))F_{\textrm{Tx},s}(\bm{\Psi}_{n,m})\\
&\textrm{exp}(j2\pi\lambda^{-1}(\hat{r}_{\textrm{Tx},n,m}\cdot\vec{d}_{\textrm{Tx},s}))\delta(\tau-\tau_{n,m})\\
&\delta(\bm{\Omega}-\bm{\Omega}_{n,m})\delta(\bm{\Psi}-\bm{\Psi}_{n,m}),
\end{split}
\end{equation}

where $a_{n,m}$ is the complex amplitude of the $m$th MPC within the $n$th cluster and where $F_{\textrm{Rx},u}$ and $F_{\textrm{Tx},s}$ are the $u$th antenna patterns at the Rx side and the $s$th antenna pattern at the Tx side \cite{Kelley1993,Balanis2016}, respectively. $\hat{r}_{\textrm{Rx},n,m}$ is the unit vector in the spherical coordinate system at the Rx side, $\vec{d}_{\textrm{Rx},u}$ is the location vector of the $u$th antenna at the Rx side, and the subscript Tx similarly represents the Tx side. $\delta(\cdot)$ is the Dirac function, $\lambda$ is the wavelength, and $M_n$ is the number of delay bins. $\bm{\Omega_{n,m}}$ and $\bm{\Psi_{n,m}}$ are the angles of arrival and departure of the MPC, respectively. They include both the elevation and azimuth angles, $\theta$ and $\phi$, for 3D MIMO systems as follows:

\begin{equation}\label{equ9}
\begin{split}
&\bm{\Omega_{n,m}}=(\theta_{n,m,\textrm{ZOA}},\phi_{n,m,\textrm{AOA}}),\\ &\bm{\Psi_{n,m}}=(\theta_{n,m,\textrm{ZOD}},\phi_{n,m,\textrm{AOD}}).
\end{split}
\end{equation}

In practical 5G systems, antennas can be designed and configured to exhibit specific radiation patterns that enhance system performance, and slant polarized antennas are typically used. Thus, the radiation patterns need to be modeled when generating the CIR. Horizontal and vertical linear polarizations and the $+45^{\circ}/-45^{\circ}$ slant polarization are examples of orthogonal polarizations \cite{Bao2013}. When dual polarization is utilized, (\ref{equ8}) is changed to (\ref{bigH}).
\begin{figure*}
	\begin{equation}
	\label{bigH}
	\begin{split}
	H_{u, s, n}(t, \tau) = \sqrt{\frac{P_n}{M}}\sum_{m=1}^{M}
	\begin{bmatrix}
	F_{\textrm{Rx}, u, \theta}(\theta_{n, m, \textrm{ZOA}}, \phi_{n, m, \textrm{AOA}})\\
	F_{\textrm{Rx}, u, \phi}(\theta_{n, m, \textrm{ZOA}}, \phi_{n, m, \textrm{AOA}})
	\end{bmatrix}^\textrm{T}
	\begin{bmatrix}
	\textrm{exp}(j\Phi_{n,m}^{\theta\theta}) & \sqrt{\kappa_{n, m}^{-1}}\textrm{exp}(j\Phi_{n, m}^{\theta\phi})\\
	\sqrt{\kappa_{n, m}^{-1}}\textrm{exp}(j\Phi_{n, m}^{\phi\theta}) & \textrm{exp}(j\Phi_{n,m}^{\phi\phi})
	\end{bmatrix}\\
	\begin{bmatrix}
	F_{\textrm{Tx}, s, \theta}(\theta_{n, m, \textrm{ZOD}}, \phi_{n, m, \textrm{AOD}})\\
	F_{\textrm{Tx}, s, \phi}(\theta_{n, m, \textrm{ZOD}}, \phi_{n, m, \textrm{AOD}})
	\end{bmatrix}
	\textrm{exp}(j2\pi\frac{\hat{r}_{\textrm{Rx}, n, m}^\textrm{T}\cdot \vec{d}_{\textrm{Rx}, u}}{\lambda})
	\textrm{exp}(j2\pi\frac{\hat{r}_{\textrm{Tx}, n, m}^\textrm{T}\cdot \vec{d}_{\textrm{Tx}, s}}{\lambda})
	\textrm{exp}(j2\pi v_{n, m}t)\delta(\tau-\tau_{n, m})
	\end{split}
	\end{equation}
\end{figure*}

In (\ref{bigH}), $\left \{{\Phi}_{n, m}^{\theta\theta}, {\Phi}_{n, m}^{\theta\phi}, {\Phi}_{n, m}^{\phi\theta}, \textrm{and} \; {\Phi}_{n, m}^{\phi\phi} \right \}$ are random initial phases corresponding to four different polarization combinations ($\theta\theta$, $\theta\phi$, $\phi\theta$, and $\phi\phi$, respectively) for ray $m$ of cluster $n$. The distribution of the initial phases is uniform within ($-\pi, \pi$). ${F}_{\textrm{Rx}, u, \theta}$ and ${F}_{\textrm{Rx}, u, \phi}$ are the field patterns of the receiving antenna element $u$ in the direction of the spherical basis vectors, which are characterized by $\theta$ and $\phi$, respectively. ${F}_{\textrm{Tx}, s, \theta}$ and ${F}_{\textrm{Tx}, s, \phi}$ are the field patterns of the transmitting antenna element $s$ in the $\theta$ and $\phi$ directions. $\kappa_{n, m}$ is the cross-polarization power ratio (XPR) on a linear scale. $v_{n, m}$ is the Doppler frequency component of ray $m$ of the $n$th cluster. $P_n$ is the normalized power of the $n$th cluster.

\section{ISAC Channel Measurement and Modeling}\label{SecISAC}

\begin{figure}[!t]
\centering
\includegraphics[width=3.4in]{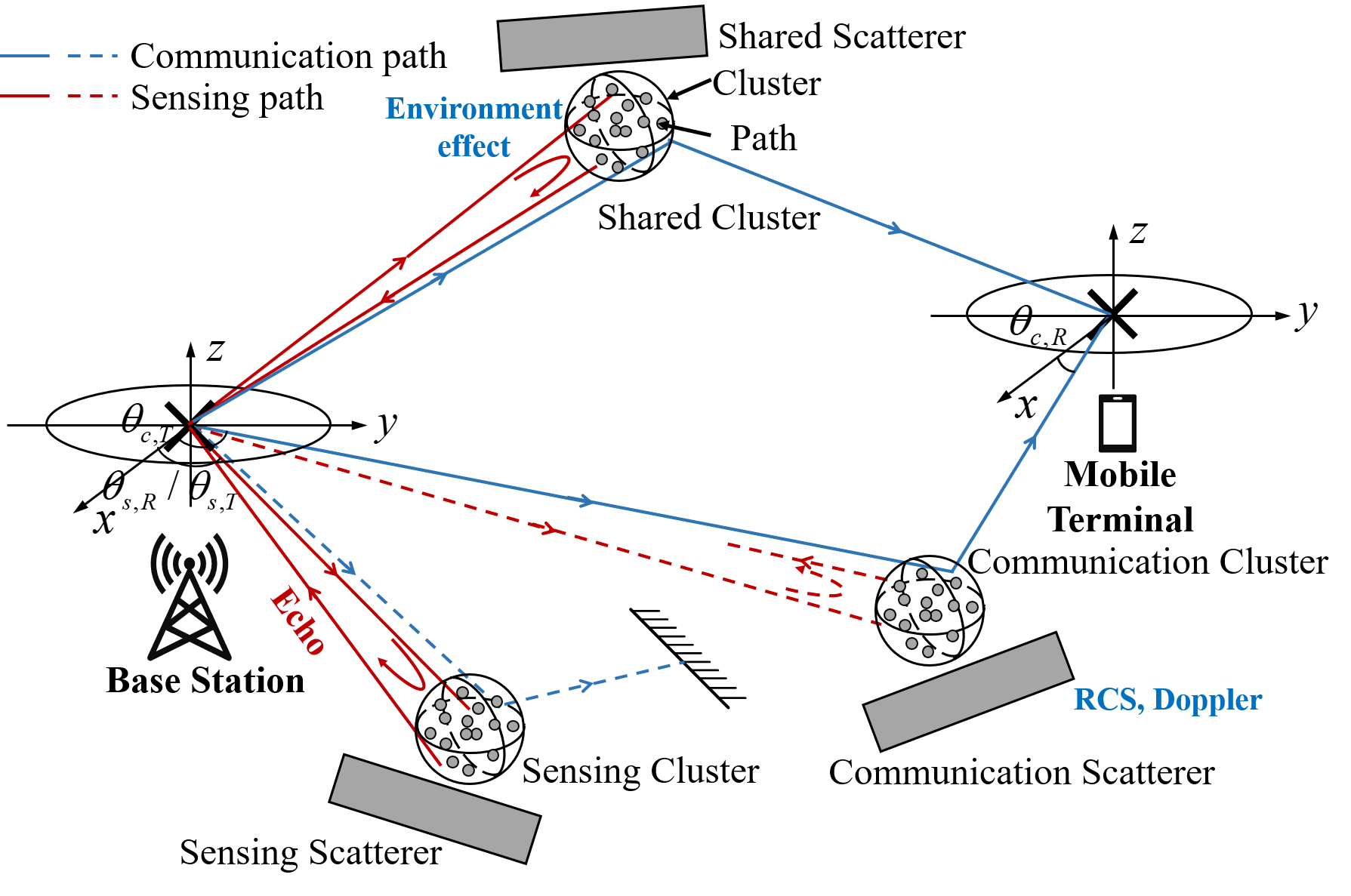}
\caption{Illustration of the ISAC channel model and new characteristics. The dashed lines indicate that the paths are interrupted \cite{ISAC18}.}
\label{fig_jcas3}
\end{figure}

At the 44th meeting of the ITU-R Working Party 5D (WP 5D), ISAC officially became one of the six typical application scenarios of 6G. ISAC technology introduces sensing into communication systems, utilizing the transmission, reflection, and scattering of radio waves to understand the physical world. In communication systems, only the communication channel is considered. For better illustration, Fig.~\ref{fig_jcas3} shows the channel model of the ISAC system to understand the differences between the communication and ISAC channels. This channel contains a typical ISAC scenario with one BS and one mobile terminal. In the subsequent descriptions, Tx is used to describe the BS. Rx\_S and Rx\_C are used to describe the receiving devices of sensing and communication, respectively. When Rx\_S and Tx are in the same position, the sensing signals are scattered by scatterers and returned to Rx at the same angle, creating an echo channel. This mode is referred to as monostatic sensing. However, when Rx\_S and Tx are not in the same position, it is referred to as bistatic sensing. The model takes monostatic sensing as an example: Tx transmits sensing and communication signals and receives sensing signals, whereas the mobile terminal is designed to receive communication signals.

\subsection{ISAC Channel Measurement Progress}
Channel measurement activities are a key step in conducting ISAC channel research, through which new channel characteristics can be discovered and regular conclusions can be summarized. Many studies have measured ISAC channels that focus on various perspectives, such as frequencies, scenarios, and sensing modes. Moreover, analysis has been conducted on ISAC channel characteristics, such as path loss, angular spread, delay spread, and RCS. Because different sensing modes lead to different measurement methods, the measurement activities of ISAC channels are summarized in terms of monostatic and bistatic sensing.

To simulate monostatic sensing, the Tx and Rx-S antennas are placed in the same position during channel measurements, and physical isolation is used to eliminate the mutual influence of the Tx and Rx-S antennas. For example, a research team used this approach to scan the surrounding environment while steering the beam pattern in different directions and observing the target reflections \cite{ISAC12}. Moreover, ISAC measurements were conducted to investigate and model the path loss in the sensing channel under the monostatic sensing mode\cite{ISAC9}. A measurement activity measures the sensing channel to determine if the received power is affected by the sensing target, and the Tx to Rx\_S and Rx\_S to Tx channels will produce self-interference, which should be portrayed in the ISAC channel model \cite{ISAC32}. The range-power profile, range-Doppler spectrum, and range-angle spectrum are obtained and analyzed in monostatic sensing measurements \cite{ISAC24, ISAC15}.

Another mode of sensing is bistatic sensing, where Tx and BS\_S are placed in different positions to sense the environment or target. In \cite{ISAC33}, the cluster power attenuation factor, which is a new parameter of the ISAC channels, was measured in bistatic sensing mode. Small-scale fading is an important characteristic of the channel that has been measured and modeled \cite{ISAC10}. The Doppler characteristics under the bistatic sensing mode have been studied and modeled using measurement data\cite{ISAC22}. To reconstruct the indoor environment, some researchers have combined the angle of departure and angle of arrival information of the sensing channel\cite{ISAC23}. Additionally, the RCS patterns of specific targets (e.g., UAVs and traffic objects) were systematically characterized in the millimeter-wave band, providing comprehensive datasets encompassing polarization, angular dependencies, and statistical distribution properties\cite{UAVRCS}\cite{TrafficRCS}. On the basis of the above measurements, researchers have begun to conduct preliminary research on sensing channel characteristics under different sensing modes. However, the current measurements focus primarily on recognizing traditional features, such as path loss, PDP, angular spread, and delay spread. 
Systematic studies of target scattering, environmental effects, the correlation between communication and sensing, and other new features in ISAC channels are still lacking.

A detailed summary of the measurement campaigns is shown in Table~\ref{JCAStable1}, which outlines the sensing modes, scenarios, and channel characteristics. Table \ref{JCAStable1} provides several key findings. First, many studies have focused on the measurement of ISAC channels at millimeter waves. Since onboard radar primarily operates in the millimeter-wave range, which has short wavelengths and large bandwidths, it enables high-precision sensing and high-speed communication. Second, with respect to measurement methods, frequency domain VNA and correlation-based time-domain channel platforms are the two primary methods used  to measure the ISAC channel \cite{ISAC3}. The pros and cons of the two methods have been well compared and explained in Section \ref{SecXLMIMO}. Third, the measurement scenarios are related primarily to the target types and applications of ISAC \cite{ISAC4}. For example, for the sensing detection and recognition of UAVs and vehicles, measurements are conducted in corresponding low-altitude and outdoor street environments. For gesture sensing, it is necessary to construct finger models and conduct studies on specific scenarios. The measurement of target scattering RCS is typically performed in laboratory environments or anechoic chambers.

\begin{table*}[htbp]
\caption{ISAC channel measurement campaigns}
\centering
\renewcommand\arraystretch{1.25}
\resizebox{\linewidth}{!}{%
\begin{tabular}{c|c|c|c|c}
\toprule\hline
\textbf{Measurement platforms} & \textbf{Frequency/BW (GHz)} & \textbf{Scenarios} & \textbf{Channel characteristics} & \textbf{Reference} \\
\hline
\multirow{6}{*}{Correlation-based}
 & 0.1--3 / 2.9 & Foliage environment & PL, PDP, SF, inter-cluster parameters & \cite{ISAC9} \\
\cline{2-5}
 & 28 / 1 & Indoor office & PL, Doppler, RCS & \cite{ISAC12} \\
\cline{2-5}
 & 70.6 / 2 & Indoor empty room & Power, delay, Doppler, RCS & \cite{ISAC32} \\
\cline{2-5}
 & 73 / 2 & V2V & Doppler, RCS, power & \cite{ISAC24} \\
\cline{2-5}
 & 24 / 0.25 & Laboratory & Doppler, PL, RCS, SnS & \cite{ISAC22} \\
\cline{2-5}
 & 60 / 0.1 & Corridor & Delay, AS, PL, DS & \cite{ISAC23} \\
\hline
\multirow{6}{*}{VNA}
 & 57.5--63.5 / 6 & Indoor lab & PL, RCS, reflection & \cite{ISAC15} \\
\cline{2-5}
 & 2--8 / 0.3 & Anechoic chamber & Inter-cluster parameters, PL, AS, DS & \cite{ISAC10} \\
\cline{2-5}
 & 8--10 & Meeting room, classroom & SnS, PDP, near-field, intra-cluster parameters & \cite{ISAC33} \\
\cline{2-5}
 & 8.2-18 / 9.8 & anechoic chamber & RCS, polarization & \cite{UAVRCS} \\
\cline{2-5}
 & 76-81 / 5 & anechoic chamber & RCS & \cite{TrafficRCS} \\
\hline\bottomrule
\end{tabular}
}
\label{JCAStable1}
\end{table*}

\subsection{ISAC New Channel Characteristics}
The ISAC channel model needs to support both sensing and communication evaluation, and new channel characteristics are emerging, posing significant challenges to channel modeling. For example, the purpose of sensing is to obtain the characteristics of the target, so the channel model needs to include the target, which is not present in communication channels. There may be some multipath overlap between the sensing and communication channels in the environment, and it is necessary to study their sharing characteristics. This section lists the new channel characteristics of the ISAC channel.

\subsubsection{Target scattering and concatenation}
Because the purpose of sensing is to obtain information about a target, target scattering parameters play an important role in sensing channel modeling. In existing research, that models target scattering, the RCS is the main parameter used to describe the target in the sensing channel, which represents the ability of the target to intercept and scatter the signal power. When the signal is incident at different angles, the scattering ability of the target varies, affecting the power of the sensing signal. In general, RCS modeling is based on measurements or deterministic methods, such as approximate solutions to Maxwell's equations and measurements in an anechoic chamber. The traditional Swerling model characterizes the random fluctuation of the target RCS through statistical methods, but it fails to reflect information such as the movement direction of the perceived target. Researchers \cite{ISAC27} have proposed a method for modeling the scattering of targets on the basis of surface roughness, and this method has been verified through actual measurements. Moreover, most RCS studies assume that the electromagnetic waves reaching the target are plane waves under far-field conditions. However, in the application of ISAC, the majority of targets are usually in the near field of the antenna. Researchers \cite{huxidongRCS} have conducted RCS measurements on a variety of sensing targets and reported significant differences in the scattering characteristics of different targets, such as the average scattering power and angular correlation. Establishing an RCS model for each type of target would significantly increase the overhead of the ISAC system. A unified RCS model, which encompasses attributes such as average scattering power and angular correlation, needs to be developed for application to diverse targets within ISAC systems.

Owing to the incorporation of sensing targets, the target channel can be divided into Tx-target and target-Rx subchannels, enabling separate analysis of their propagation characteristics. This overall Tx-target/target-Rx channel can be modeled by the concatenation of the two, which has been empirically validated in \cite{zhang2024cascaded}.

\subsubsection{Environment object}
\begin{figure}[!h]
\centering
\subfigure[]{
\centering
\includegraphics[width=8.5cm]{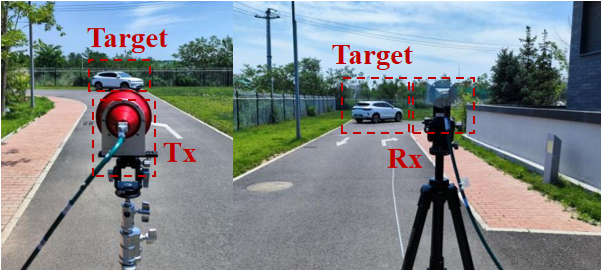}
\label{fig:a}}
\subfigure[]{
\centering
\includegraphics[width=8.5cm]{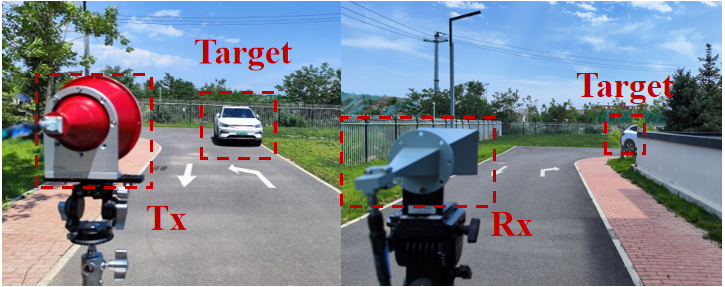}
\label{fig:b}}
\caption{The measurement campaign for urban streets using a car as a sensing target: (a) Case 1 scenario; (b) Case 2 scenario.}
\label{fig:two_figures}
\end{figure}

ISAC enables simultaneous communication and environmental sensing, which is crucial for autonomous driving's high-precision positioning through dynamic maps and positioning in urban environments. However, conventional positioning methods such as time of arrival (ToA) and time difference of arrival (TDoA) perform well under line-of-sight (LoS) conditions \cite{kuutti2018survey}, and their accuracy degrades significantly in urban vehicular scenarios because of multipath propagation caused by scatterers such as buildings and trees. To address this challenge, effectively utilizing non-line-of-sight (NLoS) paths for positioning becomes essential. However, not all NLoS components are equally beneficial, whereas single-bounce reflections from large scatterers such as buildings can significantly improve positioning accuracy. Owing to their strong and deterministic channel characteristics, higher-order multipath components often introduce diffuse scattering and interference. Empirical studies have revealed that NLoS paths contribute substantially to received signal power, playing a crucial role in ISAC systems where sensing targets further complicate channel dynamics. For example, \cite{chen2024empirical} conducted indoor measurements at 105 GHz and demonstrated that human detection is significantly affected by wall-reflected NLoS paths. Extending to outdoor scenarios, \cite{jiang2025novel} investigated NLoS effects in vehicular ISAC positioning through urban street measurements at 26 GHz (Fig. \ref{fig:two_figures}), revealing that environmental objects contribute significantly to received power through strong reflections. These findings emphasize the necessity of explicitly modeling EOs in the target channel.

\subsubsection{Macro- and micro- Doppler shift}
\begin{figure}[!h]
\centering
\subfigure[]{
\centering
\includegraphics[width=0.45\textwidth]{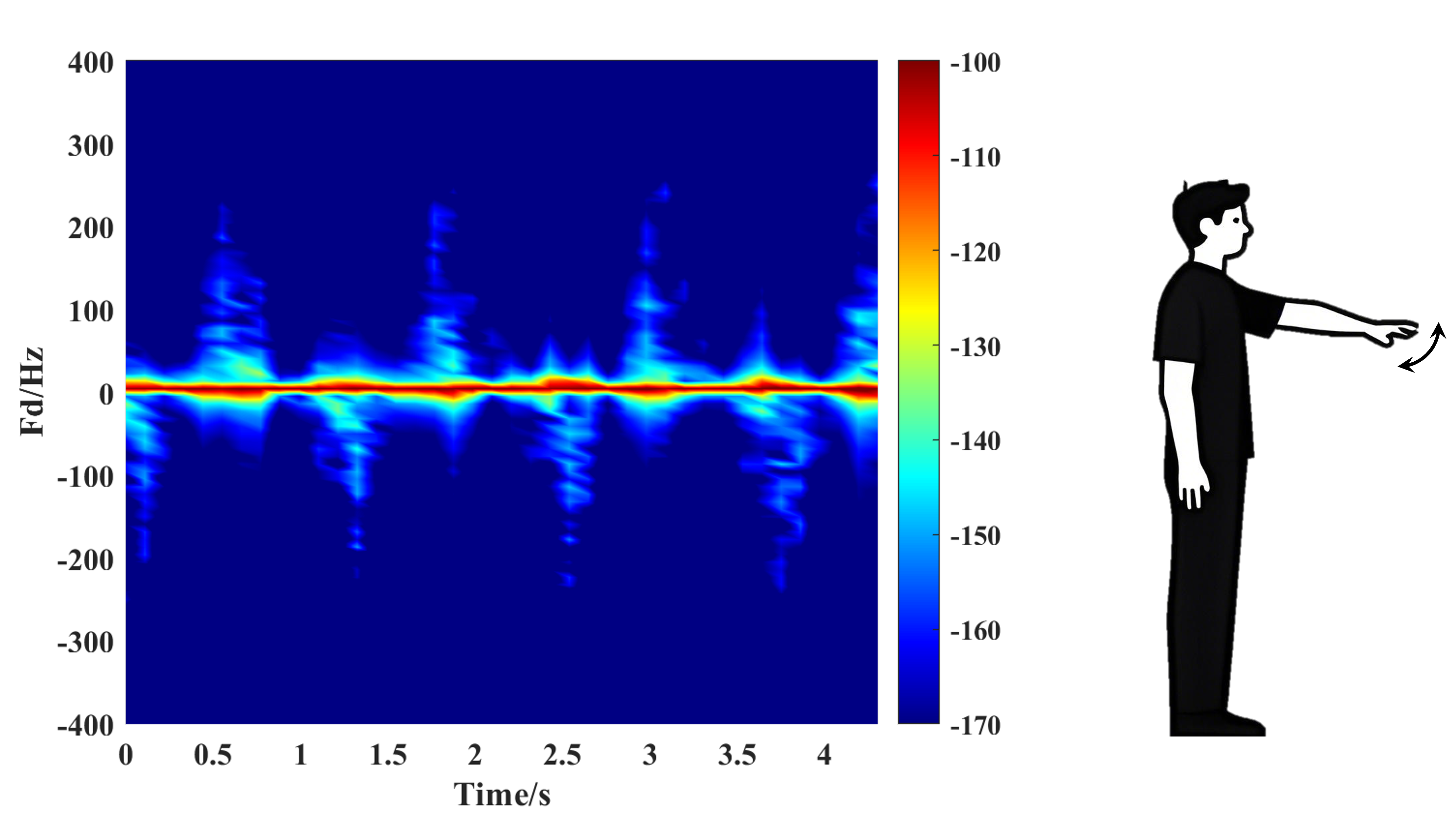} 
\label{Doppler_arm}}
\subfigure[]{
\centering
\includegraphics[width=0.45\textwidth]{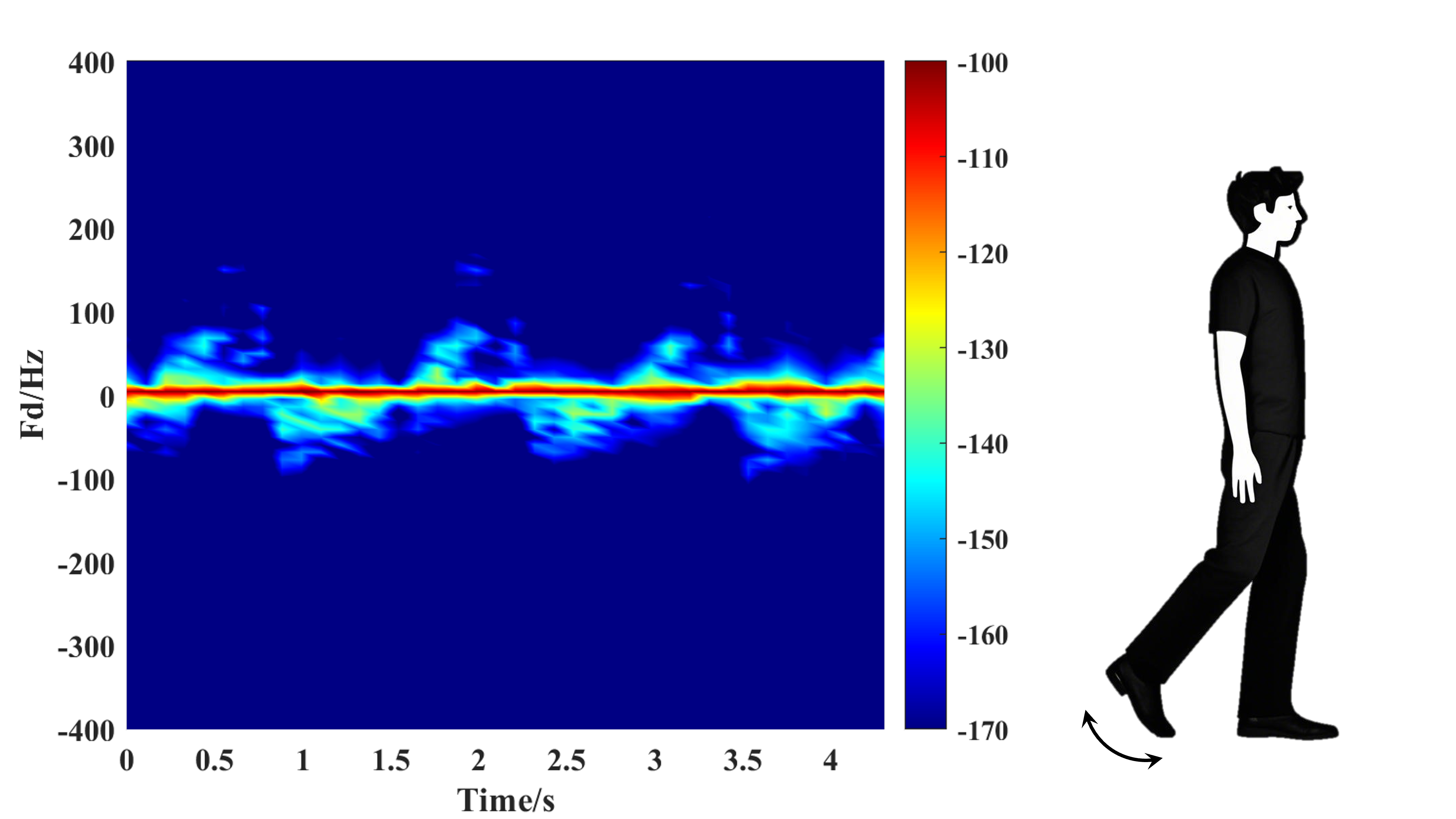} 
\label{Doppler_leg}}
\caption{Illustration of a micro-Doppler shift for human arm and leg vibration \cite{R1-2410659}: (a) micro-Doppler of the human arm; (b) micro-Doppler of the human leg.}
\label{Doppler_human}
\end{figure}

To achieve high-precision sensing, the ISAC system requires support from high carrier frequencies. However, the Doppler effect, which is proportional to the carrier frequency, becomes even more stringent in ISAC systems, particularly in high-mobility scenarios. 
Generally, the Doppler effect can be categorized into the macro-Doppler effect and the micro-Doppler effect. The macro-Doppler effect results from the target's overall motion, producing a frequency shift. The majority of research has focused on outdoor scenarios. 
For example, studies such as \cite{noh2016doppler} and \cite{cha2019doppler} concentrated on V2V scenarios, and investigated the mobility of vehicles and its impact on channel modeling.
Conversely, the micro-Doppler effect arises from subtle vibrations or rotations of structures on the target, leading to fluctuations in the frequency shift.
This effect is pivotal for target sensing in personal area networks (PANs)—short-range networks interconnecting nearby devices—particularly when utilizing mmWave and THz sensing signals.
For example, \cite{shen2022indoor} presented an effective feature-driven method for indoor human activity recognition via mmWave radar. \cite{razavian2022micro} presented a noncontact vibration sensing method using a silicon-based THz pulse. 
Furthermore, significant micro-Doppler shifts can be observed during field measurements. As shown in Fig. \ref{Doppler_human}, measurements of human single-arm and single-leg devices conducted in an anechoic chamber are provided in \cite{R1-2410659}. Here, the horizontal axis denotes time, whereas the vertical axis indicates the corresponding Doppler shifts.
These findings confirm the viability of micro-Doppler effect-based sensing for PAN applications.
Nevertheless, practical ISAC deployments face critical challenges. When implemented in widely adopted orthogonal frequency division multiplexing (OFDM) systems, high-Doppler propagation in sensing channels disrupts subcarrier orthogonality, causing intercarrier interference and compromising speed estimation accuracy.

\begin{figure*}[!t]
\centering
\includegraphics[width=6in]{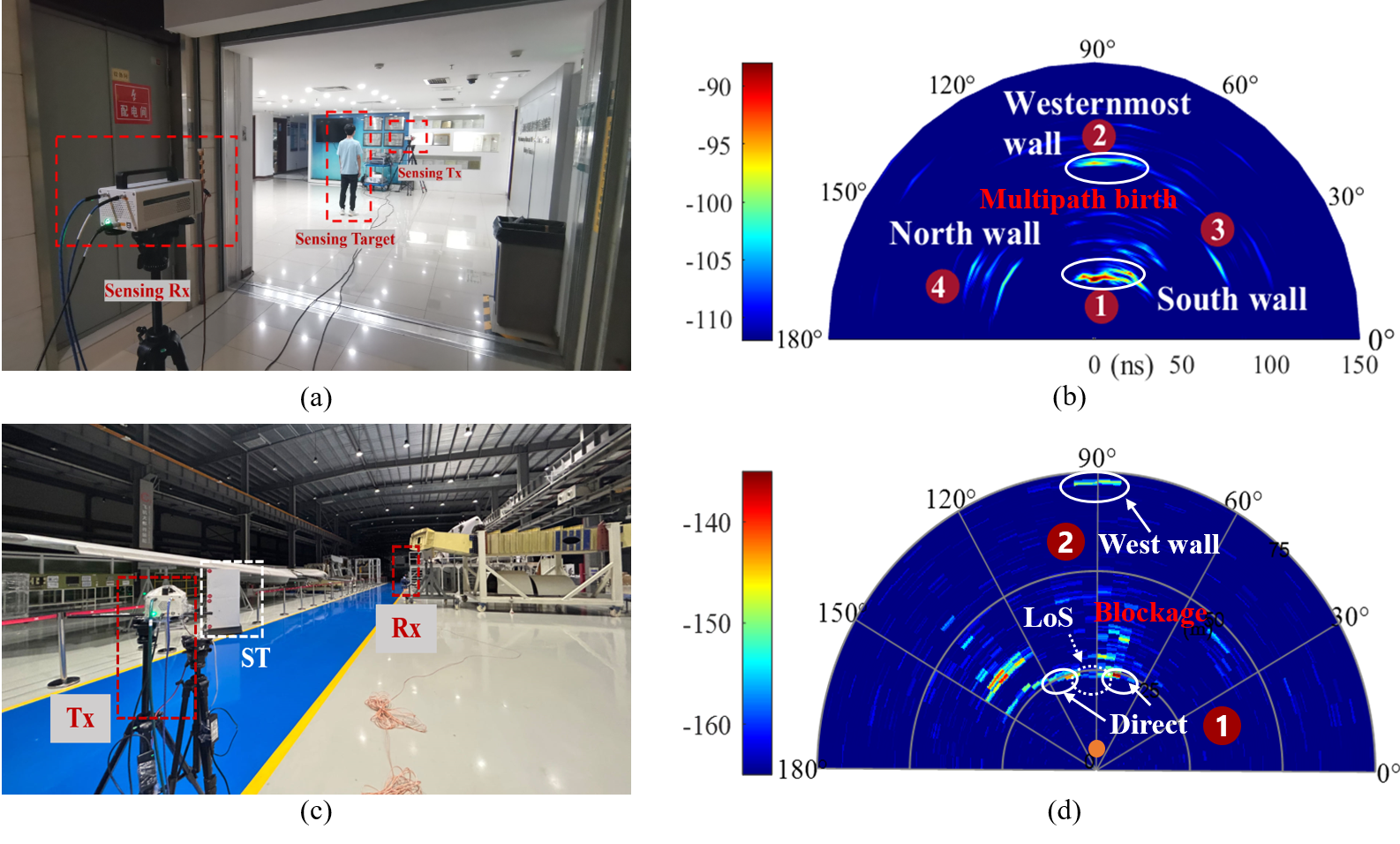}
\caption{Measurements and results of background effect: (a) Measurement environment in an indoor hall; (b) ISAC channel PADP with a human as the target \cite{chen2024empirical}; (c) Measurement scenario in an indoor factory; (d) ISAC channel PADP with an AGV as the target \cite{liu2025coupling}.}
\label{fig_isac_env}
\end{figure*}

\begin{figure}[!t]
\centering
\subfigure[]{
\centering
\includegraphics[width=7.5cm]{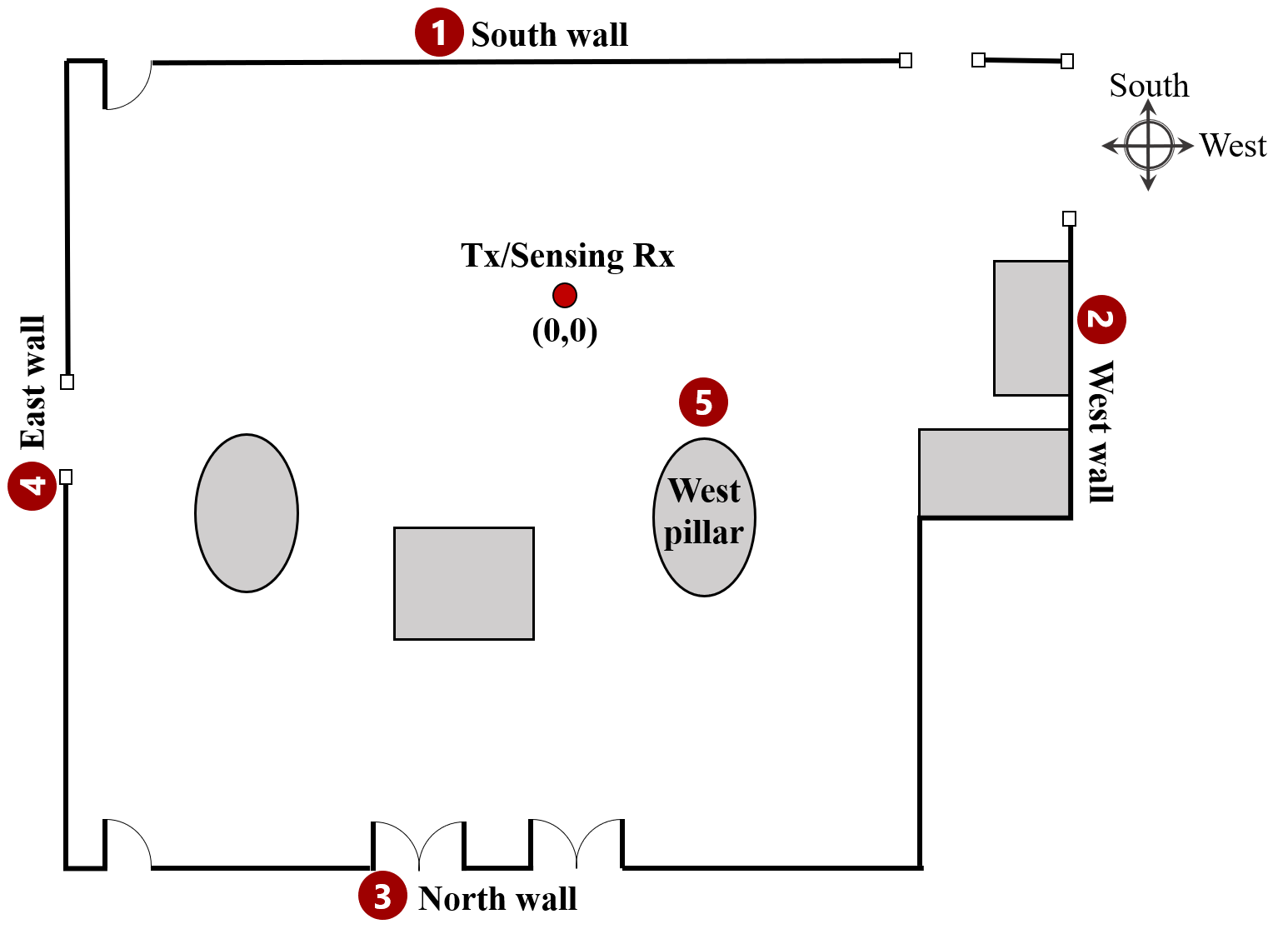} 
\label{ISAC2}}
\subfigure[]{
\centering
\includegraphics[width=7.5cm]{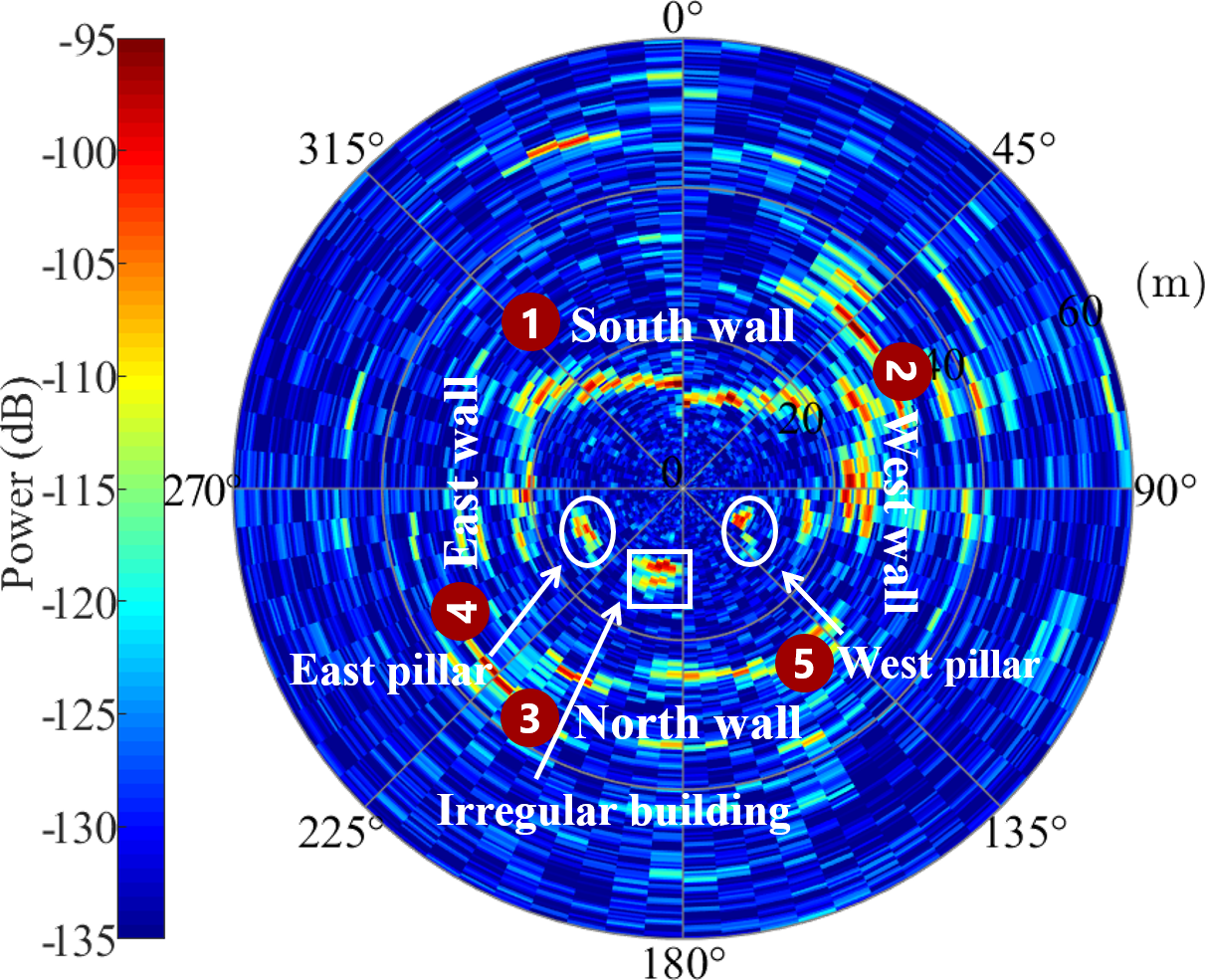} 
\label{ISAC2.1}}
\subfigure[]{
\centering
\includegraphics[width=7.5cm]{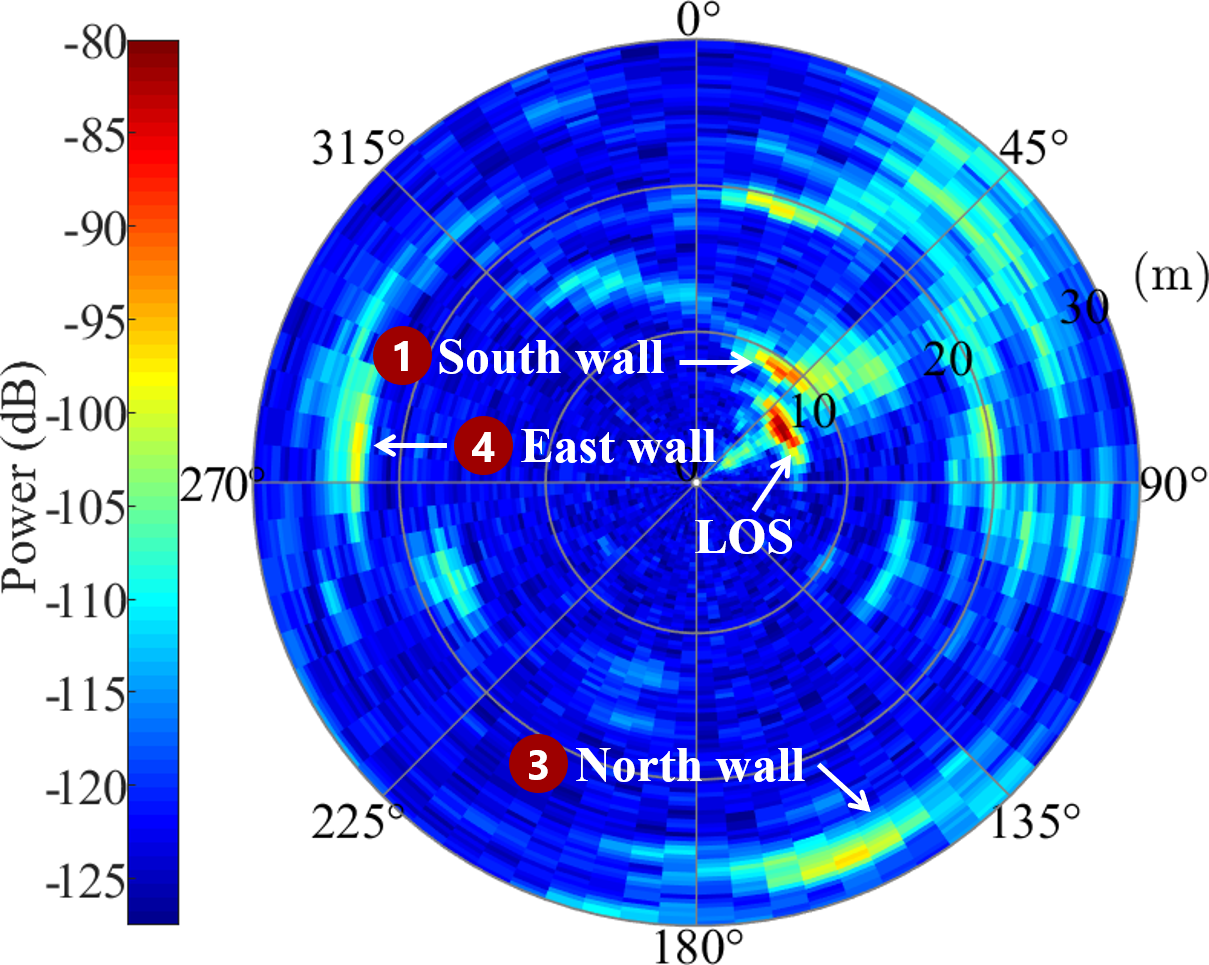} 
\label{ISAC2.2}}
\caption{Measurements and results of shared scatterers: (a) Sensing and communication measurement environment in an indoor hall; (b) Sensing PADP; (c) Communication PADP\cite{ISAC18}.}
\label{fig_jcas2}
\end{figure}

\subsubsection{Background effect}

The ISAC background channel consists of sensing multipaths contributed solely by environmental scatterers, which can affect the sensing performance of the target. Therefore, it is essential to model the background channel and account for the impact of the sensing targets on it. In \cite{chen2024empirical}, indoor channel measurements at 105 GHz were conducted to investigate the background channel with human target. At the Tx, a horn antenna was rotated 180$^\circ$ in 5$^\circ$ increments for measurement, whereas the Rx used an omnidirectional antenna to receive the signal. The measurement setup and power-angular-delay-profile (PADP) are shown in Fig. \ref{fig_isac_env}(a) and (b).
By comparing the measurements with and without the human, variations in the quantity and power of background multipaths are observed. To quantify this impact, a power control factor (PCF) is proposed to evaluate the target-induced changes in background path loss.
Moreover, \cite{liu2025coupling} conducts ISAC channel measurements in an indoor factory scenario at 105 GHz (as shown in Fig. \ref{fig_isac_env}(c)), where the multipath power variations due to the interaction between an automated guided vehicle (AGV) and the background channel were clearly observed, as shown in Fig. \ref{fig_isac_env}(d). Then, a coupled ISAC channel modeling framework was proposed to jointly characterize the target and background channels.
This model was validated through a similarity comparison with measured data, demonstrating 30\% and 5\% higher accuracy for both the LoS and NLoS scenarios than the uncoupled model. These results underscore the importance of characterizing background effects in accurate ISAC channel modeling.

\begin{figure*}[h]
\begin{subequations}
\label{eqn_1}
\begin{align}
\label{eqn_1a}
h_{c}(\theta_{c,R},\theta_{c,T}, \tau_{c})=&\underbrace{\sum\limits_{n_0=1}^{N_0}\sum\limits_{m_0=1}^{M_0} a_{c,n_0,m_0}\delta(\theta_{c,R}-\theta_{c,n_0,m_0,R})\delta(\theta_{c,T}-\theta_{c,n_0,m_0,T})\delta(\tau_{c}-\tau_{c,n_0,m_0})}_\text{Shared\ Communication\ Sub-Clusters}\notag\\&+\sum\limits_{n_1=1}^{N_1}\sum\limits_{m_1=1}^{M_1} a_{c,n_1,m_1}\delta(\theta_{c,R}-\theta_{c,n_1,m_1,R})\delta(\theta_{c,T}-\theta_{c,n_1,m_1,T})\delta(\tau_{c}-\tau_{c,n_1,m_1}),\\
\label{eqn_1b}
h_{s}(\theta_{s,R},\theta_{s,T}, \tau_{s})=&\underbrace{\sum\limits_{n_0=1}^{N_0}\sum\limits_{m_0=1}^{M_0} a_{s,n_0,m_0}\sigma_{n_0,m_0}\delta(\theta_{s,R}-\theta_{s,n_0,m_0,R})\delta(\theta_{s,T}-\theta_{s,n_0,m_0,T})\delta(\tau_{s}-\tau_{s,n_0,m_0})}_\text{Shared\ Sensing\ Sub-Clusters}\notag\\&+\sum\limits_{n_2=1}^{N_2}\sum\limits_{m_2=1}^{M_2} a_{s,n_2,m_2}\sigma_{n_2,m_2}\delta(\theta_{s,R}-\theta_{s,n_2,m_2,R})\delta(\theta_{s,T}-\theta_{s,n_2,m_2,T})\delta(\tau_{s}-\tau_{s,n_2,m_2}).
\end{align}
\end{subequations}
\end{figure*}

\subsubsection{Sharing feature}

In the environment, a segment of scatterers introduces perturbations that can exert a notable influence on the propagation of communication signals, thus potentially impairing transmission quality. Furthermore, these scatterers might concurrently impact the accuracy of the target detection and recognition processes. Therefore, the overlapping scatterers contributing to both the communication and sensing channels are defined as shared scatterers, as shown in Fig.~\ref{fig_jcas3}. These shared scatterers are validated and analyzed through realistic ISAC channel measurements conducted in \cite{ISAC18}. Fig.~\ref{fig_jcas2}(a) shows the layout of the ISAC channel measurement scenario, with primary scatterers identified by red sequence numbers. On the basis of the CIRs obtained through measurements, the PADPs of the sensing and communication channels are calculated, as shown in Fig.~\ref{fig_jcas2}(b) (c). As shown in Fig.~\ref{fig_jcas2}, the sensing multipaths corresponding to the scatterers exhibit a realistic environment, whereas the communication multipaths indicate the dominant scatterers that affect communication signal propagation. A comparison of Fig.~\ref{fig_jcas2}(b) with Fig.~\ref{fig_jcas2}(c) shows that some scatterers are shared between the sensing and communication channels, i.e., the south wall, north wall, and east wall. Moreover, the shared clusters generated by these shared scatterers reveal realistic sharing of communication and sensing channels, which need to be adequately considered in channel modeling for ISAC systems.

\subsection{ISAC Channel Modeling}

To model channels for ISAC systems, it is necessary to extend the channel modeling method of the above 5G system. First, the ISAC channel model needs to include the concatenation of the target channel and target scattering, which affects path loss and can be introduced into the large parameter part of the model. Second, EOs, especially large scatterers with specular reflections should be modeled in the ISAC channel. Third, the Doppler parameter needs to be extended on the basis of the communication model. To determine the background effect, several relevant parameters, such as the PCF and coupling factor, are introduced into the ISAC channel model for large-scale and small-scale characteristics. For the sharing feature of the ISAC channel, a shared cluster-based stochastic channel model was proposed \cite{ISAC18}. This model involves the superposition of shared and nonshared clusters and includes RCS features. The sharing degree (SD) metric is introduced to measure the PR of the shared clusters in the channels. Some studies have analyzed the sensing and communication channels on the basis of similar AoDs \cite{ISAC30}. In addition, a calculation method for the ISAC channel based on sensing and communication clusters was proposed \cite{ISAC31}.

As illustrated in Fig.~\ref{fig_jcas3}, the clusters contributed by the shared scatterers in both the communication and sensing channels are defined as the shared clusters of the ISAC channels. The specific formulas are given in (\ref{eqn_1a}) and (\ref{eqn_1b}). In this model, $h_{c}(\theta_{c,R},\theta_{c,T}, \tau_{c})$ and $h_{s}(\theta_{s,R},\theta_{s,T}, \tau_{s})$ denote the CIRs of communication and sensing, respectively. 
The symbol \(\sigma _{n_{0},m_{0} } \) represents the RCS of the target. It can be modeled according to the method proposed in \cite{huxidongRCS}, which is expressed as \( \sigma_{n_0,m_0} = \sigma_M \sigma_D \sigma_S \). Here, \( \sigma_M \) is a constant value representing the large-scale power of the scattering point. \( \sigma_D \) is either 1 or an angle-dependent function, and \( \sigma_S \) follows a log-normal distribution.
The subscripts $c$ and $s$ denote communication and sensing propagation, respectively. $a$ and $\tau$ are employed to describe the complex amplitude and time delay of the communication and sensing MPCs, respectively. $\theta$ denotes the azimuth transceiver angle of the MPCs, where $\theta_T$ represents the angle of departure and $\theta_R$ represents the angle of arrival. A detailed explanation of this formula can be found in \cite{ISAC18}.

In the proposed ISAC channel model, the sensing and communication channels are generated independently, but they are partially coupled with each other, which provides insights for the subsequent ISAC channel model, which is based on the 5G standard channel model extension. Moreover, owing to the changes in channel topology, parameters, such as delay spread and angular spread in the ISAC channel model, need to be supplemented through channel measurement. Moreover, the aforementioned ISAC channel characteristics are accounted for in channel modeling, which is not currently addressed in communication channel models.

\subsection{Summary and Prospects}

ISAC channel measurement focuses primarily on the frequency or time domain, and measurement methods are classified into monostatic and bistatic sensing. With the integration of sensing and communication, target scattering, concatenation, EO, Doppler, background effect and sharing feature are important in an ISAC channel. To capture these characteristics of ISAC channels, a preliminary statistical channel model was constructed \cite{ISAC18}, and some modeling views were proposed. However, many aspects require further research. This subsection summarizes the current progress of ISAC channel measurement, new characteristics, and modeling and describes several future research directions for ISAC development.

\subsubsection{New measurement scheme}
Channel measurement is an effective method for extracting the characteristics of ISAC channels. Studies have primarily analyzed the characteristics of path loss, angular spread, delay spread, and RCS for ISAC channel measurements. ISAC channel measurement requires the accurate capture of sensing and communication channels in large bandwidth and high-frequency bands, which presents challenges for generating and receiving two channels simultaneously, storing large amounts of data and calibrating the system. Additionally, the shared integration mode remains challenge for future ISAC channel data acquisition, which requires more exploration.

\subsubsection{New characteristics integration}

The study of ISAC channel characteristics is based on practical measurement results. Studies on ISAC channel characteristics have focused primarily on traditional communication channel characteristics, such as delay and angle spread, which are inadequate for a comprehensive understanding of the ISAC channel. Preliminary progress has been made in the study of target scattering, concatenation, EO, Doppler shift, background effect, and sharing feature, as discussed in the previous subsection. Further in-depth analysis is needed in areas such as the frequency and material dependence of the RCS, as well as the micro-Doppler characteristics of UAV wing rotation, etc., to better support sensing performance evaluation for future 6G. Moreover, it is necessary to incorporate these features into the ISAC channel model.

\subsubsection{Channel modeling extension based on the 3GPP framework}

In terms of ISAC channel modeling, the communication geometry-based stochastic model (GBSM) channel model is well developed. Therefore, extension based on the communication channel model is a mainstream method, and 3GPP is working toward achieving this. The ISAC channel model is supported by supplementing the traditional modeling process and parameters. Environmental reconstruction-based channel modeling \cite{ISAC1}, map-based hybrid channel modeling \cite{ISAC29}, and 3GPP extension-based channel modeling \cite{EGBSM} are potential research directions in ISAC channel modeling. Moreover, the combination of ISAC channel modeling with other emerging technologies, such as RIS and non-orthogonal multiple access, is a potential direction for future research.

\section{XL-MIMO Channel Measurement and Modeling}\label{SecXLMIMO}
\begin{figure}[!t]
\centering
\includegraphics[width=0.40\textwidth]{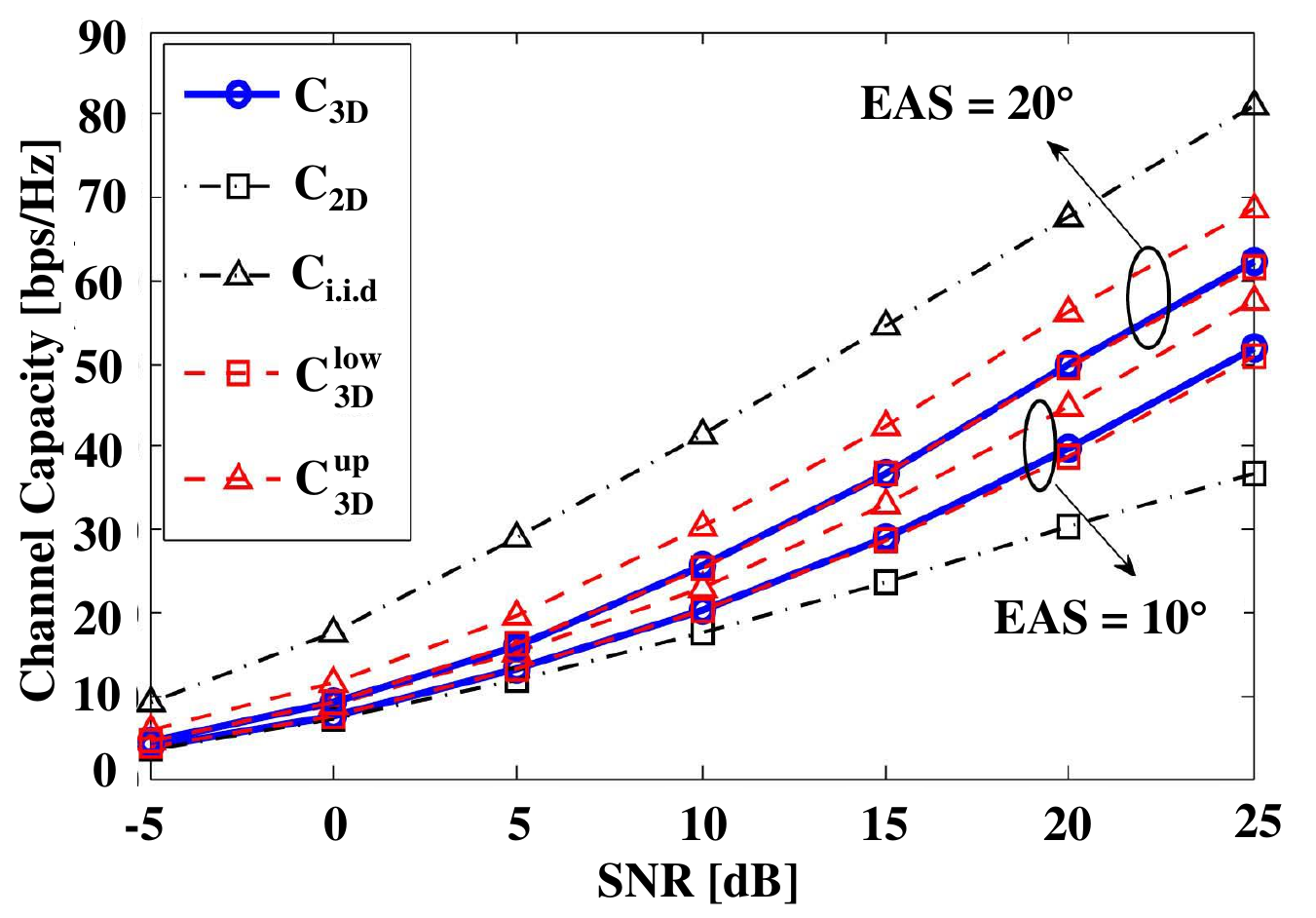}
\caption{Channel capacity: 3D channel model and 2D channel model \cite{MIMO0-3}.}
\label{fig_mimo1}
\end{figure}

The concept of large antenna arrays was proposed in \cite{MIMO0-1}. In \cite{MIMO0-2}, the number of antennas tended to infinity,  and favorable propagation was valid under the condition of a practical angle distribution. In the 5G era, 3D MIMO is considered a promising practical technology, as shown in Fig.~\ref{3Dchannelmodel}. The 3D channel model is an extension of the 2D channel model that includes elevation angles at both departure and arrival, which makes full use of the elevation domain and increases capacity \cite{Zhang3DMIMO}. It is easier to satisfy favorable propagation conditions with a large angular spread in azimuth and elevation. In addition, the spatial correlation of the 3D channel was theoretically analyzed, which showed that the spatial correlation of the 3D channel is inversely proportional to the vertical space and angular spread between the antenna elements. The capacity of the 3D channel is significantly greater than that of the 2D channel \cite{MIMO0-3}. In Fig.~\ref{fig_mimo1}, a large capacity gap between $C_\mathrm{3D}$ and $C_\mathrm{2D}$ can be observed (38.4\% at SNR = 25 dB, EAS = 10$^\circ$) and larger EAS values will expand the gap. The derived tight bounds fit well with $C_\mathrm{3D}$ for both EAS = 10$^\circ$ and 20$^\circ$. In \cite{MIMO1-17}, to further investigate 3D massive MIMO channels, field measurements were made on 32 to 256 antenna units at the transmitter. Based on the channel information extracted from the measurement data, the power angle spectrum, RMS angle spread, and channel capacity were studied. As the number of antennas increases, the angle distribution becomes more diffuse, and the channel capacity increases.

It is well known that massive MIMO technology can significantly increase the wireless transmission data rate, improve the communication quality, and further explore the wireless resources in the space domain. A massive MIMO system can have more than one hundred antennas on the BS side, forming a large-scale multiantenna array. It makes full use of the spatial degree of freedom, which can effectively improve the transmission capacity and power efficiency of wireless communication systems. In the 6G era, the number of antenna elements in XL-MIMO systems is expected to increase compared to massive MIMO and form much larger-scale antenna arrays, such as extremely large aperture array and holographic massive MIMO \cite{VISION}. In some studies, the number of XL-MIMO array elements reached the order of hundreds or thousands \cite{MIMO1-8-1}. Predictably, with rapid technological innovation and enormous commercial demand, massive MIMO will be further expanded, and the number of array elements will range from hundreds to thousands, forming an XL-MIMO array that will change the channel characteristics and models \cite{MIMO1-1-0, MIMO1-16}. Meanwhile, some new characteristics will be introduced, including near-field effect and SnS characteristics. Therefore, this will likely completely reshape the future 6G communication field.

\subsection{XL-MIMO Channel Measurement}
\begin{figure}[!t]
\centering
\includegraphics[width=3.0in]{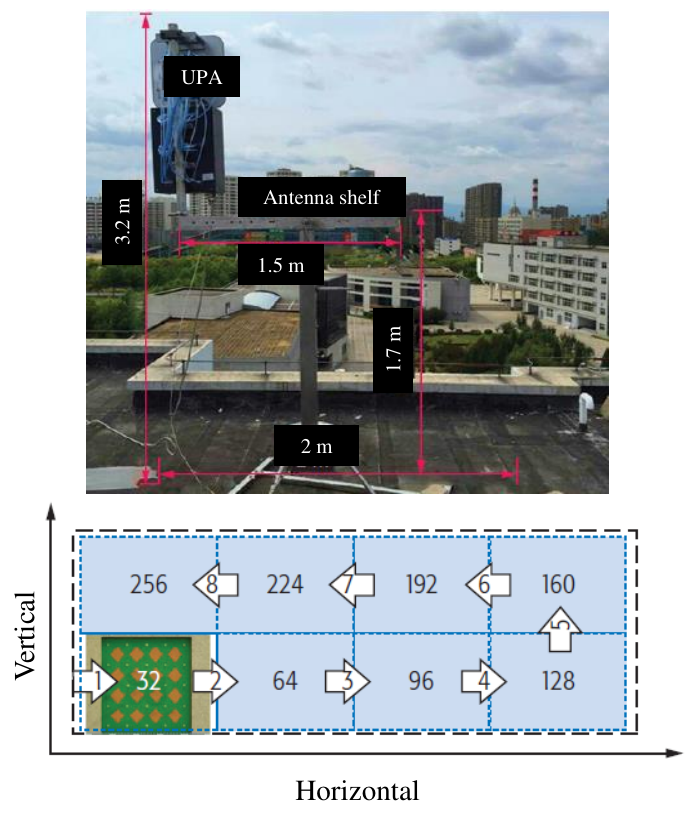}
\caption{A virtual massive MIMO channel measurement \cite{MIMO1-17}.}
\label{fig_mimo2}
\end{figure}

The methods for MIMO channel measurement can be divided into two categories. One is to use a virtual antenna array (VAA), and the other is to use a real antenna array (RAA). Table~\ref{mimotable} summarizes some channel measurements. Only a few channel measurement activities have employed RAA for measurements \cite{MIMO1-18, MIMO1-13-1, MIMO1-34-2} in Table~\ref{mimotable}. However, this method requires an expensive investment and must take into consideration the problem of massive MIMO antenna calibration. Therefore, the VAA is more common in measurements, especially for XL-MIMO with thousands of antenna array elements, as shown in Table~\ref{mimotable}. For VAA measurements, there are two ways to form a VAA. One method is an antenna array composed of a single custom antenna with a mechanical 3D turntable, which has been adopted in many studies due to its convenience and easy operation \cite{MIMO1-14, MIMO1-15, MIMO1-16}. The second method is to use a small antenna array, which is conducive to splicing more array elements in a shorter time, and the measurement quantity is relatively smaller \cite{MIMO1-17}, as shown in Fig.~\ref{fig_mimo2}. In the measurement with a virtual antenna, there are some problems, including the synchronization of the transceiver and receiver and the phase deviation caused by array movement.

In addition to the antenna configuration, there are two main channel measurement platforms, including a frequency domain measurement platform based on a VNA and a time domain measurement platform based on a time-impulse sounder. The measurement system based on VNA has the advantages of high time domain resolution, internal synchronization of the transmitter and receiver, and low complexity. However, this method must connect the transceiver with the VNA, which limits its application in remote outdoor scenarios. To solve this problem, some researchers have proposed prolonging the measurement distance of VNAs by exploiting the low loss characteristics of optical fibers \cite{MIMO1-11}. This system is usually used for XL-MIMO channel measurement in InH scenarios, including conference halls and auditoriums. For the time domain measurement system, the correlation-based channel measurement system has the advantages of instantaneous broadband measurement, fast measurement, and direct acquisition of time domain results. Although a wired connection between the Tx and Rx is not required in this method, synchronization is required between Tx and Rx \cite{Zhang3DMIMO, MIMO1-12}. The system may be suitable for outdoor long-distance XL-MIMO channel measurements. In \cite{MIMO1-16-1}, a larger-scale configuration of two antenna arrays was proposed for monitoring the birth and death of MPCs by comparing the frequency correlation signature of the arrays, thus enabling a reduction in the density of spatial sampling.

\begin{table*}[htbp]
\caption{Massive MIMO and XL-MIMO measurements and antenna configurations}
\label{tab_MIMOsum}
\begin{center}
\begin{tabular}{c|c|c|c|c}
\toprule \hline
 \multicolumn{1}{c|}{Number of array elements} & \multicolumn{1}{c|}{Frequency (GHz)}& \multicolumn{1}{c|}{Antenna} & \multicolumn{1}{c|}{Channel characteristics}  &\multicolumn{1}{c}{Reference}   
\\ \hline
Tx: 16; Rx: 64 & 5.8 &  RAA & \begin{tabular}[c]{@{}c@{}}Near-field: Power, condition number, etc. \end{tabular} &\cite{MIMO1-18} 
 \\ \hline
 Tx: 128; Rx: 1 & 2.6 & VAA, RAA & \begin{tabular}[c]{@{}c@{}}Near-field: Capacity, PAS, \\ singular value spreads, etc. \end{tabular} &\cite{MIMO1-13,MIMO1-13-1} 
 \\ \hline
  Tx: 4; Rx: 128 & 5.3 & RAA & \begin{tabular}[c]{@{}c@{}}Near-field: Correlation, \\singular value, etc.\end{tabular}  &\cite{MIMO1-34-2}
\\ \hline
  Tx: 1; Rx: 128 & 5.6 & \begin{tabular}[c]{@{}c@{}}VAA\end{tabular} & \begin{tabular}[c]{@{}c@{}}Near-field: DS\end{tabular} &\cite{MIMO1-14}
\\ \hline
  Tx: 128 or 64; Rx: 1 or 4 & 26 & VAA & \begin{tabular}[c]{@{}c@{}}Near-field: PDP, DS, \\ coherence bandwidth, etc. \end{tabular} &\cite{MIMO1-15}
  \\ \hline 
  Tx: 256; Rx: 16 & 3.5 & VAA & \begin{tabular}[c]{@{}c@{}} Far-field: PDP, DS, etc.\end{tabular} & \cite{MIMO1-12}
\\ \hline
 Tx: 256; Rx: 16 & 3.5, 6 & \begin{tabular}[c]{@{}c@{}}VAA\end{tabular} & \begin{tabular}[c]{@{}c@{}}Far-field: PAS, AS, \\capacity, etc.\end{tabular} &\cite{MIMO1-17, MIMO1-17-1}
 \\ \hline
  Tx: 1; Rx: 256 & 11 & VAA & \begin{tabular}[c]{@{}c@{}}Near-field: PAS, power, cluster lengths, \\cluster arrival interval, etc. \end{tabular} &\cite{MIMO1-34}
  \\ \hline
  Tx: 301; Rx: 1 & 90-110 & VAA & \begin{tabular}[c]{@{}c@{}}Near-field: PDP, non-linear phase\end{tabular} & \cite{THz_resubmit_14}
  \\ \hline
  Tx: 21$\times$21; Rx: 3$\times$3 & 26-30 & VAA & \begin{tabular}[c]{@{}c@{}}Far-field: PDP, DS, AS, etc. \end{tabular} & \cite{MIMO1-34-1}
    \\ \hline
  Tx: 531; Rx: 1 & 132 & VAA & \begin{tabular}[c]{@{}c@{}}Near-field: PDP, DS, AS, etc.\end{tabular} & \cite{THz_resubmit_13}
\\ \hline
  Tx: 720; Rx: 1 & 26.5-32.5 & VAA & Near-field: CIR, power  &\cite{MIMO1-25}
  \\ \hline
  Tx: 1; Rx: 896 & 158 & RAA & \begin{tabular}[c]{@{}c@{}}Near-field: PDP, PAS \end{tabular} & \cite{THz_resubmit_15}
  \\ \hline
  Tx: 1; Rx: 10$\times$10$\times$10 & 5.15 & VAA & \begin{tabular}[c]{@{}c@{}}Near-field: Spherical wave coefficient \end{tabular} &\cite{MIMO1-16-0}
  \\ \hline
  Tx: 1; Rx: 1600 & 15 & VAA & \begin{tabular}[c]{@{}c@{}}Near-field: K, \\DS, AS\end{tabular} &\cite{MIMO1-16}
    \\ \hline
  Tx: 1; Rx: 1800 & 26-30 & VAA & \begin{tabular}[c]{@{}c@{}}Near-field: CIR, PADP \end{tabular} &\cite{MIMO1-16-1}
\\ \hline 
 Tx: 1; Rx: 2400 & 100 & VAA  & \begin{tabular}[c]{@{}c@{}}Near-field: Delay, PAS, \\path gain, etc. \end{tabular}  &\cite{MIMO1-1-0}
 \\ \hline
  Tx: 64$\times$64; Rx: 1 & 290 & VAA & \begin{tabular}[c]{@{}c@{}}{Near-field: Path gain, non-linear phase}\end{tabular} & \cite{THz_resubmit_16,THz_resubmit_17}
  \\ \hline
  \midrule
      \multicolumn{5}{c}{\begin{tabular}{c} RAA: real antenna array, VAA: virtual antenna array \end{tabular}} 
    
\\ \hline \bottomrule
\end{tabular}
\end{center}
\label{mimotable}
\end{table*}

\subsection{XL-MIMO Channel Characteristics}
Table~\ref{mimotable} shows the analysis of channel characteristics in near-field and far-field region, including power, delay spread, power angular spectrum, etc. In addition, these channel characteristics exhibit the SnS along the array.

\subsubsection{Near-field effect} With the largest phase discrepancy among all BS and user sider antennas reaching $\pi/8$, the distance between the center of a BS array and the center of a user side array is defined as the Rayleigh distance. In general, the Rayleigh distance of an antenna array is defined by $2D^2/\lambda$, where $D$ and $\lambda$ are the sizes of the array and carrier wavelength, respectively \cite{MIMO1-21}, also known as the Fraunhofer distance. Since the number of antennas is not very large in the 5G massive MIMO system, the Rayleigh distance is on the level of a few meters, which is negligible. Therefore, the existing 5G communication is mainly developed from far-field communication theory and technology. However, with the significant increase in the number of antennas and carrier frequencies in future 6G systems, the near-field region of XL-MIMO will be expanded by several orders of magnitude. For example, a 7.3-meter-long virtual linear array was adopted in \cite{MIMO1-13-1}, and the corresponding Rayleigh distance was over 900 m at 2.6 GHz, which is much larger than the radius of a typical 5G cell. Therefore, near-field MIMO communication will become the basic component of the future 6G mobile network, where an accurate channel model considering near-field propagation, such as the spherical wave model, is necessary \cite{mimo_midband}.

In \cite{MIMO1-22}, the sizes of the antenna Fresnel and Fraunhofer field regions were systematically derived starting from a general phase factor representation of the scalar diffraction theory. However, with the introduction of novel technologies, the near-field range in RIS systems was determined by the harmonic mean of the BS-RIS distance and RIS to user equipment (RIS-UE) distance \cite{MIMO1-23}. In addition, as the antenna array size becomes larger with a large number of antenna elements, the distance between the receiver and transmitter may be shorter than the Rayleigh distance, and the far-field and plane-wave front assumptions in the traditional channel model no longer apply to XL-MIMO. The spherical wavefronts should be taken into consideration \cite{MIMO1-24}. The plane wave can be considered a long-distance approximation of a spherical wave. In the far-field region, the phase of the electromagnetic wave can be approximated by a linear function of the antenna index by Taylor expansion. In \cite{MIMO1-23}, with plane wavefronts, far-field beamforming can redirect the beam energy to specific angles at different distances. In the near-field region, the phase of the spherical wave is precisely derived based on physical geometry and is a nonlinear function of the antenna index. The information on the incident angle and distance in each path between the BS and UE is embedded in this nonlinear phase. In addition, in the near-field region, XL-MIMO can present non-stationary characteristics on arrays \cite{MIMO1-25, MIMO1-26, mimo_midband}, which are described in the following section.

\subsubsection{Spatial non-stationarity}
\begin{figure}[!t]
\centering
\includegraphics[width=3.4in]{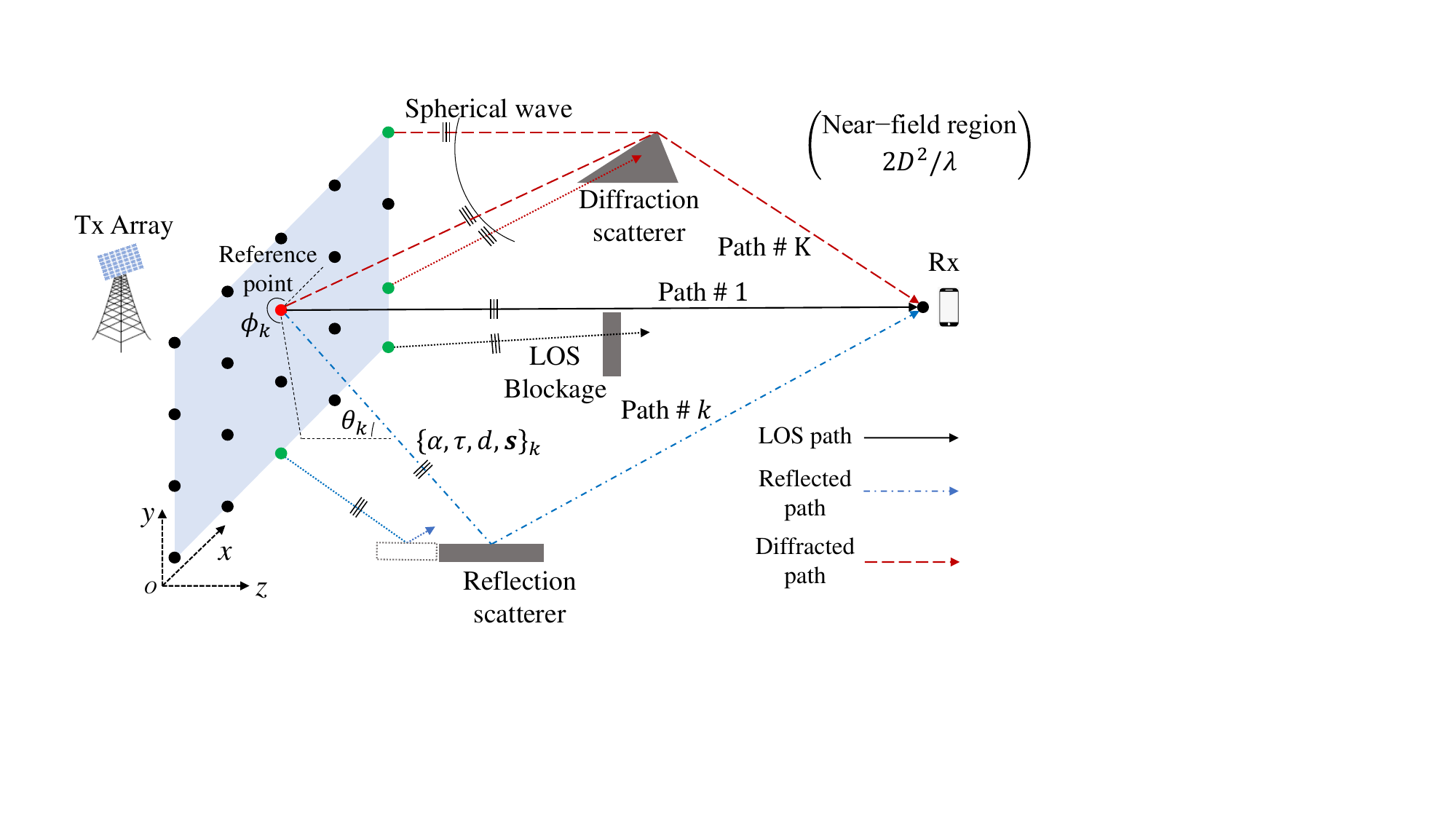}
\caption{Spherical propagation with SnS characteristics in the near-field region. Three cases that contribute to SnS propagation, i.e., LOS blockage, incomplete reflection, and diffraction, are illustrated \cite{MIMO1-25}.}
\label{fig_mimo4}
\end{figure}

Since XL-MIMO systems are equipped with large-scale antenna arrays, the different regions of the array see different propagation paths, resulting in some channel characteristics showing non-stationarity on the array, which is called the SnS characteristic. In \cite{mimo_midband}, the spatial non-stationarity of the channel characteristic parameters on the array was analyzed in the 6 GHz band. In \cite{MIMO1-13}, both linear and cylindrical arrays experienced large-scale fading over the array. In Fig.~\ref{fig_mimo4}, the SnS characteristic is shown in the XL-MIMO channels since the objects in the scenario may no longer serve as complete scatterers for the entire antenna array as its aperture increases. The SnS characteristic of multipaths has been observed in XL-MIMO channel measurements \cite{MIMO1-25}. The different channels are shown by the elements marked in green and the reference element in Fig.~\ref{fig_mimo4}. Therefore, unlike previous MIMO systems, the SnS characteristic is a new challenge in XL-MIMO scenarios and must be considered in channel modeling. The different antennas at the BS can observe different clusters at different times, which can be modeled as a birth-death process \cite{MIMO1-27}. However, for the existing statistical channel models, channel non-stationarity is not considered, or the characterization and interpretation of SnS are not sufficient. Therefore, channel non-stationarity should be handled in the XL-MIMO channel model for 6G \cite{3-2}.

\begin{figure*}[htbp]
\centering
\includegraphics[width=1\textwidth]{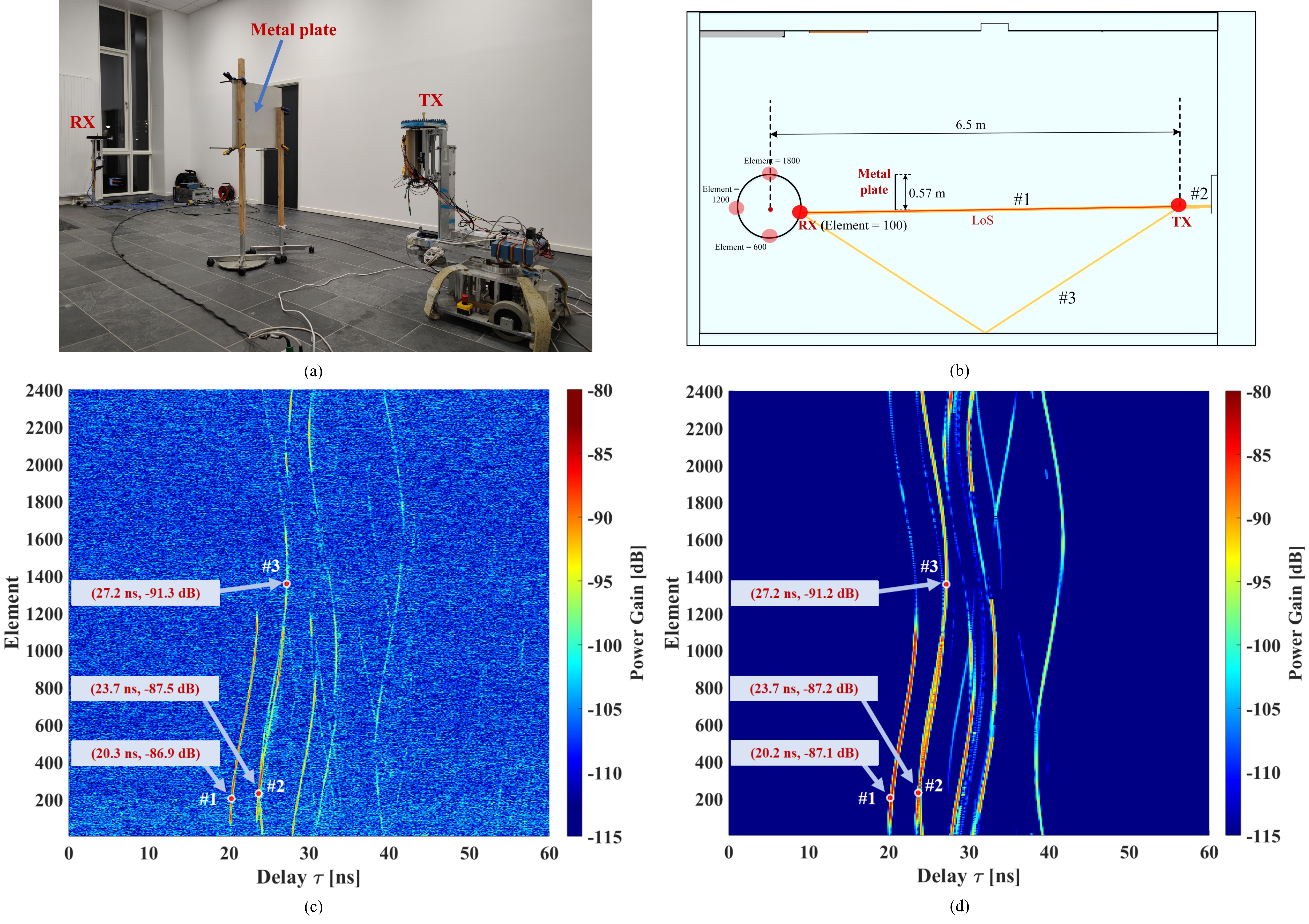}
\caption{SnS measurement and simulation: (a) Rx view in an empty-room scenario; (b) Ray trajectories; (c) Power-delay-element measurement result; (d) Power-delay-element simulation result \cite{MIMO1-1-0}.}
\label{fig_mimo5}
\end{figure*}

\subsection{XL-MIMO Channel Modeling}
The assumption of a plane wave and channel spatial stationary is used in the 3GPP 38.901 channel model of 5G, which is mainly aimed at small-scale MIMO systems \cite{38901}. However, these assumptions will be challenged in massive MIMO systems, especially XL-MIMO systems. The Rayleigh distance of XL-MIMO will become larger in a real deployment scenario. In this case, the UE or scatterer is most likely to be in the near field of the BS. Thus, the more general spherical wave model should be considered for channel modeling.

Statistical channel modeling is an important modeling approach for massive MIMO channel modeling, which can be further divided into the correlation-based stochastic model (CBSM) and geometry-based stochastic model (GBSM). Owing to its low complexity and limited precision, the CBSM is not sufficient for simulating the above non-stationary phenomena or spherical wave effects. The GBSM is more practical because of its relatively higher accuracy. In \cite{MIMO1-30}, second-order approximate channel models of spherical wavefronts (i.e., parabolic wavefronts) for massive MIMO channels were established in both spatial and temporal domains to effectively simulate near-field effects. In \cite{MIMO1-31}, the general 3D non-stationary GBSM for XL-MIMO communication systems was proposed in 6G. Based on the complex principal component analysis (PCA) method in machine learning, the principle involves maximizing the converted channel power, extracting representative channel characteristics, and using these characteristics to ultimately reconstruct the CIR \cite{MIMO1-32}. The proposed method is robust with more antennas, which provides insights for future XL-MIMO channel modeling. In addition, the COST 2100 model is a cluster-level GBSM for the MIMO system, which is the first to use the concept of the visible region (VR) \cite{MIMO1-33} to simulate SnS along the antenna array. The cluster can only be viewed through the antenna element located in the corresponding VR. However, due to the plane wave hypothesis, the modeling precision of SnS characteristics is limited in the COST 2100 model. Based on the VR concept, the GBSM with different frequency bands and scenarios is used to represent SnS and near-field effects for the large-scale array configuration \cite{MIMO1-34}. However, the validity of these models needs to be further verified by more measurements in real scenarios.

To reduce the complexity of XL-MIMO channel modeling, \cite{MIMO1-25} proposed a channel modeling framework based on the capture and observation of multipath propagation mechanisms (i.e., LOS, reflection, and diffraction). In Fig.~\ref{fig_mimo5}, it is shown that spherical propagation with the SnS characteristic is based on measurement and RT simulation. Fig.~\ref{fig_mimo5}(a) shows the measurement environments. Fig.~\ref{fig_mimo5}(b) shows the ray trajectory diagrams of the dominant path. The trajectory color changes in the order of red, orange, yellow, and green, indicating that the received power decreases sequentially. Compared with traditional statistical channel modeling, only one additional SnS parameter is added to the proposed framework, which is required for low-complexity implementation. This method has good real-time performance, low complexity, and accuracy. In \cite{MIMO1-25}, there are $K$ SnS spherical propagation paths between the Tx array and Rx. The XL-MIMO channel at frequency $f$ can be modeled as a superposition of channel frequency responses (CFRs) of the $K$ paths on the array, which is Fourier transform of CIR and can be expressed as

\begin{align}
 \begin{split}
 \mathbf{H}^{\rm sns}(f) & =  \mathbf{S} \odot  \mathbf{A}(f) \cdot  \mathbf{H}(f),
\end{split}   
\end{align}
where $\mathbf{H}^{\rm sns}(f)$ comprises $M$ complex values, i.e., $\mathbf{H}^{\rm sns}(f) \in C^{M\times 1}$, $f \in [f_L , f_U]$ is the frequency within the designed range, and $\odot$ represents the elementwise product operation. $\mathbf{H}(f) \in \mathbb{C}^{K\times 1}$ denotes CFRs at $f$ of the $K$ paths at the reference point (in Fig.~\ref{fig_mimo4}).
\begin{align}
 \begin{split}
 \mathbf{H}(f) & = [\alpha_1e^{-j2\pi f\tau_1},...,\alpha_ke^{-j2\pi f\tau_k},...,\alpha_Ke^{-j2\pi f\tau_K}]^T,
\end{split}   
\end{align}
where $\alpha_k$ and $\tau_k$ represent the complex amplitude and propagation delay of the $k$th path, respectively. $(\cdot)^T$ denotes the transpose operation. $\mathbf{A}(f) \in  \mathbb{C}^{M \times K}$ is the array manifold matrix. The manifold projected on the $m$th antenna element by the $k$th path, i.e., $ \mathbf{A} $'s $ \left( m,k \right)  $th entry $  a_{m,k} $, can be represented by the transfer difference of the $m$th element with respect to the reference point:

\begin{equation}
	a_{m,k}\left( f \right) =\frac{\left\| \boldsymbol{d}_k \right\|}{\left\| \boldsymbol{d}_{m,k} \right\|}e^{-j2\pi f\frac{\left\| \boldsymbol{d}_{m,k} \right\| -\left\| \boldsymbol{d}_{k} \right\|}{\textbf{c}}},
	\label{equ:mod_mani}
\end{equation}
where $ \left\| \cdot \right\| $ represents the Euclidean norm of the argument. $ \boldsymbol{d}_{k} $ denotes the vector pointing from the reference point to the first scattering source of the $k$th path propagation route.

A novel matrix $\mathbf{S}$ is introduced in the proposed modeling framework for the SnS characteristic. It contains $K$ nonnegative real-valued vectors, i.e.,
\begin{align}
 \begin{split}
\mathbf{S} & = [\textbf{\textit{s}}_{1},...,\textbf{\textit{s}}_{k},...,\textbf{\textit{s}}_{K}],
\end{split}   
\label{Eq_SnS}
\end{align}
where $\textbf{\textit{s}}_{k} = [\textit{s}_{1,k}, . . . , \textit{s}_{m,k}, . . . , \textit{s}_{M,k}]^T$ and $\textit{s}_{m,k}$ is the SnS property of the $k$th path on the $m$th element. In Fig.~\ref{fig_mimo5}(c) and (d), the power and delay of the MPCs vary with the element, exhibiting the SnS characteristic. In addition, the delay differences between the three dominant paths are 0.1 ns, and the power differences are within 0.3 dB, which verifies the accuracy of the channel modeling method.

Compared with statistical channel modeling, deterministic channel modeling can accurately capture electromagnetic wave propagation characteristics. The RT method can accurately capture channel characteristics along XL-MIMO antenna arrays, which is a promising channel modeling method for XL-MIMO systems \cite{MIMO1-36, MIMO1-37}. However, only a few studies have implemented RT to characterize channels with large array configurations. The reason is that the arrays on the Tx and Rx sides are usually obtained by strong simulation of each Tx and Rx antenna pair in RT. The computational complexity is particularly challenging for XL-MIMO systems because it increases rapidly with the number of array elements. Recently, the map-based model has been proposed to implement the RT method, which can effectively reduce the computational complexity by simplifying the geometry and electromagnetic description of the environment and restricting the sequence of interactions between rays and objects \cite{MIMO1-38}. The model may have poor accuracy in predicting the channel at specific locations, but it is fully capable of simulating the non-stationarity and transition of large arrays. However, the strategy has difficulty capturing SnS between array elements (i.e., ray birth-death process). In addition, the hybrid modeling approach that combines statistical and deterministic modeling is expected to reduce the complexity of the model because only the dominant path has RT characteristics, but the key issue is the simplicity and accuracy of channel modeling.

\subsection{Summary and Prospects}
The XL-MIMO channel measurement mainly adopts the frequency domain or time domain channel measurement platforms based on the VAA. With the increase in the 6G spectrum, near-field effects and spatially non-stationary characteristics are more likely to appear. To address these characteristics, scientists have carried out statistical and deterministic modeling, and some achievements have been made. However, there are still many unsolved problems and challenges. This section summarizes the channel research method, new characteristics, and modeling method and discusses several future research directions.

\subsubsection{Channel measurements for more scenarios and bands} There are many dynamic application scenarios in 6G, but the current VAA measurement method has difficulty meeting the measurement requirements of the XL-MIMO system in a dynamic environment, and the phase sensitivity of large array antennas should be considered. Besides, 6G has more frequency bands, including mid-band, mmWave, and THz and so on. Channel measurements with larger antenna arrays are needed at these frequency bands.

\subsubsection{Channel characteristics combined with new technology} SnS is an important channel characteristic of XL-MIMO. Improving the description of the SnS characteristic on arrays is also an important topic in 6G channel research. In addition, the impact of novel technologies in near-field communication (such as RIS-assisted near-field communication, THz communication) has not been fully studied. For some performance metrics, such as capacity, the classical Rayleigh distance may not capture the performance loss of these metrics well.

\subsubsection{Channel modeling based on ITU/3GPP framework} In real systems, both far-field and near-field signals are usually present in the communication environment. Compared with the existing GBSM model based on plane waves, the spherical wave model can be introduced to solve the near-field effect, but the model is too complicated. The deterministic modeling approach can accurately capture massive MIMO channel characteristics, but the computational complexity may be high. To balance the relationship between complexity and accuracy or to choose channel modeling methods according to the needs of the system, corresponding solutions should be provided in future studies.

\section{Multi-Frequency Channel Measurements and Models}\label{SecTHz}
\begin{figure*}[]
\centering
\includegraphics[width=18cm]{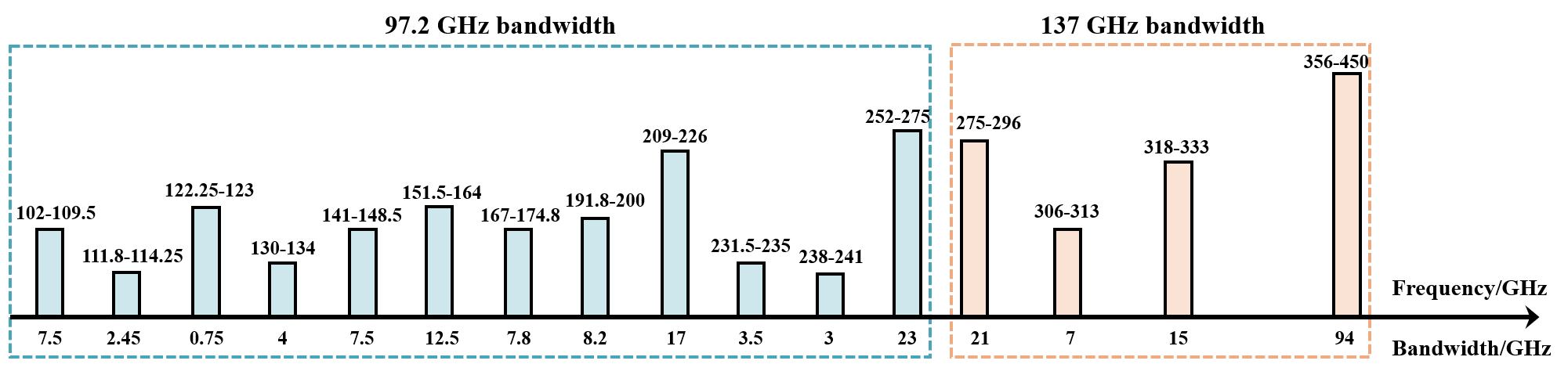}
\caption{Accessible bandwidths for fixed service and land mobile applications from 100 to 450 GHz.} 
\label{THz_spectrum}
\end{figure*}

As presented in Section \ref{section1}, 6G uses a higher frequency spectrum than previous generations \cite{VISION} to achieve multi-Gbps data rate services such as immersive communication. Recently, new frequency bands beyond 5G, such as mid-band \cite{THz_mhy} and THz-band \cite{THz_Sparsityczw}, have been researched to meet the requirements of the high data rates. Among these frequency bands, the new WRC cycle will focus on the following bands for the IMT: 4400-4800 MHz, 7125-8400 MHz, and 14.8-15.35 GHz. In December 2023, the 3GPP TSG RAN Rel-19 discussed the research work in 7--24 GHz band. In addition, the THz band has been identified as a promising spectrum in 6G due to the wide bandwidth and low interference characteristics, and the ITU has shown strong support for the research and development of THz technologies. As shown in Fig.~\ref{THz_spectrum}, the ITU has allocated a total of 234.2 GHz in the 100-450 GHz frequency band for fixed services and land mobile applications. Specifically, 12 bands (total 97.2 GHz bandwidth) within the 100-275 GHz frequency range \cite{THz_ITU_RR} have been allocated by the ITU. Notably, the largest contiguous bandwidth available in the frequency bands is 23 GHz. Below 200 GHz, the largest contiguous bandwidth currently allocated for fixed or mobile applications is 12.5 GHz. The limited availability of contiguous bandwidths poses a challenge for developing high-bandwidth wireless communication systems in 6G. Therefore, according to the ITU Radio Regulations 2020, the WRC 2019 (WRC-19) identified 275-296 GHz, 306-313 GHz, 318-333 GHz, and 356-450 GHz, for a total of 137 GHz, which can be used for fixed service and land mobile applications \cite{THz_275_450}. Moreover, the maximum available continuous bandwidth is 94 GHz. 

The WRC is not the only organization planning for the THz spectrum. Additionally, spectrum planning and allocation regulators in different countries and regions have released parts of the THz band to facilitate scientific research. For example, Japan's Ministry of Information and Communication (MIC) designated the frequency range of 116-134 GHz as ``Commercial Telecommunications Services'' in 2015, and the 152-164 GHz and 287.5-312.5 GHz bands were made available for experimental testing in 2020 \cite{THz_MIC}. In 2018, the European Conference of Postal and Telecommunications Administrations (CEPT) and the European Radiocommunications Committee (ERC) released a recommendation on ``short range devices,'' also known as ``unlicensed devices,'' which specified provisions for these devices to operate within the frequency ranges of 122-122.25 GHz and 244-246 GHz \cite{THz_Eucap}. The Federal Communications Commission (FCC) decision in March 2019 authorized the unlicensed use of 116-123 GHz, 174.8-182 GHz, 185-190 GHz, and 244-246 GHz in the United States \cite{THz_FCC}. In conclusion, high priority has been placed on THz spectrum development and utilization in various countries and organizations.

\subsection{New Mid-band Channel Characteristics} In \cite{midband_survey}, a comprehensive summary of multi-scenario and multi-frequency channel characteristics based on new mid-band (6-24 GHz) channel measurements, including urban macrocell (UMa), urban microcell (UMi), and O2I, was presented. In \cite{THz_mhy}, the presence of some obstacles (e.g., walls, trees, street light poles, and vehicles) in the environment caused the PLEs in the LoS condition to exceed the FSPL model ($n$ = 2). Nevertheless, these values still align well with the PLE ($n$ = 2.1) of UMi-LoS in the 3GPP. A comparison of the PLEs across various LoS conditions and frequency bands reveals that the PLEs are greater in the NLoS condition than in the LoS condition and that they further increase with increasing frequency. This observation indicates that path loss increases at a faster rate with increasing frequency in the NLoS condition. 

Under LoS conditions, although the PLE difference among the four bands is very small, the shadow fading of the 15 GHz band is relatively low. Under NLoS conditions, the shadow fading in the 15 GHz band is relatively low, indicating that the rate of increase in path loss is more comparable than that in the 3.3 GHz band, thus demonstrating good coverage capability. Furthermore, it is obvious that although the coverage capability at 6.5 GHz and 15 GHz is clearly weaker than that at 3.3 GHz, these frequency bands are more capable of achieving stronger signal coverage than the 28 GHz band.

The extensive UMi outdoor propagation measurement campaign at 6.75 and 16.95 GHz was conducted via a 1 GHz bandwidth sliding correlation channel sounder\cite {shakya2024urbanoutdoorpropagationmeasurements}. The path loss index, RMS DS and the mean value of RMS AS were obtained. The authors of \cite{Sun2016} conducted channel propagation measurements ranging from 2 to 73.5 GHz in three typical outdoor scenarios: UMi Street Canyon, UMi open-air Square, and UMa. The results show that, compared with the three-parameter ABG model, the goodness of fit of the single-parameter CI model in the LoS and NLoS environments is very similar.

\begin{figure}[!htbp]
	\xdef\xfigwd{\columnwidth}
	\centering
	\begin{tabular}{c}
        \includegraphics[width=2.6in]{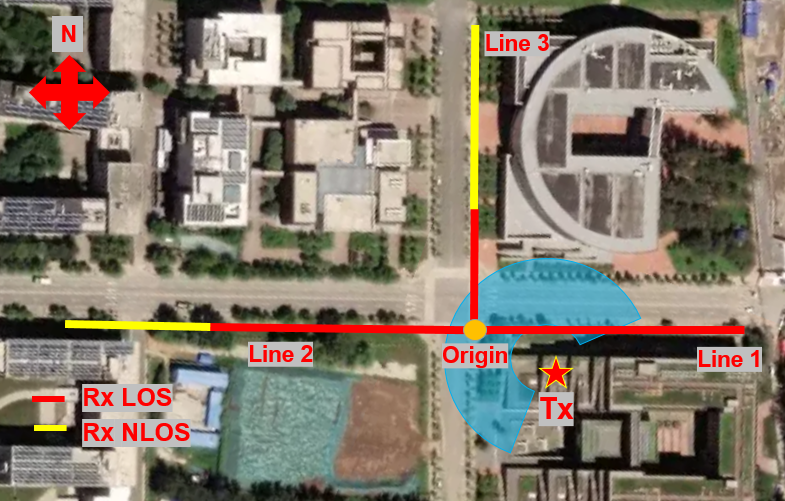}\\
		{\footnotesize\sf (a)} \\[3mm]
        \includegraphics[width=2.6in]{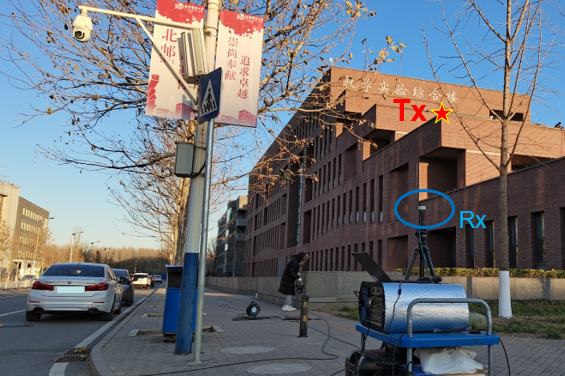}\\
		{\footnotesize\sf (b)} \\
	\end{tabular}
	\caption{The photography of UMi scenario. (a) Layout of UMi scenario. (b) Transceiver antennas location.\cite{THz_mhy}}
	\label{fig:umi}
\end{figure}

\begin{figure}[!htbp]
	\centering
    \includegraphics[width=0.45\textwidth]{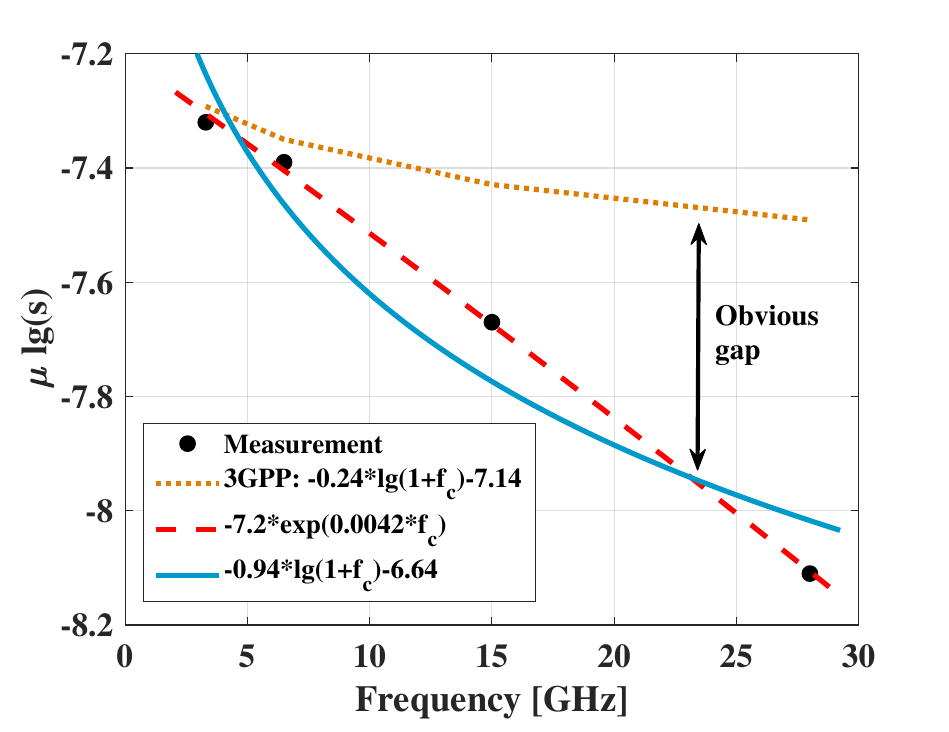}
    \caption{Frequency dependence model for mean of delay spread.\cite{mid-band}}
	\label{figure_rms_delay}
\end{figure}
\par Fig. \ref{fig:umi} and Fig. \ref{figure_rms_delay} show the distributions of the RMS delay spread probability density in the 3.3, 6.5, 15 and 28 GHz frequency bands in the UMi scenario. Fig \ref{figure_rms_delay} shows that the higher the frequency is, the smaller the delay. This is because as the frequency increases, the path loss also increases, resulting in a significant attenuation of multipath power. Furthermore, as the carrier frequency increases, the multipath density decreases and the degree of dispersion increases. Therefore, the channel is controlled by a few effective paths, and most of the energy of the MPCs is concentrated in a relatively small area of the delay domain. In the UMi scenario, the RMS delay spread of 28 GHz is approximately several nanoseconds, which is consistent with the measurement results given in \cite{ying2020analysis}. The average RMS delay spread obtained in the low-frequency band (3.3 and 6.5 GHz) is greater (\textgreater 20 ns). Owing to the relatively large number of obstacles, it indicates that the effective long multipath in the UMi scenario is more abundant. This result is consistent with the propagation characteristics of low-frequency signals, which exhibit a stronger diffraction ability. The UMi scene contains many obstacles, such as trees and buildings. Furthermore, the delay spread difference between the two frequency bands is minimal (\textless 5 ns), indicating similar multipath delay characteristics. Therefore, the 5G system framework structure designed on the basis of channel characteristics of the 3.5 GHz band is applicable to the 6 GHz band in the 5G-Advanced system.

\subsection{THz Channel Measurements}

It is important to conduct extensive THz measurements to build a channel model for THz communication \cite{THz_SCIs}. To date, many research organizations and institutions have conducted multiple THz channel measurement campaigns in the abovementioned frequency bands. We have classified and summarized some representative THz channel measurement campaigns by frequency bands, striving to comprehensively cover all currently measured THz frequency bands (100-1100 GHz), as shown in Table~\ref{THz_label_measurement_tab}. 

In the THz spectrum allocated for land mobile applications, as illustrated in Fig.~\ref{THz_spectrum}, THz measurement campaigns are relatively concentrated around the frequencies of 100, 140, 220, and 300 GHz \cite{THz_VTC,THz_Ref_3,MIMO1-1-0, THz_Ref_2, THz_Ref_4, THz_Ref_11, THz_Ref_12, THz_Ref_13, THz_Ref_14,THz_Ref_19,THz_Ref_20,THz_Ref_21,THz_Ref_22,THz_Ref_23,THz_Ref_24,THz_Ref_25,THz_Ref_26,THz_Ref_27,THz_Ref_28,THz_Ref_29,THz_Ref_30,THz_Ref_31,THz_Ref_37,THz_Ref_43,THz_Ref_44,THz_Ref_45,THz_Ref_46,THz_Ref_47,THz_Ref_48,THz_Ref_49,THz_Ref_50,THz_Ref_54,THz_Ref_59,THz_Ref_62,THz_Ref_60,THz_Ref_5,THz_Ref_6,THz_Ref_7,THz_Ref_8,THz_Ref_9,THz_Ref_10,THz_Ref_15,THz_Ref_16,THz_Ref_17,THz_Ref_18,THz_Ref_32,THz_Ref_33,THz_Ref_34,THz_Ref_35,THz_Ref_36,THz_Ref_37,THz_Ref_38,THz_Ref_39,THz_Ref_40,THz_Ref_41,THz_Ref_42,THz_Ref_51,THz_Ref_52,THz_Ref_53,THz_Ref_55,THz_Ref_62,THz_resubmit_1,THz_resubmit_4,THz_resubmit_6,THz_resubmit_7,THz_resubmit_8,THz_resubmit_9,THz_resubmit_10,THz_resubmit_11,THz_resubmit_12}. The frequency of 100 GHz represents the upper limit of the ITU channel model. The primary objective of most studies is to investigate the applicability of existing channel models in the THz frequency band. Due to the non-linear factors of atmospheric absorption, the selection of actual frequencies primarily considers the ``atmospheric window'' frequency bands. In the THz frequency band, the ``atmospheric window'' refers to special frequency bands, such as 140 GHz and 220 GHz, where propagation attenuation is relatively low. Hence, these bands have been the focus of extensive research. Additionally, 300 GHz falls within the lower boundary of the THz frequency range defined by the ITU radio regulations, making it a frequency band of significant interest. However, in the 450-1100 GHz band \cite{THz_Ref_42,THz_Ref_56,THz_Ref_57,THz_Ref_58}, there are fewer channel measurement campaigns limited by hardware capabilities, such as instruments and equipment. Due to limitations in transmitted power, the majority of studies focus only on reflection, scattering, and attenuation characteristics.

Second, the measurement platforms in the summarized THz channel measurement campaigns are introduced. Most measurement platforms widely used in THz channel measurement campaigns are frequency domain vector network analyzer (VNA)-based platform \cite{THz_VNA}. In the VNA-based platform, the Tx and Rx are directly connected to the VNA. Owing to its simple structure, synchronized Tx and Rx, large bandwidth, and high dynamic range, this measurement platform is mainstream. However, due to the high loss of the cables, the measurement distance is usually short but can be extended to 100 m by radio-over-fiber (RoF) \cite{THz_Ref_62}. To conduct long-distance channel measurements, a correlation-based time-domain channel measurement platform is used, which operates with the strong autocorrelation of pseudo-random noise (PN) sequence. It has also been used in several THz channel measurements. Since the Tx and Rx are independent of each other and do not require cable connections, the platform is preferable for measurements in outdoor environments \cite{THz_Ref_2}. In addition, THz time-domain spectroscopy system (THz-TDS)-based \cite{THz_TDS} channel measurement platforms are limited by the output power and are commonly used for measuring the reflection and diffraction properties of materials.

Third, the scenarios of the channel campaigns are presented. THz band channel measurements focus on desktops, meeting rooms, offices, data centers, computer motherboards, and indoor factory scenarios in indoor scenarios, and UMi, canyons, and streets in outdoor scenarios, covering vision scenarios, such as wireless local area networks (WLANs), wireless personal area networks (WPANs), data center networks (DCNs), and intradevice communication (IDCs) \cite{THz_scenario}. However, wireless backhaul/fronthaul scenarios have been poorly studied \cite{THz_backhaul}. Most of the channel measurement campaigns in these scenarios are in the 100-450 GHz frequency band due to the limited hardware performance in the high-frequency bands. Furthermore, studies include channel characteristics, which will be described in detail in the next subsection.

\begin{table*}[]
    \centering
    \caption{Summary of main THz channel measurement campaigns}
    \begin{tabular}{ c| c |c |c |c }
    \toprule
    \hline
         {\begin{tabular}{c} Frequency\\/BW  (GHz)\end{tabular} } & {\centering \begin{tabular}{c} Measurement \\ platforms \end{tabular}} & {Scenarios} & {\centering Channel characteristics} & {\centering Reference} \\ \hline
         \midrule  
        \multicolumn{5}{c}{\begin{tabular}{c} 100 - 450 GHz \end{tabular}}
        \\ \hline
        
        \begin{tabular}{c} 100/1.2 \end{tabular}  &  {\begin{tabular}{c}Correlation- \\ based \end{tabular}} & {\begin{tabular}{c} Office \end{tabular}} &  {\begin{tabular}{c} PDP, PL, DS \end{tabular}} & \cite{THz_VTC}
         \\ \cline{1-5}
        
        {\begin{tabular}{c} 100/6\end{tabular}} & VNA+RoF & Empty room & SnS & \cite{MIMO1-1-0} 
        \\ \cline{1-5}
         {\begin{tabular}{c} 120/40\end{tabular}} & VNA & Computer motherboard & PL, DS, reflection/penetration loss & \cite{THz_Ref_21}
         \\ \hline
         {\begin{tabular}{c} 132/1.2\end{tabular}} & {\begin{tabular}{c}Correlation- \\ based \end{tabular}} & Indoor factory, UMi & {\begin{tabular}{c}Cross-polarization discrimination,  \\ PL, DS, K, AS\end{tabular}} & \cite{THz_resubmit_11,THz_Sparsityczw}
         \\ \hline
         {\begin{tabular}{c} 135/10 \end{tabular}} & \multirow{3}{*}[-1ex]{\begin{tabular}{c} VNA \end{tabular}} & Data center & {\begin{tabular}{c} PL, SF, DS, AS, cluster numbers, \\ intra/inter-cluster parameters
        \end{tabular}} & \cite{THz_Ref_20}
        \\ \cline{1-1} \cline{3-5}
         {\begin{tabular}{c} 140/20 \end{tabular}} &  & Indoor conference room & {\begin{tabular}{c}Antenna impact on PL, DS, and K\end{tabular}} & \cite{THz_resubmit_10}
        \\ \cline{1-1} \cline{3-5}
         {\begin{tabular}{c} 140/60 \end{tabular}} &  & Indoor, outdoor rooftop & {\begin{tabular}{c} PL, SF, DS, snow attenuation\end{tabular}} & \cite{THz_Ref_4}, \cite{THz_resubmit_7}
         \\ \hline
        \begin{tabular}{c} 140/1.536 \end{tabular}  &  {\begin{tabular}{c}Correlation- \\ based \end{tabular}} & {\begin{tabular}{c} Laboratory  \end{tabular}} &  {\begin{tabular}{c} PL, SF, K\end{tabular}} & \cite{THz_resubmit_1}
         \\ \cline{1-5}
          \multirow{3}{*}[-3ex]{\begin{tabular}{c} 142/1 \end{tabular}} & \multirow{3}{*}[-3ex]{\begin{tabular}{c}Correlation- \\ based \end{tabular}} & {\begin{tabular}{c} Office, conference, \\ classrooms, hallways, \\ indoor factory \end{tabular}}& {\begin{tabular}{c} PL, SF, DS, AS, cluster numbers, \\  intra/inter-cluster parameters\end{tabular}} & {\begin{tabular}{c}\cite{THz_Ref_44,THz_Ref_45}, \\ \cite{THz_resubmit_9}\end{tabular}}
          \\ \cline{3-5}
             &   & {\begin{tabular}{c} Building materials \end{tabular}} &  {\begin{tabular}{c} Reflection, scattering, \\ transmission, PL\end{tabular}} & \cite{THz_Ref_46}
         \\ \cline{3-5}
            &   & {\begin{tabular}{c} UMi \end{tabular}} &  {\begin{tabular}{c} PL, foliage attenuation, \\ cluster numbers, DS, AS, \\spatial consistency, coverage\end{tabular}} & \cite{THz_Ref_47,THz_Ref_48,THz_Ref_49,THz_Ref_50}
            \\ \cline{1-5}
           \begin{tabular}{c} 142/4 \end{tabular}  & \multirow{3}{*}[-2ex]{\begin{tabular}{c} VNA \end{tabular}}  & {\begin{tabular}{c} Entrance hall, \\ residential street \end{tabular}} &  {\begin{tabular}{c} Power proportion \\ per reflection order \end{tabular}} & \cite{THz_resubmit_12}
         \\ \cline{1-1} \cline{3-5}
           \begin{tabular}{c} 143.1/4 \end{tabular}  &   & {\begin{tabular}{c} Shopping mall, \\ airport check-in hall \end{tabular}} &  {\begin{tabular}{c} PADP, PDP, PAS, PL, DS, AS\end{tabular}} & \cite{THz_Ref_22,THz_Ref_23}
         \\ \cline{1-1} \cline{3-5}
           \begin{tabular}{c} 144.75/7.5 \\ 140.5/1 \\145.5/1  \end{tabular}  &   & {\begin{tabular}{c} Outdoor (courtyard, \\linear route, \\crossroad, canyon) \end{tabular}} &  {\begin{tabular}{c} PDP, PL, SF, DS, AS, \\ Power distribution of MPC, \\Qtapnumber\end{tabular}} & {\begin{tabular}{c}\cite{THz_Ref_25,THz_Ref_26} \\ \cite{THz_Ref_28,THz_Ref_30} \end{tabular}}
         \\ \cline{1-5}

        \begin{tabular}{c}  205/8 \end{tabular}  &  VNA & {\begin{tabular}{c} Indoor (meeting room, \\ office room) \end{tabular}} &  {\begin{tabular}{c} PL, SF, DS, AS, \\K, cluster numbers, \\ intra/inter-cluster parameters\end{tabular}} & {\begin{tabular}{c}\cite{THz_Ref_14}  \end{tabular}} 
        \\ \hline
        \begin{tabular}{c} 220/1.536 \end{tabular}  &  {\begin{tabular}{c} Correlation- \\ based \end{tabular}} & {\begin{tabular}{c} Laboratory, campus \end{tabular}} &  {\begin{tabular}{c} PL, SF, K, DS, AS\end{tabular}} & \cite{THz_resubmit_1}
        
        \\ \cline{1-5}    
           \begin{tabular}{c} [220:10:330]/2 \end{tabular}  &  {\begin{tabular}{c}Correlation- \\ based \end{tabular}} & {\begin{tabular}{c} Desktop \end{tabular}} &  {\begin{tabular}{c} PL, celluar performance, \\ reflection\end{tabular}} & \cite{THz_Ref_2,THz_Ref_3}
         \\ \cline{1-5}

           \begin{tabular}{c} 275/110 \end{tabular}  &  VNA+RoF & {\begin{tabular}{c} Hall \end{tabular}} &  {\begin{tabular}{c} PDP, PAS, PADP\end{tabular}} & \cite{THz_Ref_62}
         \\ \cline{1-5}       
        \begin{tabular}{c} 280/120 \end{tabular}  &  VNA & {\begin{tabular}{c} Download kiosk \end{tabular}} &  {\begin{tabular}{c} DS, XPR, K, \\Coherence bandwidth\end{tabular}} & \cite{THz_Ref_37}
        \\ \hline
        \begin{tabular}{c} 300/2 \end{tabular}  &  {\begin{tabular}{c}Correlation- \\ based \end{tabular}} & {\begin{tabular}{c} UMi (courtyard),\\ industrial environment\end{tabular}} &  {\begin{tabular}{c} PDP, PL, DS, AS\end{tabular}} & \cite{THz_Ref_51,THz_Ref_53}
        \\ \hline
        \begin{tabular}{c} 300/2 \end{tabular}  &  {\begin{tabular}{c}VNA+RoF \end{tabular}} & {\begin{tabular}{c} Spacious hall \end{tabular}} &  {\begin{tabular}{c} PDP\end{tabular}} & \cite{MIMO1-1-0}
        \\ \hline
        \begin{tabular}{c} 304.2/8 \end{tabular}  &  {\begin{tabular}{c}Correlation- \\ based \end{tabular}} & {\begin{tabular}{c} V2V, data center \end{tabular}} &  {\begin{tabular}{c} Blockage, PDP, PL, DS, AS, \\ scattering\end{tabular}} & \cite{THz_Ref_33}, \cite{THz_resubmit_8}
        \\ \hline
        \begin{tabular}{c} 305/10 \end{tabular}  & \multirow{6}{*}[-3ex]{\begin{tabular}{c}VNA \end{tabular}} & {\begin{tabular}{c} Desktop,\\ building material \end{tabular}} &  {\begin{tabular}{c} PL, absorption coefficients\end{tabular}} & \cite{THz_Ref_32}
        \\ \cline{1-1} \cline{3-5}
        \multirow{3}{*}[0ex]{\begin{tabular}{c} 310/20 \end{tabular}}  &   & {\begin{tabular}{c} Desktop \end{tabular}} &  {\begin{tabular}{c} PDP, PL, SF, RMS DS, TCF\end{tabular}} & \cite{THz_Ref_5}
        \\  \cline{3-5}
           &   & {\begin{tabular}{c} Computer motherboard \end{tabular}} &  {\begin{tabular}{c} PL, SF, reflection\end{tabular}} & \cite{THz_Ref_6}
        \\  \cline{3-5}
          &   & {\begin{tabular}{c} Data center \end{tabular}} &  {\begin{tabular}{c} PL, reflection\end{tabular}} & \cite{THz_Ref_7}
        \\  \cline{1-1} \cline{3-5}
          \multirow{2}{*}[-2ex]{\begin{tabular}{c}313.5/15 \end{tabular}}&   & {\begin{tabular}{c} Indoor (hallway, \\corridor, lobby) \end{tabular}} &  \multirow{2}{*}[0ex]{\begin{tabular}{c} PL, SF, DS, AS, K, \\cluster numbers, \\ intra/inter-cluster parameters\end{tabular}} & {\begin{tabular}{c}\cite{THz_Ref_15,THz_Ref_16},\\ \cite{THz_resubmit_4}\end{tabular}}
         \\  \cline{3-3} \cline{5-5}
           &   & {\begin{tabular}{c} Outdoor (campus \\street, atrium) \end{tabular}} &   & \cite{THz_Ref_17,THz_Ref_18}
        \\  \hline
        \bottomrule
        
    \end{tabular}
    \label{THz_label_measurement_tab}
\end{table*}

\begin{table*}[]
    \centering
    \caption*{Table \ref{THz_label_measurement_tab} (Continued): Summary of main THz channel measurement campaigns}
    \begin{tabular}{ c| c |c |c |c }
    \toprule
    \hline
    {\begin{tabular}{c} Frequency\\/BW  (GHz)\end{tabular} } & {\centering \begin{tabular}{c} Measurement \\ platforms \end{tabular}} & {Scenarios} & {\centering Channel characteristics} & {\centering Reference} \\ \hline
         \midrule 
        \multicolumn{5}{c}{\begin{tabular}{c} 450 - 1100 GHz \end{tabular}}
        \\ \hline
        \begin{tabular}{c} 450/500 \end{tabular}  &  THz-TDS & {\begin{tabular}{c} Building material \end{tabular}} &  {\begin{tabular}{c} Diffuse scattering\end{tabular}} & \cite{THz_Ref_42}
        \\ \hline
        \begin{tabular}{c} 625/250 \end{tabular}  &  \multirow{3}{*}[0ex]{VNA} & {\begin{tabular}{c} Desktop \end{tabular}} &  {\begin{tabular}{c} Atmospheric effects, PDP\end{tabular}} & \cite{THz_Ref_58}
        \\  \cline{1-1} \cline{3-5}
        \begin{tabular}{c} 925/350 \end{tabular}  &   & {\begin{tabular}{c} Building material \end{tabular}} &  {\begin{tabular}{c} Penetration loss\end{tabular}} & \cite{THz_Ref_57}
        \\  \cline{1-1} \cline{3-5}
        \begin{tabular}{c} 1000/200 \end{tabular}  &   & {\begin{tabular}{c} Desktop \end{tabular}} &  {\begin{tabular}{c} Ground reflections\end{tabular}} & \cite{THz_Ref_56}
        \\  \hline
        \midrule
                \multicolumn{5}{c}{\begin{tabular}{c} 
                BW: bandwidth, UMi: urban microcell, V2V: vehicle-to-vehicle, PADP: power angular delay profile, \\ 
                  PDP: power delay profile, PL: path loss, SF: shadow fading, K: Ricean K-factor, DS: delay spread, \\ 
                 AS: angular spread, PAS: power angular spectrum, XPR: cross-polarization ratio, TCF: temporal correlation function \end{tabular}} 
        \\  \hline
        \bottomrule
    \end{tabular}
\end{table*}

\subsection{Important Channel Propagation Characteristics in THz Bands}

Current research on THz characteristics can be categorized into two main types \cite{MIMO1-1-0}. The first focuses on the material characteristics at THz bands, which actually have an impact on radio propagation (e.g., reflection) \cite{THz_kurner2021thz}. The second is dedicated to newly emerging channel characteristics in THz bands, such as atmospheric loss and channel sparsity.

\subsubsection{Reflection characteristics}

Research on propagation mechanism characteristics is the key foundation and fundamental premise for THz channel characteristics research. Reflections are fundamental propagation mechanisms that should be thoroughly characterized to establish possible THz wireless links due to the short wavelength introduced in Section \ref{section1}. Therefore, this subsection focuses on the analysis and modeling of the reflection propagation characteristics of THz waves.  

In the THz band, the characteristics of reflection in the case of ideally smooth and homogeneous surfaces can be described by the well-known Fresnel reflection coefficient $\Gamma$ \cite{THz_Reflection_Fresnel}. The Fresnel reflection coefficient expressions for vertical ($\Gamma _\textrm{v}$) and parallel ($\Gamma _\textrm{h}$) polarizations for such smooth surfaces \cite{THz_ljx,THz_ljx_ref} are expressed as:

\begin{equation}
\begin{aligned}
\Gamma _\textrm{v} = \frac{{Z\cos {\theta_\textrm{in}} - {Z_0}\cos {\theta _\textrm{tr}}}}{{Z\cos {\theta _\textrm{in}} + {Z_0}\cos {\theta _\textrm{tr}}}},
\end{aligned}
\end{equation}

\noindent
and

\begin{equation}
\begin{aligned}
\Gamma _\textrm{h} = \frac{{Z\cos {\theta _\textrm{tr}} - {Z_0}\cos {\theta _\textrm{in}}}}{{Z\cos {\theta _\textrm{tr}} + {Z_0}\cos {\theta _\textrm{in}}}},
\end{aligned}
\end{equation}

\noindent
where $\theta _\textrm{in}$ is the incident angle, $\theta_\textrm{tr}$ is the transmitted wave angle, $Z_{0}=\sqrt{\frac{\mu_{0}}{\varepsilon_{0}}} $ denotes the free space impedance, and $Z$ is the wave impedance of the reflected material, calculated as

\begin{equation}
\begin{aligned}
Z = \sqrt {\frac{{{\mu _0}}}{{{\varepsilon _0}({n^2} - {{(\frac{{\alpha \mathrm{c}}}{{4\pi f}})}^2} - j\frac{{2n\alpha \mathrm{c}}}{{4\pi f}})}}}.
\end{aligned}
\end{equation}

\noindent
Here, $\mu_0$ and $\varepsilon _0$ are the free space permeability and permittivity, respectively. $n$ is the complex refractive index of the smooth surface, and $\alpha$ is the absorption coefficient of the incident surface material.

At low frequencies, the surface of a material is considered primarily as a smooth surface. However, as the frequency increases to the THz band, the wavelength is short enough to be very close to the surface roughness of typical materials. Therefore, in the THz band, the material cannot be considered a smooth surface. The modified reflection coefficient is obtained by introducing a roughness factor $\rho$. The Rayleigh roughness factor can be expressed as \cite{THz_ljx_ref}

\begin{equation}
\begin{aligned}
\rho  = \exp ( - \frac{1}{2}{(4\pi {\sigma _h}\frac{{\cos {\theta _\textrm{in}}}}{\lambda })^2}),
\end{aligned}
\end{equation}

\noindent
where $\sigma _h$ denotes the standard deviation height of the material surface roughness. The modified reflection coefficient $\Gamma '$ is expressed as

\begin{equation}
\begin{aligned}
\Gamma ' = \rho \Gamma .
\end{aligned}
\end{equation}

\begin{figure}[htbp]
\centering
\includegraphics[width=8.5cm]{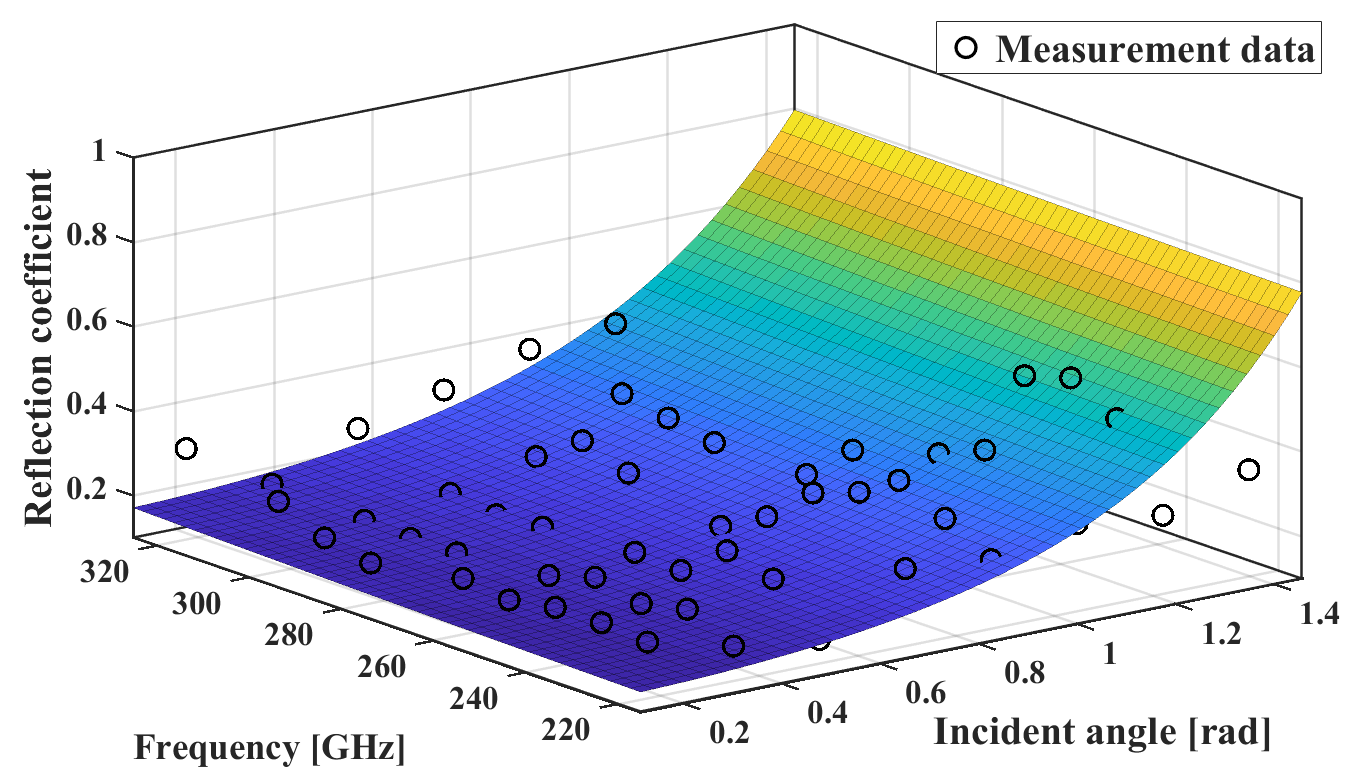}  
\caption{Reflection coefficient measurement data and fitting results for plasterboard from 220 to 320 GHz \cite{THz_Ref_2}.}
\label{THz_CZW}
\end{figure}

The modified reflection coefficient model is a function of the frequency and angle of incidence, and in reference \cite{THz_Ref_2}, this phenomenon was verified by experiments. As an example, based on the modified Fresnel model, frequency-angle two-dimensional reflection coefficient modeling for vertical polarization is proposed. The model of non-metallic materials is written as: 

\begin{equation}\label{e_19}
\begin{split}
\Gamma_{\mathrm{v,FA}}=&e^{(-10^af^2{\rm{cos}}^2\theta_\textrm{in})}\\
&\cdot \left(\frac{{\rm{cos}}\theta_\textrm{in}-\sqrt{1+\frac{10^b}{10^c-df^2-jf}-{\rm{sin}}^2\theta_\textrm{in}}}{{\rm{cos}}\theta_\textrm{in}+\sqrt{1+\frac{10^b}{10^c-df^2-jf}-{\rm{sin}}^2\theta_\textrm{in}}}\right),
\end{split}
\end{equation}

\noindent
where $a$, $b$, $c$, $d$ are the parameters to be fitted, and their means are shown in \cite{THz_Ref_2}. The plasterboard results are shown in Fig.~\ref{THz_CZW}. The analysis revealed a low root-mean-square (RMS) error of approximately 0.1 for the glass, tile, board, and plasterboard materials, indicating a fit between the points of the measured data and the modified reflection coefficient model. The reflection coefficient of the material and its growth rate increased with increasing angle of incidence. Therefore, the model characterizes the reflection coefficient as a function of the frequency and angle of incidence.

\subsubsection{Path loss}
\begin{figure}[]
\centering
    \centering
    \includegraphics[width=7.5cm]{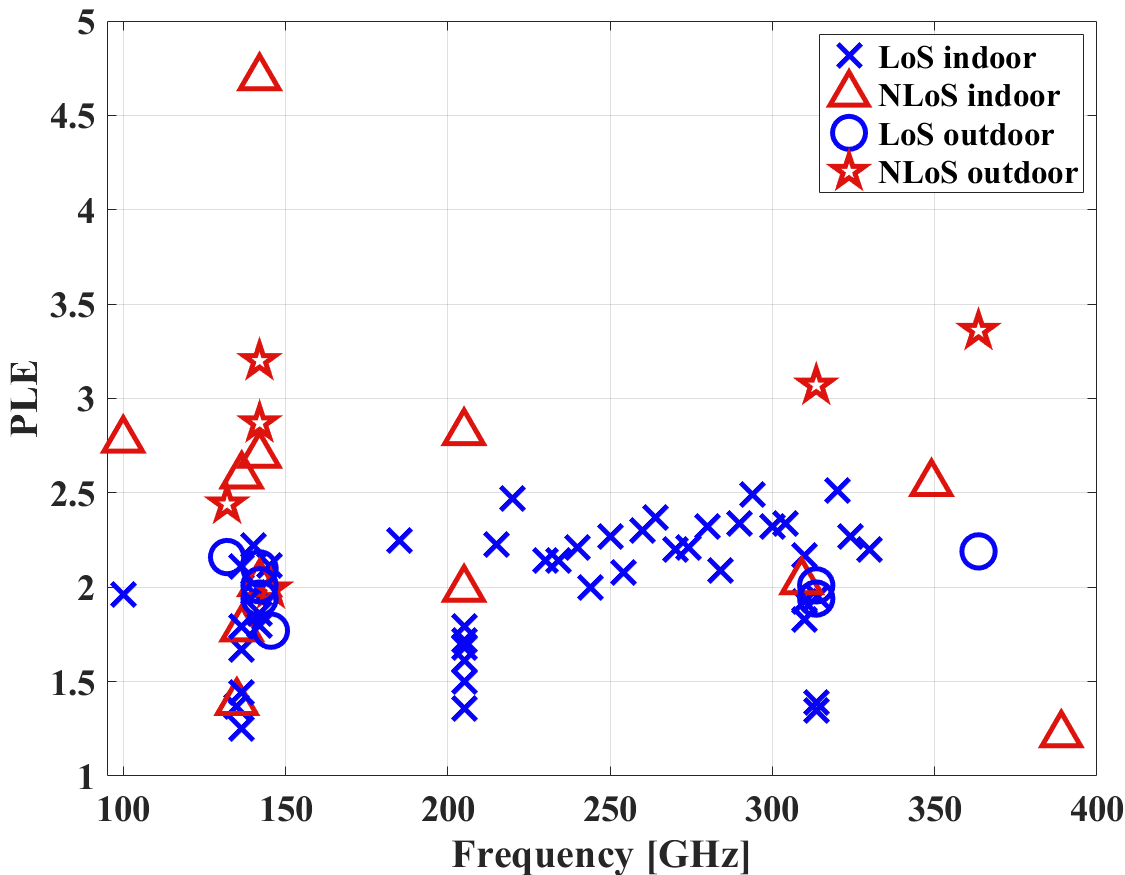}  
\caption{Representative results of PLE based on measurements of THz channels.}
\label{THz_PLE}
\end{figure}

THz signals transmitted over channels have distinct characteristics compared to lower frequencies, leading to novel and urgent challenges \cite{38901}. Representatively, the spreading loss increases quadratically with frequency. Besides, with the THz band operating at exceptionally high frequencies, signals transmitted over a THz link are severely affected by the spreading loss \cite{THz_Ref_3}. 
The channel characteristics of high propagation loss can be characterized by path loss. A novel index is essential for characterizing and modeling path loss in the THz channel.
The modeling of path loss in the THz band follows 5G modeling methods, such as the FI model, ABG model, and CI model shown in \eqref{equadd2}, \eqref{equ3}, and \eqref{equadd1}, respectively, in Section II.A. Based on the references in Table~\ref{THz_label_measurement_tab}, the PLE of the CI model $\alpha_{\mathrm{CI}}$ for indoor and outdoor scenarios is summarized in Fig.~\ref{THz_PLE}. The PLE of the THz band is close to 2 in most cases, especially in LOS scenarios. In addition, the PLEs of the NLOS scenarios are markedly higher than those of the LOS scenario, even up to 4.6. This is because the blocking losses caused by walls, cars, etc., can be as large as tens of dB in NLOS scenarios, which means that severe attenuation can occur.

\begin{figure}[htbp]
\centering
\includegraphics[width=8.5cm]{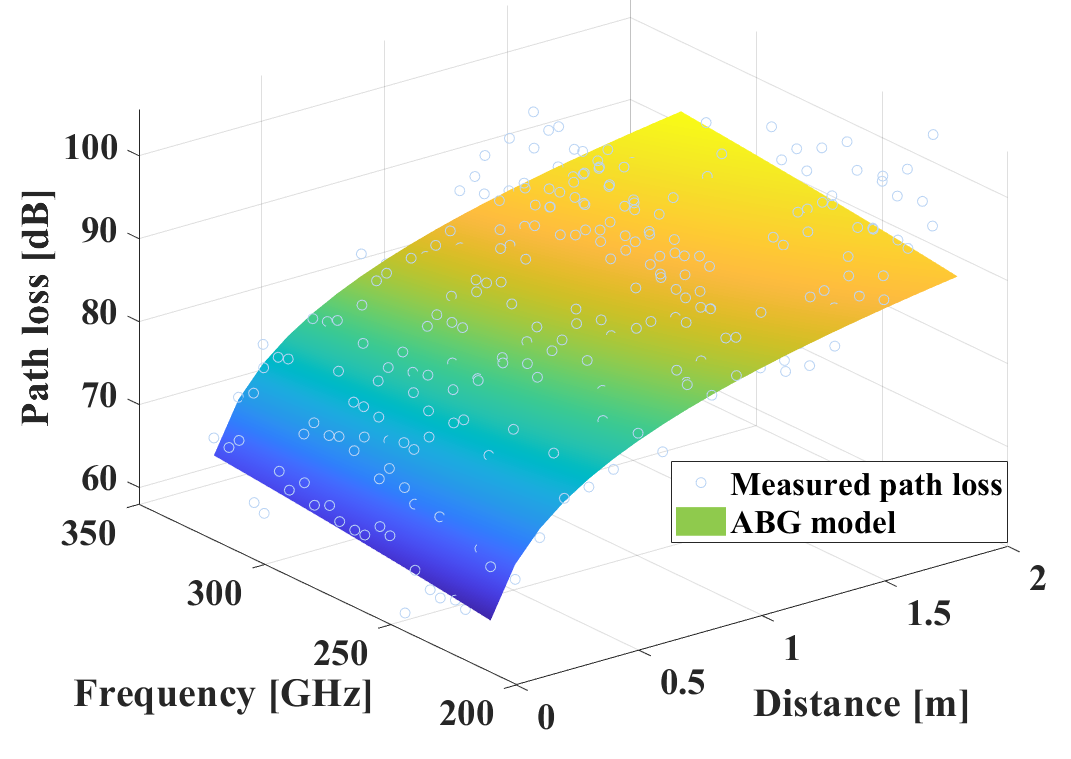}  
\caption{Multi-frequency measurement data and path loss modeling results from 220 to 330 GHz \cite{THz_Ref_3}.}
\label{THz_TP}
\end{figure}

Due to the wide span of the THz band, it will show a stronger frequency dependence. To further investigate the frequency dependence of path loss in THz bands, \cite{THz_Ref_3} used the ABG model to fit the measured path loss. The measured path loss and ABG model are plotted in Fig.~\ref{THz_TP}. The frequency dependence factor $\gamma_{\mathrm{ABG}}$ and distance dependence factor $\alpha_{\mathrm{ABG}}$ are 2.1 and 1.93, respectively. This finding indicates that the frequency dependence is slightly stronger than that in the free space \cite{38901}.

\subsubsection{Atmospheric loss}
In outdoor environments, THz signal propagation is strongly affected by atmospheric attenuation and weather-dependent scattering \cite{THz_resubmit_22}. The primary component of atmospheric loss arises from molecular absorption by oxygen and water vapor, introducing both frequency-selective and distance-dependent loss characteristics into the channel \cite{SAGIN_ITU3}. Oxygen absorption is prominent in certain sub-100 GHz bands, such as the well-known 60 GHz peak, whereas water vapor absorption dominates above 100 GHz. For terrestrial deployments beyond 100 GHz, such as UMi and short-range backhaul links within 200 m, gas attenuation must be incorporated into the link budget. Specifically, for the 100-300 GHz band, the attenuation ranges from 0.106-1.293 dB at 200 m and 0.53-5.72 dB at 1,000 m. In the 300-450 GHz band, the corresponding values increase significantly, reaching 1.29-68.76 dB and 6.45-343.8 dB, respectively \cite{Above100G}. At higher altitudes, such as in airborne or satellite communications, attenuation becomes strongly altitude-dependent and typically decreases with elevation due to reduced pressure and temperature.

In addition to gaseous attenuation, various weather conditions, including rain, fog, and snow, introduce additional propagation variability. Rain causes relatively stable attenuation across frequencies and is well-characterized by existing droplet models, making it more predictable in system-level design \cite{THz_resubmit_23,THz_resubmit_24}. Fog exhibits a stronger frequency dependence and is more sensitive to ambient humidity and temperature. Snow presents the greatest uncertainty due to its highly variable particle shapes and the marked differences between wet and dry conditions \cite{THz_resubmit_7}. These weather-induced effects can impact both signal strength and link stability, and thus must be carefully considered in the design of THz systems that target long-range applications such as vehicular, aerial, or backhaul communication links.

\subsubsection{Channel sparsity}
When THz waves impinge upon ``rough'' surfaces, they give rise to diffuse scattering phenomena. Compared with low-frequency specular reflections, diffuse scattering-generated multipath energy is weaker. This leads to a considerable reduction in the quantity and power of NLOS MPCs received at the receiver, resulting in most of the power being concentrated on a few MPCs, also known as sparsity.

The channel sparsity can be expressed in both the angular and delay domains and in the low rank of the channel matrix, which refers to a small number of coefficients containing a large proportion of the energy \cite{THz_SparsityHN2009}. It is often claimed that mmWave channels are ``sparse,'' i.e., have few entries in the delay angle bins (ITU-R M.2412 \cite{2412}). The sparse mmWave channel at 28 GHz was discovered and studied in 2018 \cite{THz_wangchao}. Therefore, due to its weak diffraction ability, the THz channel is more likely to be sparse \cite{3-2, Priebe300}. \cite{THz_Sparsitylxm} quantified the sparsity of a channel via the Gini index. Channel measurements conducted in sub-6 GHz, mmWave, and THz bands confirmed the sparsity characteristic in the THz channel. Then, an intracluster power allocation model within clusters was proposed to model the sparsity in the delay domain. \cite{THz_Sparsityczw} analyzed the number of clusters of delay, angle and power domains. The number of clusters was reduced, which reflected the sparsity of the THz channel.

\subsection{THz Channel Modeling}
Existing THz channel modeling approaches can be broadly classified into stochastic, deterministic, and hybrid approaches \cite{THz_model_general}. THz stochastic channel models are obtained mainly on the basis of measurements and describe the type of environment rather than a specific location. While parameterizing such models from measurements requires significant effort, implementing channel modeling from stochastic models is much less complex than implementing deterministic models \cite{THz_VLC_TY}. 

THz deterministic approaches solve (approximatly) Maxwell's system of equations in a given environment and can achieve high accuracy \cite{THz_deterministic}, but they require detailed information about the geometric and electromagnetic properties of the environment and have high computational complexity, such as the RT method \cite{THz_RT_GuanKe}. RT-based THz channel modeling has been extensively utilized in sub-6 GHz and mmWave bands \cite{THz_RT_mmwave}. However, channel modeling using the RT approach for the THz band is different from that for the sub-6 GHz and mmWave bands. These differences result in new opportunities for RT-based channel modeling in THz bands. For the inherent characteristics of RT, RT based on geometric optics can be explained by the high-frequency approximation of Maxwell's equations. Consequently, the quasioptical characteristics make the RT results in the THz band more reliable and suitable than those in the sub-6 GHz band. Additionally, for THz channel characteristics, the high propagation loss and sparsity characteristics of the THz channel increase the need for site-specific analysis, which can be well captured by RT simulations. 

\begin{figure*}[]
\centering
\includegraphics[width=14cm]{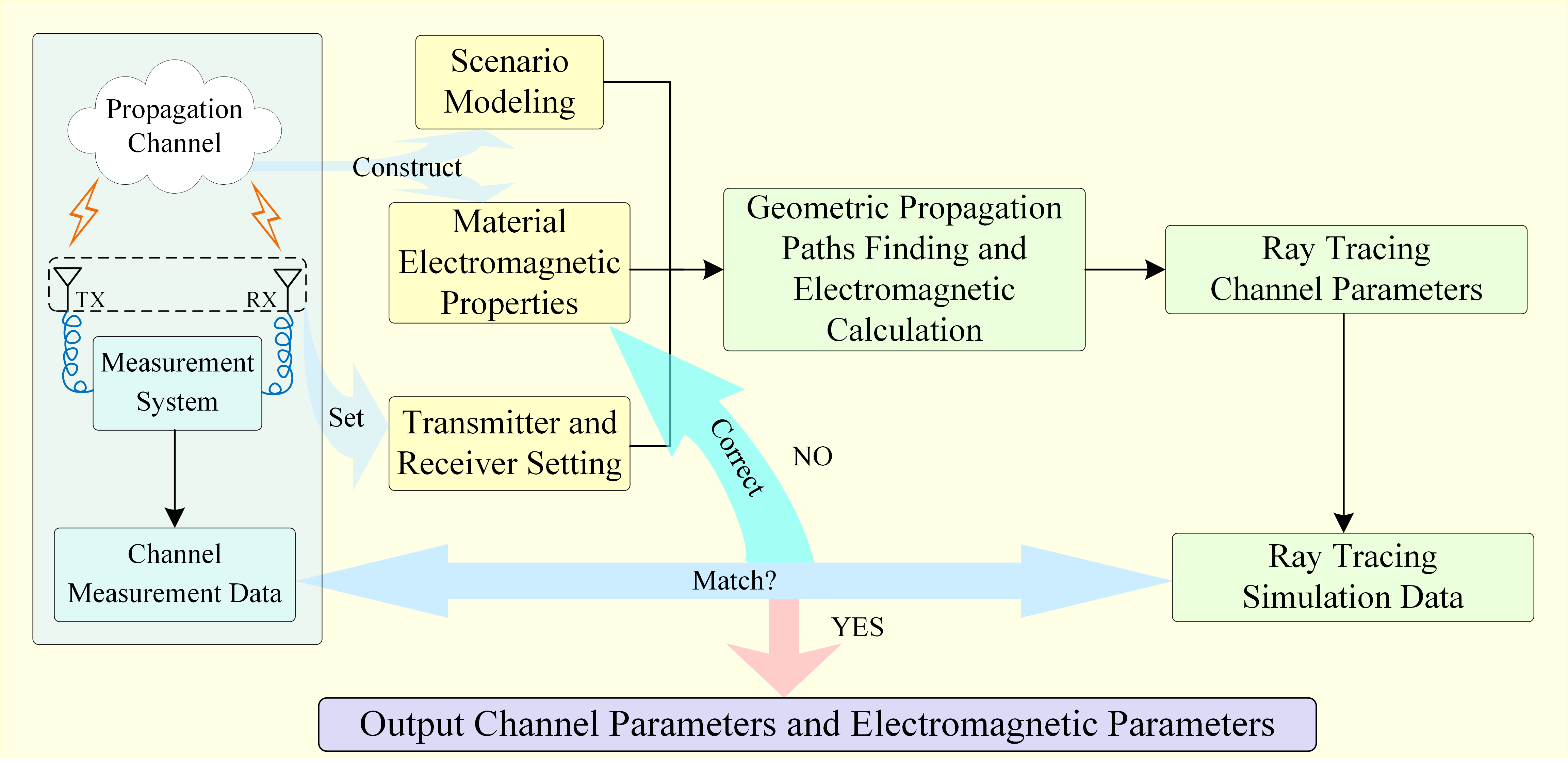}  
\caption{Data and model dual-driven THz channel modeling flowchart \cite{MIMO1-1-0}.}
\label{THz_modeling}
\end{figure*}

However, there are still some challenges for RT-based channel modeling in THz bands. Most notably, there is no complete electromagnetic property of the material in THz bands \cite{THz_EM_1}. Additionally, there is a lack of adequate measurements for validating and calibrating RT \cite{Priebe300}. For real-time applications, the RT channel model is difficult to implement. Therefore, a data and model dual-driven channel modeling method is presented, which uses channel measurement data to calibrate simulation parameters \cite{MIMO1-1-0}. The data and model dual-driven channel simulation workflow is shown in Fig.~\ref{THz_modeling}. The material EM properties in RT simulations are tuned to reach the best agreement in terms of power and delay for the dominant propagation paths. Initially, the RT simulation outputs the results corresponding to the initial EM parameters, which are compared with the measurement data. If the results of the power and delay parameters for the dominant paths match (with the objective of minimizing the RMS error), the simulation results are output. If the results do not match, the relative permittivity and conductivity of the materials are updated, and the simulation results are output again. This operation is repeated until the best agreement between the simulation and measurement results is achieved.

\begin{figure*}[]
\centering
\subfigure[]{
\centering
\includegraphics[width=8.5cm,height = 4cm]{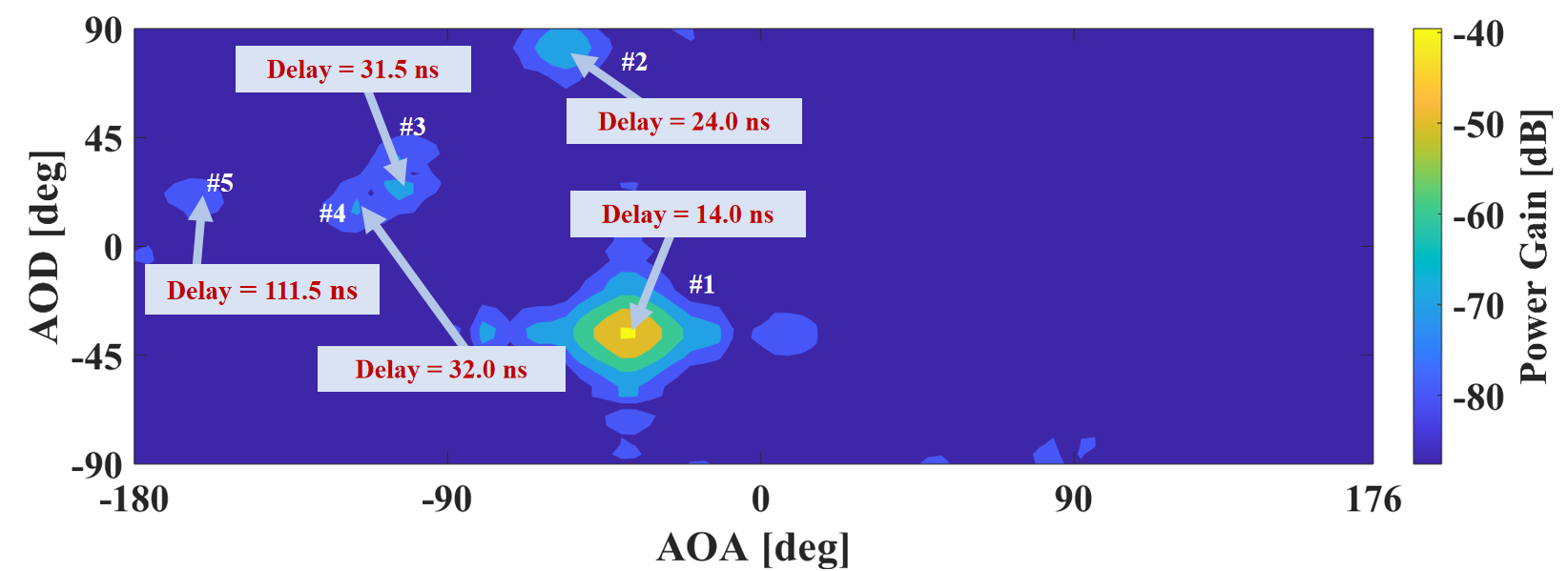} 
\label{THz_RT_1}
}
\subfigure[]{
\centering
\includegraphics[width=8.3cm,height = 4cm]{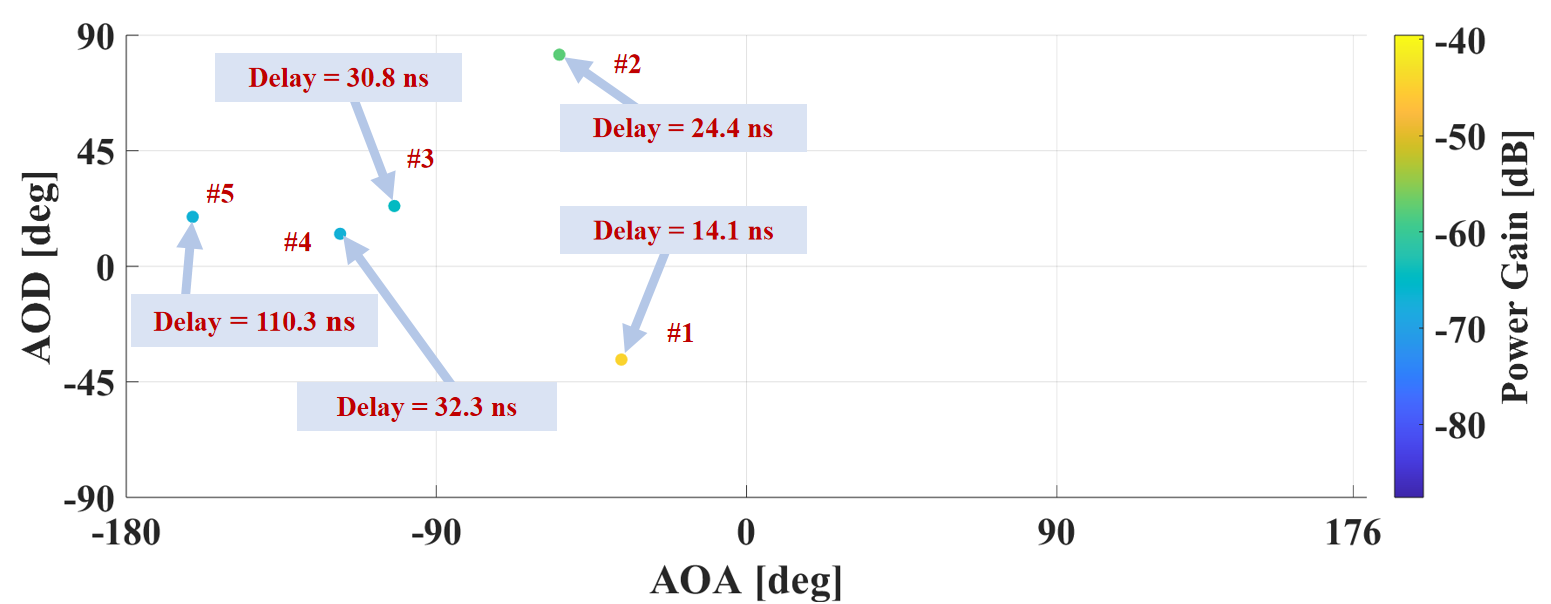} 
\label{THz_RT_2}
}
\caption{Data and model dual-driven channel modeling method results: a) power, angle of arrival, and angle of departure measurement results; b) power, angle of arrival, and angle of departure simulation results \cite{MIMO1-1-0}.}
\label{THz_RT}
\end{figure*}

The performance of the data and model dual-driven THz channel modeling method is demonstrated through a comparison between simulations and channel measurements. Fig.~\ref{THz_RT_1} shows the measurement power angle profile in which the five dominant paths are marked by numbers. Furthermore, the delays of the five dominant paths are plotted for comparison. Fig.~\ref{THz_RT_2} shows the results of RT simulations for the five dominant paths. By comparing Fig.~\ref{THz_RT_1} with Fig.~\ref{THz_RT_2}, it can be seen that the MPCs have good agreement in both the spatial and delay domains between the measurement and RT results. For example, the differences between the measurements and RT simulations for the delay of the five paths are within $\pm$2 ns. The maximum differences in the angle of arrival, angle of departure, delay, and power gain of the dominant paths are $-1.8^{\circ}$, $-1.5^{\circ}$, 0.7 ns and -0.7 dB, respectively. Therefore, the data and model dual-driven channel modeling method can effectively describe the propagation characteristics, i.e., the delay and spatial dispersion, with reduced simulation complexity in the THz bands \cite{MIMO1-1-0}.

THz hybrid channel models combine deterministic and stochastic channel modeling approaches, providing a balance between accuracy and efficiency, such as in map-based channel models \cite{THz_hybrid_map} and hybrid ray and graph models \cite{THz_hybrid_graph}. For example, due to the sparsity of the THz channel, the main propagation paths of the THz channel (e.g., the LOS path and primary reflection path) are generated by the deterministic model, whereas the other multipaths are generated by the stochastic model, which ensures the accuracy of the main propagation paths and quickly obtains the remaining multipath information \cite{THz_hybrid}. Therefore, the hybrid channel model is a compromise approach for THz channel modeling.

\subsection{Summary and Prospects}
New mid-band channel measurement campaigns have been carried out in typical scenarios, such as UMa, UMi, and InH. The channel characteristics of the new mid-band are different from those of other frequency bands such as sub-6 GHz, mmWave and THz bands. The path loss is less than that in high-frequency bands such as mmWave and THz, which is conducive to coverage. The delay spread is less than sub-6 GHz, and a smaller protection interval can be set in the frame structure. The new mid-band combines the advantages of coverage and capacity, making it the major candidate frequency band for future 6G applications.

Existing studies indicate that the THz band exhibits different channel characteristics based on THz band channel measurements and characteristic comparisons, including characteristics such as scattering, reflection, high path loss, and sparsity. Stochastic, deterministic, and hybrid channel modeling approaches have been employed in THz channel modeling. Among them, the data and model dual-driven THz channel modeling methods show promising prospects. Despite the research on THz channels in recent years, many open issues motivate future research efforts in the following areas \cite{THz_Ref_61}.

\subsubsection{Channel measurement platforms with higher performance} Current channel measurement platforms are unable to achieve a balance between measurement accuracy, measurement speed, and measurement distance, which constrains the study of THz channel characteristics. In addition, existing virtual MIMO channel measurements based on rotating horn antennas are time-consuming. Therefore, it is necessary to build a multi-frequency XL-MIMO array and phased array to achieve accurate parameter estimation and fast beam sweeping.

\subsubsection{Channel characteristics} An increasing number of frequency bands, ranging from sub-6 GHz to 10 THz, will be deployed and coexist in the future. In multiband parallel communication systems, it is possible to use the channel characteristics extracted from one frequency band to assist in establishing links in another band. However, the frequency separation between the bands under consideration is large, and many characteristic channel variations must be considered. Moreover, new channel characteristics, such as sparsity, near-field propagation, and non-stationary characteristics, need to be modeled, characterized, and incorporated into the channel model.

\subsubsection{Standardized and frequency-dependent channel model} The standardization work for the 7--24 GHz channel model in 3GPP Rel-19 has been completed at present. The THz band spans up to 10 THz, far beyond the scope of current 5G channel models. To enable standardized modeling, future THz channel models must account for strong frequency dependence in both large- and small-scale parameters, and address unique propagation effects such as frequency-selective molecular absorption, high directionality, and frequency non-stationarity across wide bandwidths. 

\subsubsection{Sensing for high-resolution localization and imaging} The ultra-wide bandwidth and high carrier frequency of the THz band enable fine-grained sensing capabilities, supporting millimeter-level range resolution and sub-centimeter localization accuracy. These advantages make THz sensing highly attractive for applications such as gesture recognition, object imaging, and environmental mapping. However, several challenges hinder its practical deployment. The highly directional nature of THz beams makes the sensing performance extremely sensitive to alignment errors and significantly restricts the field of view, especially in dynamic or cluttered scenarios. In addition, conventional sensing models often neglect THz-specific propagation effects such as frequency-selective scattering, strong molecular absorption, and weak surface reflections, which can degrade the sensing accuracy in real-world environments.

\subsubsection{Near-field propagation}
In THz communication systems, large antenna apertures are required to achieve the high directional gain needed to overcome severe path loss, which inherently amplifies near-field effects \cite{THz_resubmit_21}. Under the near-field condition, the phase and amplitude distributions at the Tx determine the beam's shape and curvature, while the Rx may capture only part of the beam or experience highly non-uniform phase distributions across its aperture, resulting in constructive or destructive interference that can significantly affect the received power \cite{THz_resubmit_19}. These characteristics introduce several design challenges, including the selection of appropriate beamshapes for varying mobility conditions, hardware limitations in generating precise and reconfigurable wavefronts, sensitivity to positional shifts that cause beam misalignment, and beam squint in ultra-wideband systems \cite{THz_resubmit_18}. At the same time, near-field propagation opens up new opportunities through advanced wavefront engineering-enabling self-healing Bessel beams and curving Airy beams for blockage mitigation and NLoS support, improved spatial energy concentration, enhanced physical-layer security via wavefront hopping, and high-capacity spatial multiplexing via orbital angular momentum \cite{THz_resubmit_20}.

\section{RIS Channel Measurement and Modeling}\label{SecRIS}

\begin{figure}[!t]
\centering
\includegraphics[width=3.0in]{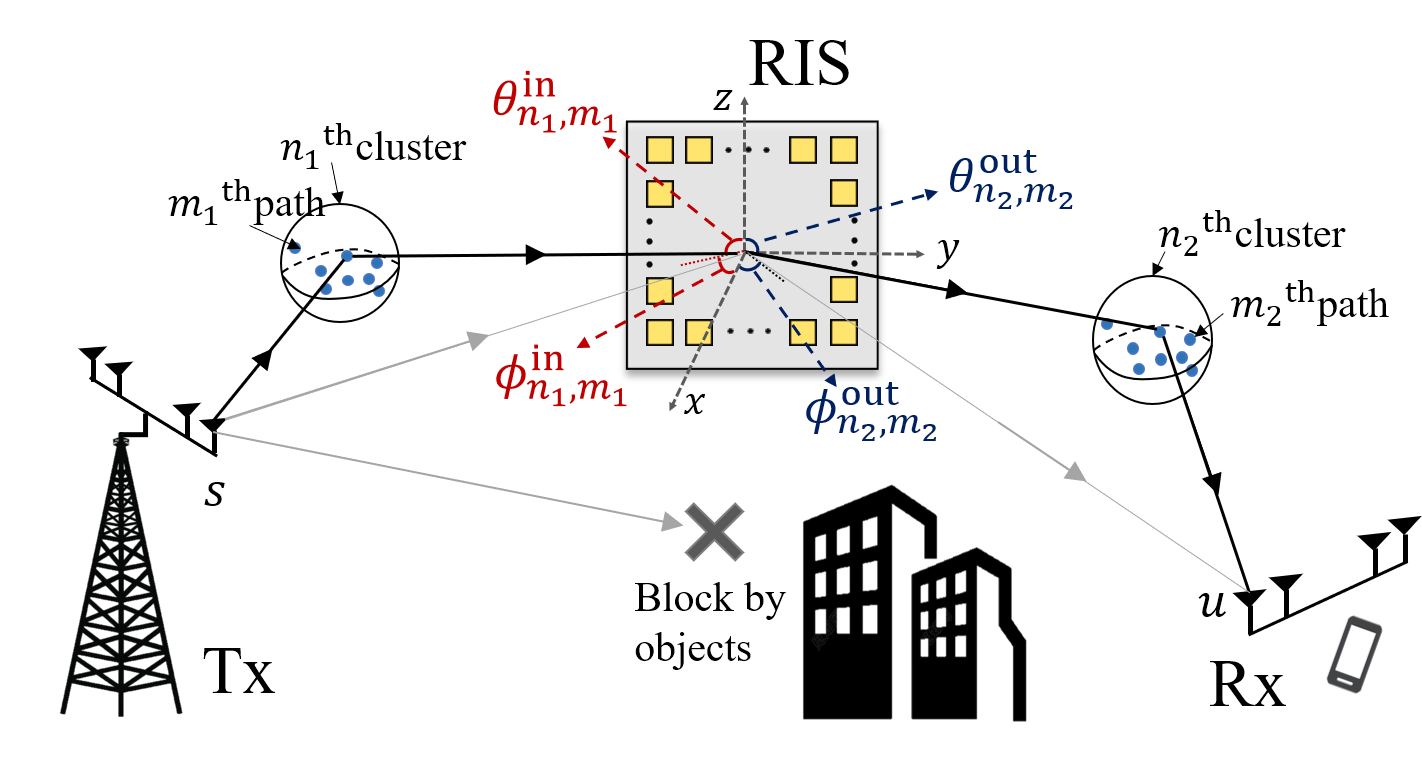}
\caption{Typical RIS-assisted communication scenario \cite{ghw}.}
\label{fig_RISChannel}
\end{figure}

Different from traditional communication technologies, an RIS realizes active control of a wireless signal propagation environment by passively reflecting electromagnetic waves. Fig.~\ref{fig_RISChannel} shows a typical RIS-assisted communication scenario. In this scenario, the LOS path is obstructed, and an RIS is employed to enhance the received signal. Obviously, traditional channel modeling methods are not suitable for this kind of channel, which requires new research on the RIS channel model. Notably, in Fig.~\ref{fig_RISChannel}, the RIS channel was characterized by a Tx-RIS-Rx concatenated channel, which was composed of a Tx-RIS subchannel, RIS-Rx subchannel, and RIS response. In this concatenated channel, the subchannels required independent modeling and were combined in the RIS response. The subsequent RIS channel modeling described in this paper is based on this method.

\subsection{RIS Channel Measurements}
Currently, a few researches on RIS channels based on channel measurements have been carried out. Table~\ref{RISStable1} summarizes some RIS channel measurement campaigns. The current focus of RIS measurement research is on the large-scale characteristics. Most measurement campaigns are aiming at obtaining received signal power to analyze the power gain caused by RIS. In terms of measurement frequency, measurement campaigns are mostly carried out in sub-6 GHz and millimeter band. These measurements are typically conducted in microwave anechoic chambers as well as practical scenarios, with a primary emphasis on the large-scale characteristics of RIS channel. In \cite{RIS6}, authors carry out a path loss measurement in a microwave anechoic chamber with three different RISs in different scenarios. This work investigates the free-space path loss in the RIS-assisted broadcasting and beamforming scenarios and compares it with the theoretical model proposed by the author. In addition, in \cite{RIS_yinhaifan}, a measurement campaign is carried out in a microwave anechoic chamber to obtain the radiation pattern of RIS.

\begin{figure}[!t]
\centering
\includegraphics[width=3.0in]{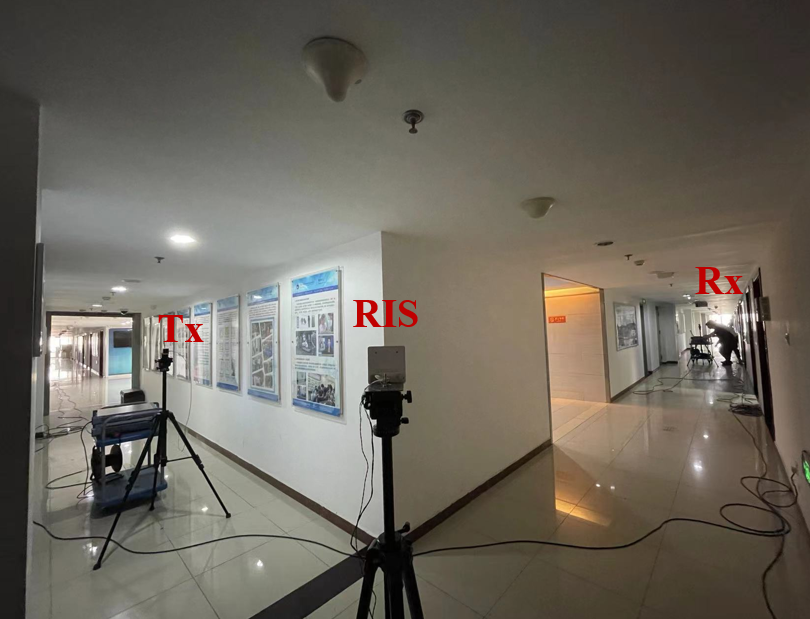}
\caption{RIS channel measurement in a corridor scenario \cite{RIS4}.}
\label{fig_RISChannelmeasc}
\end{figure}

Several other measurement campaigns have been conducted in practical environments to investigate RIS-assisted channels. For example, near-field RIS channel measurements were performed in both indoor and outdoor settings to study the path loss characteristics of RIS channels \cite{RIS5}. To support system-level simulations, \cite{RIS22} proposes a physical model for reflective RIS and conducts corresponding indoor measurements. This work aims to abstract the RIS physical properties for integration into practical channel modeling frameworks.
In \cite{RIS4}, a RIS-assisted channel measurement platform based on time-domain correlation principles is developed, which can be applied to a variety of complex deployment scenarios. Utilizing this platform, the authors carry out a measurement campaign at $26$~GHz in a typical indoor corridor environment, as shown in Fig.~\ref{fig_RISChannelmeasc}. Based on the collected data, they analyze the path loss characteristics of RIS-assisted links under corridor conditions.
Furthermore, the study in \cite{RIS_reciprocity} investigates channel reciprocity in RIS-assisted communication networks under TDD systems. Through measurements conducted in indoor scenarios with two representative RIS types, the authors demonstrate that the channel reciprocity theorem generally holds.
In addition, \cite{RIS_multiSce} and \cite{RIS_multiSce_k} conduct RIS channel measurements at $2.6$~GHz across indoor, outdoor, and O2I scenarios. The former analyzes the PDP, path loss, SF, and AS under different RIS configuration strategies, while the latter focuses on variations in the Rician $K$-factor introduced by RIS deployment.

\begin{table*}[htbp]
\caption{RIS measurement campaigns}
\begin{center}
\resizebox{\linewidth}{!}{
\begin{tabular}{c|c|c|c|c}
\toprule
    \hline
       {\begin{tabular}{c} Phase resolution  \end{tabular}} & Frequency (GHz) &  Scenario & Channel characteristics  &  Reference \\ \hline
       1 bit & 5.8 &  Indoor, outdoor & Power gain & \cite{RIS5} \\ \hline
       1 bit & 5.8 &  {\begin{tabular}{c}Microwave anechoic chamber,\\indoor, outdoor\end{tabular}} & {\begin{tabular}{c} Power gain,\\ radiation pattern,\\ channel reciprocity\end{tabular}} & \cite{RIS_yinhaifan} \\ \hline
       1 bit & 26 & Indoor & PL & \cite{RIS4} \\ \hline
       1 bit & 26.9 & Indoor & Radiation pattern & \cite{RIS22} \\ \hline
       Continuously & 4.25, 10.5 & Microwave anechoic chamber & PL & \cite{PL_Tang} \\ \hline 
       1 bit & 27, 33 & Microwave anechoic chamber & {\begin{tabular}{c} PL,\\ radiation pattern\end{tabular}} & \cite{RIS6} \\ \hline
       Continuously,  1 bit & 4.25, 27 & Indoor & Channel reciprocity & \cite{RIS_reciprocity} \\ \hline
       1 bit & 2.6 & Indoor, outdoor, O2I &  PDP, PL, SF, DS, K & \cite{RIS_multiSce,RIS_multiSce_k} \\ \hline
       \bottomrule
    \end{tabular}
    }
\end{center}
\label{RISStable1}
\end{table*}

\subsection{RIS Channel New Characteristics}
In the RIS channel, there are three types of characteristics that are worth paying attention to, namely the reflection coefficient characteristics of RIS element, RIS radiation pattern characteristics, and RIS concatenated loss characteristics. Among them, RIS radiation pattern characteristics consists of RIS overall radiation pattern characteristics and RIS element radiation pattern characteristics. The reasons are as follows.

\begin{figure}[!t]
\centering
\includegraphics[width=3.4in]{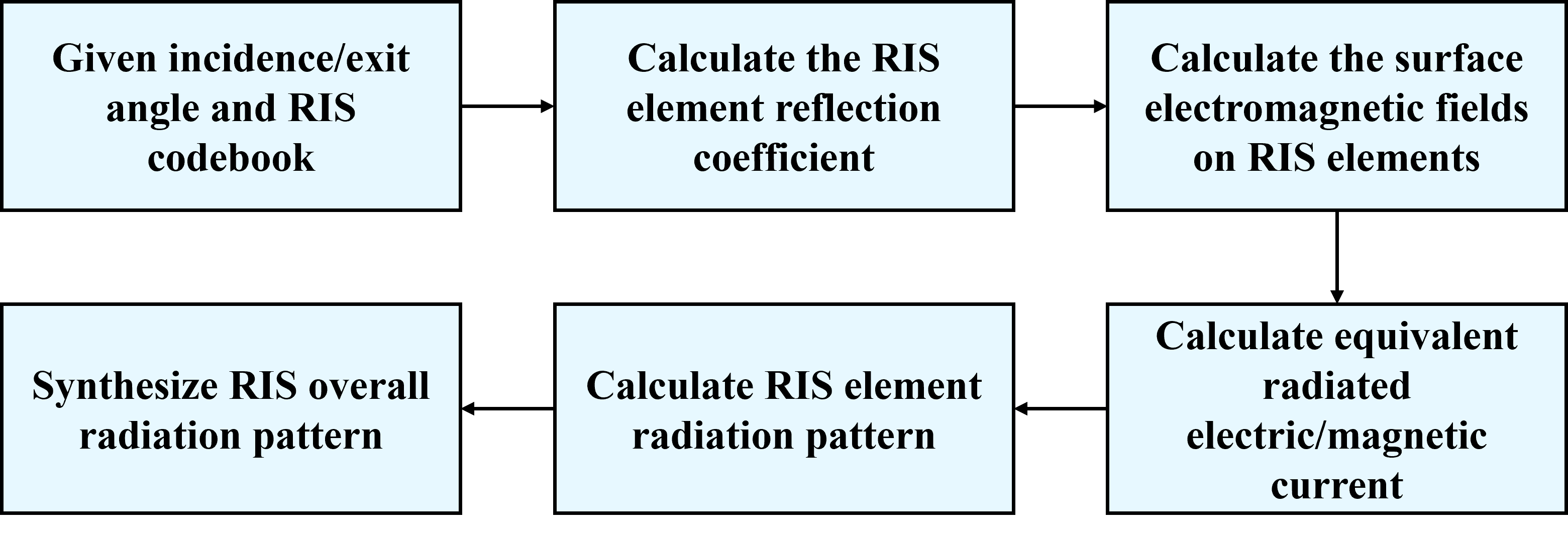}
\caption{The process of calculating RIS radiation pattern.}
\label{RIS_Reflection}
\end{figure}

Firstly, the RIS is one of the most important nodes in the RIS channel, and its response are an important component of the RIS channel. In order to calculate the RIS response, it is necessary to calculate the reflection coefficient of each RIS element in advance. Thus, the RIS element reflection coefficient characteristics are very important. Secondly, the final form of RIS response is RIS overall radiation pattern or RIS elemenmt radiation pattern, therefore, their characteristics can not be ignored either. Fig.~\ref{RIS_Reflection} shows a feasible process of calculating the RIS radiation pattern, corresponding to (\ref{RISequ1})-(\ref{location}). Finally, since the RIS concatenated channel is obtained by cascading Tx-RIS sub-channel, RIS response, and RIS-Rx sub-channel, its loss characteristics are different from traditional communication channel. Thus, RIS concatenated loss characteristics are worth studying.

\subsubsection{Reflection coefficient characteristics}
\label{reflection_coefficient}
The reflection coefficient characteristics of the RIS element are derived via RIS element hardware models. In the initial research on RIS-assisted communication, the hardware model of the RIS element was typically the ideal regulator model \cite{RIS20}. This implies that the RIS element has the same reflection coefficient regardless of the angle and polarization of the incident signal. However, by comparing it with the Fresnel reflection coefficient, it can be found that the incident angle and polarization dependence of the RIS element reflection coefficient may be ignored. The Fresnel reflection coefficient is widely used to characterize the reflective properties of 3D materials. Generally, the Fresnel reflection coefficient is related to the polarization and angle of the incident signal, as demonstrated by (9) and (10). In fact, for the 2D RIS element, its reflection coefficients also exhibit dependence on the incident angle and polarization.

To provide a more accurate characterization of the reflection coefficient characteristics of the RIS element, researchers have proposed electromagnetic hardware models, including the transmission line model\cite{surfaceEM}, equivalent circuit model\cite{RIS18,RIS20}, generalized sheet transmission conditions (GSTCs) model \cite{RIS10}, and load impedance model \cite{RIS16}.

\begin{figure}[!t]
\centering
\includegraphics[width=3.0in]{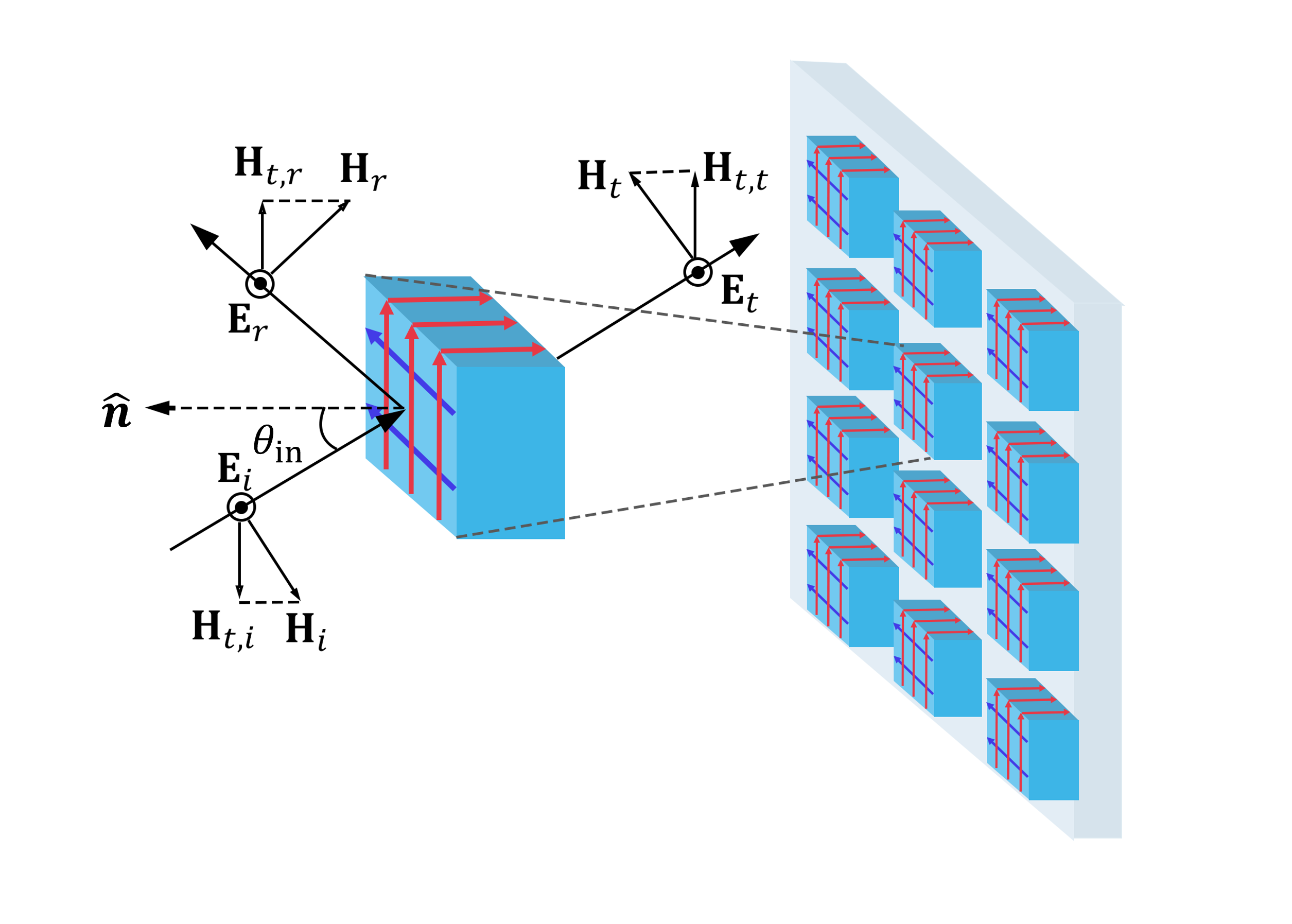}
\caption{Load impedance model of RIS \cite{RIS16}.}
\label{load impendence model}
\end{figure}

In the transmission line model, the tangential components of the electric and magnetic fields are equivalent to the voltage and current; then, the relationship between the reflection coefficient and incident angle and polarization direction can be deduced according to the Kirchhoff formula\cite{surfaceEM}. The equivalent circuit model is similar to the transmission line model, where the RIS element is equivalent to a circuit. Since the impedance in the circuit is related to the incident angle, the reflection coefficient is also naturally related to the incident angle \cite{RIS18}. The GSTCs model, commonly used in the metasurface field, employs the surface susceptibilities tensor as a characteristic parameter of the RIS element. By combining the GSTCs condition, the reflection coefficient corresponding to the incident signal with any incident angle and polarization direction can be calculated \cite{RIS10}. The load impedance model considers the RIS element as a thin surface with a certain electric and magnetic impedance. Its principle is shown in Fig.~\ref{load impendence model}; that is, upon excitation of the incident signal, the induced electric and magnetic currents are generated on the RIS element, acting as new radiation sources that radiate electromagnetic waves outward and making the RIS element exhibit a certain reflection coefficient. Since the induced electric and magnetic currents are jointly determined by the electric impedance, magnetic impedance, and the incident signal, the reflection coefficient is affected by the incident angle and polarization direction of the incident signal\cite{RIS16}.

In the load impedance model, the reflection coefficient for the vertically polarized signal is written as
\begin{equation}\label{RISequ1}
\Gamma_\textrm{v}=\frac{-Z_{0}}{2Z_{e}\cos\theta_\textrm{in}+Z_{0}} + \frac{Z_{m}\cos\theta_\textrm{in}}{Z_{m}\cos\theta_\textrm{in}+2Z_{0}}.
\end{equation}

Correspondingly, if the incident signal is parallel polarized, the reflection coefficient is written as
\begin{equation}\label{RISequ2}
\Gamma_\textrm{h}=\frac{Z_{0} \cos\theta_\textrm{in}}{2Z_{e}+Z_{0} \cos\theta_\textrm{in}} - \frac{Z_{m}}{Z_{m}+2Z_{0} \cos\theta_\textrm{in}},
\end{equation}
where $Z_{e}$ and $Z_{m}$ are the electrical impedance and magnetic impedance, respectively.

\subsubsection{RIS radiation pattern characteristics}
Firstly, introduce RIS element radiation pattern characteristics. Currently, there are two commonly used methods for modeling the RIS element radiation pattern, namely, mathematical function fitting and electromagnetic scattering theory calculations.
In \cite{RIS6} and \cite{RIS12}, the RIS element radiation pattern was fitted by the $\textrm{cos}^{q}\theta_{\textrm{out}}$ function. In \cite{RIS16} and \cite{RIS3}, the physical optics method (PO) was used to calculate the radiation pattern of the RIS element. In \cite{RIS5} and \cite{RIS7}, the radiation pattern of the RIS element was modeled based on calculating its RCS. In \cite{ghw}, the electromagnetic equivalence principle was employed to calculate the RIS element radiation pattern.

Taking the model presented in \cite{ghw} as an example, we introduce the radiation pattern modeling process of the RIS element below. Assuming that a signal with both vertical and parallel polarization components is incident on an RIS element at a horizontal angle $\phi_\textrm{in}$ and a vertical angle $\theta_\textrm{in}$, its electric field is $\mathbf{E}_{i}$, and its corresponding magnetic field is $\mathbf{H}_{i}$.
According to the equivalence theorem, the equivalent current $\mathbf{J}_{s}$ and equivalent magnetic current $\mathbf{M}_{s}$ on the RIS element surface are calculated as
\begin{equation}
\begin{aligned}
\mathbf{J}_{s} =&\hat{\boldsymbol{n}} \times (\mathbf{H}_{i}+\mathbf{H}_{r}), \\
\mathbf{M}_{s} =&-\hat{\boldsymbol{n}} \times (\mathbf{E}_{i}+\mathbf{E}_{r}),
\end{aligned}
\label{JM1}
\end{equation}
where $\hat{\boldsymbol{n}}$ represents the normal vector of the RIS element, $\mathbf{E}_{r}$ and $\mathbf{H}_{r}$ represent the reflected electric and magnetic fields, which are obtained from the reflection coefficient of the RIS element. We select an observation point $o$ in the direction of $\phi_\textrm{out}, \theta_\textrm{out}$, and the scattering field $\mathbf{E}_{s}$ at point $o$ can be calculated by calculating the radiation of electric and magnetic currents. By removing the effect of the distance of the observing point, the radiation pattern of the RIS element can be obtained as
\begin{equation}
F_\textrm{element}^{p_{1}p_{2}}(\phi_\textrm{in},\theta_\textrm{in}, \phi_\textrm{out}, \theta_\textrm{out})=\frac{4\pi}{\lambda}\frac{r}{e^{-j \frac{2 \pi}{\lambda} r}}\frac{(\mathbf{E}_{s})_{p_{2}}}{(\mathbf{E}_{i})_{p1}},
\label{V-in-element}
\end{equation}
where $r$ represents the distance between the center of the RIS element and point $o$.
$\frac{4\pi}{\lambda}$ is caused by the conversion between the electric field and the amplitude of the path.
$(\mathbf{E}_{s})_{p_{2}}$ represents the $p_{2}$ polarization component of $\mathbf{E}_{s}$.
${p_{1}p_{2}}$ represents the combination of different polarizations of the incident and reflected waves. For example, $F_\textrm{element}^{vh}$ refers to the horizontal polarization component of the reflected wave excited by the vertical polarization component of the incident wave at the RIS element.

 Then, introduce RIS overall radiation pattern characteristics. In the case where transmitter and receiver are located in the far field of RIS, there exists an RIS overall radiation pattern. The RIS overall radiation pattern is obtained by coherent superposition of the radiation pattern of each RIS element and is calculated as
\begin{equation}
\begin{aligned}
& F_\textrm{ris}^{p_{1}p_{2}}(\phi_\textrm{in},\theta_\textrm{in}, \phi_\textrm{out}, \theta_\textrm{out})  = \\
&\sum_{x,y}^{X,Y} F_\textrm{element}^{p_{1}p_{2}}(\phi_\textrm{in}, \theta_\textrm{in}, \phi_\textrm{out}, \theta_\textrm{out})e^{\frac{2\pi}{\lambda }(\mathbf{r}_\textrm{in} \cdot \mathbf{d}_{x,y})}e^{\frac{2\pi}{\lambda }(\mathbf{r}_\textrm{out} \cdot \mathbf{d}_{x,y} )},
\end{aligned}
\label{panel}
\end{equation}
where $X, Y$ is the number of rows and columns of RIS, and $\mathbf{r}_\textrm{in}$ and $\mathbf{r}_\textrm{out}$ denote the direction vector of the incident and outgoing waves, respectively.
$\mathbf{d}_{x,y}$ is the position vector of the $(x,y)^\textrm{th}$ RIS element in the panel. If we take the center point of the RIS as the reference point, $\mathbf{d}_{x,y}$ can be expressed as
\begin{equation}
\mathbf{d}_{x,y}=\left[\begin{array}{c}
(x-\frac{1+X }{2})  \\
(y- \frac{1+Y }{2})\\
0
\end{array}\right]d_\textrm{element},
\label{location}
\end{equation}
where $d_\textrm{element}$ is the interval between RIS elements.

\subsubsection{RIS concatenated loss characteristics}
Multiple studies have been conducted on the RIS concatenated loss of the RIS channel based on theoretically derived and measurements. They provide rich RIS concatenated loss modeling methods and combined reveal the RIS concatenated loss characteristics of the RIS channel.

During a period, the relationship between the RIS concatenated loss of the RIS channel and the distances $d_{1}$ and $d_{2}$ is very relevant, where $d_{1}$ is the distance between the transmitter and RIS, and $d_{2}$ is the distance between the RIS and receiver. A theoretical RIS concatenated loss model in free space under the far-field condition was proposed in \cite{RIS3}. The theoretical analysis demonstrated that the RIS concatenated loss is proportional to $(d_{1}d_{2})^{2}$.
Similar conclusions were also obtained in measurements of a corridor scene. Authors \cite{RIS4} carried out a channel measurement in a typical indoor corridor scenario and studied the RIS concatenated loss characteristic of the RIS channel in corridor scenarios. The results indicate that a modified version of the multiplicative distance model can effectively fit the measurement data.


Furthermore, some different conclusions can be made.
In \cite{RIS6}, a general RIS concatenated loss model and some special models in free space were proposed. These special models apply to RIS-assisted far-field beamforming, RIS-assisted near-field focusing, and RIS-assisted near-field broadcasting cases. The RIS-assisted far-field beamforming and RIS-assisted near-field broadcasting models were verified by several measurements conducted in a microwave chamber at 27 and 33 GHz. Different from above, the authors found that the RIS concatenated loss is proportional to $(d_{1}+d_{2})^{2}$ in the RIS-assisted near-field broadcasting case. This indicates that the RIS concatenated loss characteristics of the RIS channel are related to the encoding method of the RIS and depend on whether the terminal is in the far field or near field.

\subsection{RIS Channel Modeling}
Considering the prevalent use of reflective RIS, it is often placed between the transmitter and receiver, resulting in a concatenated channel known as the Tx-RIS-Rx channel. The main objective of this subsection is to construct an effective concatenated RIS channel model to accurately capture the response characteristics of the RIS and channel properties, while balancing the model complexity and accuracy.

\begin{figure*}[htbp]
\begin{equation}
\begin{aligned}
 h_{u,s}^{\textrm{ris}}(t,\tau)  = 
&\sum_{n_{1},m_{1}}^{N_{1},M_{1}}\sum_{n_{2},m_{2}}^{N_{2},M_{2}}\sqrt{P_{n_{1},m_{1}}P_{n_{2},m_{2}}}\left[\begin{array}{c}
        F_{\textrm{Rx}, u,\theta}\left(\theta_{n_{2},m_{2}}^{\textrm{Rx}}, \phi_{n_{2},m_{2}}^{\textrm{Rx}}\right) \\
        F_{\textrm{Rx}, u,\phi}\left(\theta_{n_{2},m_{2}}^{\textrm{Rx}}, \phi_{n_{2},m_{2}}^{\textrm{Rx}}\right)
    \end{array}\right]^{\mathrm{T}} \cdot
\left[\setlength{\arraycolsep}{0.4pt}\begin{array}{cc}
\exp \left(j \Phi_{n_{2},m_{2}}^{\theta\theta}\right) & \sqrt{\kappa_{n_{2},m_{2}}^{-1}} \exp \left(j \Phi_{n_{2},m_{2}}^{\theta\phi}\right) \\
\sqrt{\kappa_{n_{2},m_{2}}^{-1}} \exp \left(j \Phi_{n_{2},m_{2}}^{\phi\theta}\right) & \exp \left(j \Phi_{n_{2},m_{2}}^{\phi\phi}\right)
\end{array}\right]\\
&\cdot\left[\begin{array}{cc}
F_{\textrm{ris}}^{\theta\theta}(\phi_{n_{1},m_{1}}^{\textrm{in}}, \theta_{n_{1},m_{1}}^{\textrm{in}}, \phi_{n_{2},m_{2}}^{\textrm{out}}, \theta_{n_{2},m_{2}}^{\textrm{out}})  &
F_{\textrm{ris}}^{\theta\phi}(\phi_{n_{1},m_{1}}^{\textrm{in}}, \theta_{n_{1},m_{1}}^{\textrm{in}}, \phi_{n_{2},m_{2}}^{\textrm{out}}, \theta_{n_{2},m_{2}}^{\textrm{out}}) \\
F_{\textrm{ris}}^{\phi\theta}(\phi_{n_{1},m_{1}}^{\textrm{in}}, \theta_{n_{1},m_{1}}^{\textrm{in}}, \phi_{n_{2},m_{2}}^{\textrm{out}}, \theta_{n_{2},m_{2}}^{\textrm{out}}) &
F_{\textrm{ris}}^{\phi\phi}(\phi_{n_{1},m_{1}}^{\textrm{in}}, \theta_{n_{1},m_{1}}^{\textrm{in}}, \phi_{n_{2},m_{2}}^{\textrm{out}}, \theta_{n_{2},m_{2}}^{\textrm{out}})
\end{array}\right]\\&
\cdot\left[\begin{array}{cc}
\exp \left(j \Phi_{n_{1},m_{1}}^{\theta\theta}\right) & \sqrt{\kappa_{n_{1},m_{1}}^{-1}} \exp \left(j \Phi_{n_{1},m_{1}}^{\theta\phi}\right) \\
\sqrt{\kappa_{n_{1},m_{1}}^{-1}} \exp \left(j \Phi_{n_{1},m_{1}}^{\phi\theta}\right) & \exp \left(j \Phi_{n_{1},m_{1}}^{\phi\phi}\right)
\end{array}\right]
\cdot
\left[\begin{array}{c}
        F_{\textrm{Tx}, s,\theta}\left(\theta_{n_{1},m_{1}}^{\textrm{Tx}}, \phi_{n_{1},m_{1}}^{\textrm{Tx}}\right) \\
        F_{\textrm{Tx}, s,\phi}\left(\theta_{n_{1},m_{1}}^{\textrm{Tx}}, \phi_{n_{1},m_{1}}^{\textrm{Tx}}\right)
    \end{array}\right]\\
&\cdot\exp \left(j \frac{2\pi}{\lambda}\left(\hat{r}_{\textrm{Rx},n_{2},m_{2}}
\cdot \vec{d}_{\textrm{Rx},u}+\hat{r}_{\textrm{Tx},n_{1},m_{1}} \cdot \vec{d}_{\textrm{Tx},s}\right)\right)
\cdot \exp (j 2\pi f_{n_{2},m_{2}} t)
\cdot\delta{(\tau-\tau_{n_{1},m_{1}}-\tau_{n_{2},m_{2}})} .
\end{aligned}
\label{htotal}
\end{equation}
\end{figure*}

The current main channel modeling method is statistical channel modeling based on GBSM, as mentioned in Section II.B. This method considers the presence of many clusters and multipaths in the wireless channel and describes their delay, angle, power, and other parameters through specific statistical distributions.
Given the potential use of RISs in various frequency bands and service scenarios in future wireless networks, it is necessary to develop an RIS channel model that is suitable for full frequency bands and all scenarios. 
The GBSM-based statistical channel model has the advantage of scalability and can be easily extended to different communication scenarios. Therefore, it is expected to be an effective approach for modeling RIS channels \cite{ghw}.

In \cite{ghw}, the authors propose a GBSM-based channel modeling framework for the RIS concatenated channel, which takes into account the incident angle dependence of the RIS element reflection coefficient. The model was applied to an RIS-assisted MIMO communication scenario, as shown in Fig.~\ref{fig_RISChannel}. The CIR of the RIS concatenated channel between the $s^\textrm{th}$ transmitter and the $u^\textrm{th}$ receiver is expressed as \eqref{htotal}.
Similar to the channel model presented in \eqref{bigH}, most of the channel parameters in the RIS concatenated channel model are defined similarly. However, the RIS concatenated channel model differs in that it includes two subchannels and the RIS response. The specific parameters of the RIS concatenated channel model are listed below.
\begin{itemize}
    \item Two sets of cluster and path parameters are included in the equation. $n_{i},m_{i}, i \in \{1,2\}$ indicate the identifiers of the cluster and path in the Tx-RIS and RIS-Rx subchannels.
    \item The relevant SSPs encompass parameters such as power $P_{n_{1},m_{1}}, P_{n_{2},m_{2}}$, doppler shift $f_{n_{2},m_{2}}$, delay $\tau_{n_{i},m_{i}}$, $i \in\{1,2\}$, and angles $\phi_{n_{1},m_{1}}^{\textrm{in}}, \theta_{n_{1},m_{1}}^{\textrm{in}},\theta_{n_{1},m_{1}}^{\textrm{Tx}},\phi_{n_{1},m_{1}}^{\textrm{Tx}}$, representing the AOA, EOA, EOD, and AOD, respectively, of the $(n_{1},m_{1})^\textrm{th}$ path in the Tx-RIS subchannel. Similarly, $\phi_{n_{2},m_{2}}^{\textrm{Rx}}, \theta_{n_{2},m_{2}}^{\textrm{Rx}},\theta_{n_{2},m_{2}}^{\textrm{out}}, \phi_{n_{2},m_{2}}^{\textrm{out}}$ denote the AOA, EOA, EOD, and AOD of the $(n_{2},m_{2})^\textrm{th}$ path in the RIS-Rx subchannel.
    $\kappa_{n_{i},m_{i}}$, $i \in\{1,2\}$, represents the XPR of the corresponding path. Additionally, $\Phi_{n_{i},m_{i}}^{p_{1}p_{2}}$, $i \in\{1,2\}$, is the random phase of the $(n_{i},m_{i})^\textrm{th}$ path departing in the $p_{1}$ direction and arriving in the $p_{2}$ direction.
    \item $F_\textrm{ris}^{p_{1}p_{2}}$ represents the RIS overall radiation pattern, with $p_{1},p_{2} \in \{\theta,\phi\}$ denoting vertical and horizontal polarization, respectively. The notation $p_{1}p_{2}$ signifies that the RIS has distinct effects on the cluster or path incident in the $p_{1}$ polarization direction and emitted in the $p_{2}$ polarization direction.
\end{itemize}

Implementation of the proposed model is based on the 3GPP standardized channel model with some improvements, including three parts: 1) general and large-scale parameters, 2) small-scale parameters, and 3) the CIR for RIS concatenated channel. The model generates large-scale parameters (LSPs) and SSPs for both the Tx-RIS and RIS-Rx subchannels. The overall radiation pattern of the RIS is calculated using a method proposed in \cite{RIS16} to account for the incident angle-dependent characteristics of the RIS reflection coefficient. For further details on the specific implementation processes, readers may refer to \cite{ghw}.

The existing RIS channel modeling approaches can be broadly categorized as statistical approaches \cite{RIS11,RIS12}, deterministic approaches \cite{RIS16,RIS21}, and hybrid approaches \cite{RIS17}, in addition to the previously discussed GBSM-based method. These approaches differ in their trade-offs between model accuracy and complexity. Statistical approaches generally have lower model complexity but may sacrifice some accuracy, while deterministic approaches tend to provide higher accuracy but have higher computational complexity. Hybrid models aim to strike a balance between the two, leveraging the advantages of both approaches.

In \cite{RIS11}, an RIS-assisted communication channel modeling method for the sub-6 GHz band was proposed based on the GBSM principle, and related work was extended to multiple RIS cases in \cite{RIS12}. 
In \cite{RIS16}, the authors use impedance boundary conditions and equivalent impedance to derive the relationship between the reflection coefficient of the RIS element and the incident angle and then combine the physical optics and RT methods to achieve deterministic channel modeling of the RIS. To validate the accuracy of the proposed model, the authors carried out a path loss measurement activitie in the corridor scenario, and compared the results with RT simulation. The results showed a high level of agreement between the measurement and simulation.
Different from the above cluster-based modeling methods, \cite{RIS21} introduced a circuit-based communication model for RIS-assisted wireless systems. This model takes into account the mutual impedance between all radiation elements (transceiver antenna, RIS element) and introduces RIS into the communication model by presenting the influence of RIS on the equivalent current and equivalent voltage of the transceiver antenna.
The authors in \cite{RIS17} propose an RIS channel modeling method based on a map-based hybrid channel model (MHCM). In that study, the multipath information in complex scenes was generated by the RT method, and the RIS surface was modeled as a virtual BS excited by different incoming waves. The gain of the RIS for different multipaths was calculated, and finally, the channel coefficients of the entire RIS channel were generated by multipath merging.


\subsection{Summary and Prospects}
The RIS channel measurement is primarily aimed at measuring the power gain caused by RIS. The characteristics of the RIS channel primarily include reflection coefficient characteristics of the RIS element, RIS radiation pattern characteristics, and RIS concatenated loss characteristics. These characteristics are worth considering in RIS channel modeling. Based on these findings, we foresee some challenges that RIS channel modeling research could encounter.

\subsubsection{RIS channel measurement} RIS channel measurement plays a crucial role in understanding RIS channel characteristics and validating RIS channel models. However, due to the high transmission loss of the RIS-assisted channel caused by the Tx-RIS-Rx concatenated link, it imposes greater requirements on the transmission power and dynamic range of the measurement platform. In addition, the measurement platform needs a complete architecture, such as multiple channels, a high sampling rate, and transmit-receive separation, to capture channel characteristics in various complex scenarios.

\subsubsection{New characteristics modeling} The reflection coefficient characteristics are important  RIS channel characteristics. Various hardware models are previously presented to capture the reflection coefficient characteristic of RIS elements. However, due to the diversity of hardware implementations in RIS elements, it is not feasible to include all types of RIS elements in these hardware models. Considering that it is too expensive to build a hardware model for each type of RIS element, it is a major challenge to propose an accurate and universal RIS hardware model in the future.

\subsubsection{RIS channel modeling} The passive characteristic makes RIS have the advantage of low cost and low power consumption. But this also leads to a problem where RIS requires a large area to provide sufficient gain. Therefore, the RIS channel may exhibit near-field effect and SnS like XL-MIMO channel. In this case, the RIS response can not be characterized by the RIS overall radiation pattern. Thus, modeling near-field effects and SnS in the RIS channel is a major challenge.

\begin{table*}[htbp]
\caption{SAGIN measurement campaigns}
\begin{center}
\renewcommand{\arraystretch}{1.2} 
\begin{normalsize} 
\resizebox{\linewidth}{!}{
\begin{tabular}{c|c|c|c}
\toprule
    \hline
       Scenario  & Frequency (GHz)  &  Channel characteristics &   Reference \\ \hline
       Space-to-ground &0.86, 1.55, 39.4 &   PL  & \cite{SAGIN_SG_PL1,SAGIN_SG_PL2} \\ \hline
       
       Space-to-ground &0.87, 1.5, 2, 12, 14, 16, 28.2 &   SF  & \cite{SAGIN_SG_SF1, SAGIN_SG_SF2, SAGIN_SG_SF3, SAGIN_SG_SF4, SAGIN_SG_SF5, SAGIN_SG_SF6, SAGIN_SG_SF7} \\ \hline
       
       Space-to-ground & 2 &   Power, SF, capacity, K & \cite{SAGIN_SG_other1} \\ \hline

        Space-to-ground &19, 19.7, 39.04, 39.4  &  Meteorological effects & \cite{SAGIN_SG_ME1, SAGIN_SG_ME2,  SAGIN_SG_other2} \\ \hline

        Air-to-ground &0.92  &  PL, DS, Doppler, spatial correlation & \cite{SAGIN_AG_3} \\ \hline

        Air-to-ground &0.97, 5  &  PL, DS, K, spatial correlation & \cite{SAGIN_AG_4,SAGIN_M_SUN,SAGIN_AG_K1} \\ \hline

        Air-to-ground &2.05  &  PL, DS, excess delay, K & \cite{SAGIN_AG_2} \\ \hline
        
       Air-to-ground &2.3  &  PL & \cite{SAGIN_AG_1} \\ \hline


        Air-to-ground &4.3  &  PL, DS, excess delay, SF & \cite{SAGIN_AG_5} \\ \hline

        Air-to-ground &5  &  PL, Doppler, K & \cite{SAGIN_AG_K2} \\ \hline

        Air-to-ground &6.5  &  Power, excess delay, DS, K & \cite{SAGIN_AG_6} \\ \hline

        Air-to-air &2.4  &  Power, K & \cite{SAGIN_AA_1} \\ \hline
       Air-to-sea & 5 & ACF, CCF, DPSD, RMS & \cite{9460824} \\ \hline

       Air-to-sea & 2.2 &   End-to-end delay & \cite{10132977} \\ \hline
       \bottomrule
    \end{tabular}
    }
\end{normalsize}
\end{center}
\label{SAGINtable1}
\end{table*}

\section{SAGIN Channel Measurements and Models}\label{SecSAGIN}
One of the 6G usage scenarios is ubiquitous connectivity by addressing the challenges of connectivity, coverage, capacity, data rate and the mobility of terminals \cite{VISION}. One focus of this usage scenario is to address presently uncovered or scarcely covered areas, and SAGIN is considered a resolution to achieve worldwide coverage. Recently, the Working Party 4B of the ITU-R Study Group 4 (ITU-R SG4) held its 56th plenary session, where it discussed the time planning for satellite technology standards for 6G and determined the initial work plan \cite{SG4}. SAGIN is an integration of satellite systems, aerial networks, and terrestrial communications. Thus in SAGIN, channel characteristics in the communication scenarios of different segments (space, air and ground) should be considered. However, due to the long propagation distance and high node mobility, SAGIN channel characteristics are distinctly different from those of the traditional terrestrial channels \cite{LiuJJ2018}. Thus in this section, the present research on channel measurements, characteristics, and models of SAGIN is reviewed. Since terrestrial systems have been discussed in the above sections, this section is focused more on the SG and AG scenarios. In addition, a brief summary and potential future research directions are provided.

\subsection{SAGIN Channel Measurements}
For clarity, SAGIN channel measurements are classified into SG, AG, air-to-air (AA) and air-to-sea (AS), and summarized them in Table \ref{SAGINtable1}. For SG channels, research work has been started in the early years. Earlier, researchers \cite{SAGIN_SG_PL1} conducted measurements using the ATS-6 satellite to study the propagation loss of SG links at 860 and 1550 MHz. In \cite{SAGIN_SG_PL2}, propagation loss measurements were carried out at several geographical ground locations in the Q/V frequency band, and a low earth orbit (LEO) satellite channel model was developed using machine learning and statistical methods. A series of measurements of tree attenuation were conducted and studied in various frequency bands such as L-band and K-band \cite{SAGIN_SG_SF1, SAGIN_SG_SF2, SAGIN_SG_SF3, SAGIN_SG_SF4, SAGIN_SG_SF5, SAGIN_SG_SF6, SAGIN_SG_SF7}. In \cite{SAGIN_SG_other1}, a satellite MIMO channel study was conducted to investigate the channel capacity in an urban scenario. In \cite{SAGIN_SG_other2}, meteorological effects, such as rain, clouds, and scintillation, were studied based on field channel measurements in a tropical area. In addition, researchers in \cite{SAGIN_SG_ME1, SAGIN_SG_ME2} conducted measurements to study rain attenuation at the Ka/Q band. According to Table~\ref{SAGINtable1}, it can be found that most of the channel measurements of SG were carried out at L-band, S-band, K-band and Q-band \cite{431Nomenclature}, which are the common frequency bands for satellite communications. For channel characteristics, these measurement campaigns mainly focus on the path loss, shadow fading (especially shadow fading caused by tree attenuation), and meteorological effects, and less attention is paid to the small scale characteristics of the channel.

For aerial channels like AG and AA, due to the rise of UAV technology, the corresponding channel measurements have gradually conducted. In \cite{SAGIN_AG_1}, the path loss of the UAV channel was studied at 2.3 GHz in urban and nonurban scenarios. In \cite{SAGIN_AG_4, SAGIN_M_SUN, SAGIN_AG_K1}, the National Aeronautics and Space Administration (NASA) Glenn Research Center sponsored a comprehensive AG channel measurement study of the C-bands. These measurements covered nearly all the typical local ground environments, including overwater, hilly, mountainous, suburban, and near-urban areas. Researchers in \cite{SAGIN_AG_2} also conducted wideband measurements using a direct-sequence spread-spectrum channel measurement platform to characterize propagation at 2.05 GHz. In \cite{SAGIN_AG_3}, a channel measurement for a low-altitude AG channel was conducted at 915 MHz. In \cite{SAGIN_AG_5}, AG channel measurements were conducted in open and suburban areas at 4.3 GHz. In \cite{SAGIN_AG_6}, authors studied time-varying AG channels based on the measurement data at 6.5 GHz, with a particular focus on the clustering and tracking of multipath components. In \cite{SAGIN_AG_K2}, the measurement of the AG channel in a near-airport scenario at 5 GHz was conducted. In addition, AA channels for UAVs were investigated based on laboratory measurements and several experimental flights in \cite{SAGIN_AA_1}. \cite{9460824} measured the short-range AS channel at 5 GHz in the Pacific Ocean using a channel sounder based on Direct Sequence-Spread Spectrum (DS-SS) technology, and verified the proposed non-stationary channel model for unmanned aerial vehicles to ships. \cite{10132977} measured the AS channel at 2.2 GHz using NI USRP software-defined radio equipment and evaluated the communication performance of the system. According to Table~\ref{SAGINtable1}, it can be found that measurements of aerial channels are conducted mainly in AG scenarios, channel measurements in AA and other scenarios are scarce. Besides, measurement campaigns were conducted in relatively low frequency bands, which are from 0.92 to 6.5 GHz. For channel characteristics, in addition to path loss and shadow fading, AG channel measurements also focus on small scale characteristics such as delay spread, Ricean K-factor (K-factor), Doppler spread and others.

\subsection{SAGIN Channel Characteristics}
For SAGIN scenarios, especially for SG and AG channels, propagation loss is one of the most important channel characteristics due to the long communication distance. Shadow fading, which is caused by trees or buildings on the ground, could introduce signal attenuation on the order of tens of dB in the received signal. Small-scale characteristics, such as delay spread and Doppler spread, have attracted considerably more attention, especially for LEO satellite wideband communications. In addition, meteorological effects, e.g., rain, snow and fog attenuation, have a greater impact on AG or SG channels compared with terrestrial communication systems. A detailed analysis of the above channel characteristics is given below.

\subsubsection{Propagation loss and shadow fading}
Because of the long propagation distance in the SG channel, the basic propagation loss of the SG is considered free space propagation, which can be described by the Friis equation \cite{SAGIN_Friis}. The signal will suffer a large propagation loss of several hundreds of dB. In SG channels, shadow fading is mainly caused by landforms or large objects on the ground, like buildings and trees. The analysis in \cite{SAGIN_SG_SF3} demonstrated that tree shadowing is a key factor dictating the fade margin for land-mobile satellite systems compared with fading caused by multipath fading. Many studies on shadow fading in SG channels focus on tree shadowing. For example, in \cite{SAGIN_SG_SF1, SAGIN_SG_SF2}, the authors found that a single tree led to an additional attenuation ranging from 10 to 20 dB with an average median attenuation of 12 dB. In \cite{SAGIN_SG_SF4}, the additional attenuation values caused by trees with and without leaves were 23 and 7 dB, respectively.

\subsubsection{Small-scale characteristics}
Small-scale characteristics, including delay spread, Doppler spread, and the K-factor are important for AG channels. In \cite{SAGIN_AG_4}, research on AG overwater channels was conducted. The authors found that the delay spread of the AG channel in an overwater scenario is typically small, approximately 10 ns, because the propagation channel is dominated by the LOS path and the reflected path from the water surface. Researchers in \cite{SAGIN_M_SUN} found that in a hilly terrain suburban environment, the values of delay spread are mainly within the range of 10-20 ns, but a larger value of nearly 1000 ns also occasionally occurs. For Doppler effect, researchers in \cite{SAGIN_AG_3} conducted channel measurements at 915 MHz and found that a max Doppler shift of 100 MHz at 915 MHz occurs when the UAV is moving directly away from the land station \cite{SAGIN_AG_3}. In \cite{SAGIN_AG_K1}, the K-factor was measured in urban areas for the L and C-band. The mean value of the K-factor was 12 dB in the L-band and 27.4 dB in the C-band. In addition, the correlation relationships between the K-factor and elevation angle, as well as the K-factor and link distance were investigated in \cite{SAGIN_AG_2} and \cite{SAGIN_AG_K2}.

\subsubsection{Meteorological effects}
Compared with terrestrial communication, SG channels could suffer significant meteorological effects, such as atmospheric absorption, rain, fog, and ionospheric effects. These effects could change with frequency, elevation angle, atmospheric pressure, temperature, etc. In \cite{SAGIN_SG_other2}, the authors conducted experimental channel measurements in the Ku-band in a tropical area to study the impact of various meteorological effects, such as rain, clouds, and tropospheric scintillation. According to the measurement data, the rain attenuation reached approximately 2 dB when the accumulated rainfall (AR) was less than 0.2 mm. When the AR was 0.6 mm, the normalized mean and maximum signal fade levels were 5 and 6.4 dB, respectively.

A series of models have been suggested to predict meteorological effects by ITU-R. In ITU-R P.676 \cite{SAGIN_ITU3}, an atmospheric attenuation model is provided, which primarily considers the absorption effects of oxygen and water vapor molecules below 350 GHz. Additionally, ITU-R P.618 offers a method for predicting rain attenuation, which is non-negligible in the mmWave frequency band \cite{SAGIN_ITU2}. In addition, ionospheric effects lead to rapid fluctuations in the amplitude and phase, changes in the polarization angle, and a spread of the delay of the received signal. Models for ionospheric effects, such as ionospheric scintillation and Faraday rotation, are provided in ITU-R P.531 \cite{SAGIN_ITU4}.

\subsection{SAGIN Channel Modeling}
As mentioned in Section II.B, the current main channel modeling method is statistical channel modeling based on GBSM. In the early years, the SAGIN channels, especially SG channels, were modeled by statistically modeling the envelope of the received signal. Generally, the SG channel model can be expressed as the sum of a log-normally distributed random phasor and a Rayleigh phasor \cite{SAGIN_M_CLOO}:
\begin{equation}
    r = ze^{j\phi _{0} } + we^{j\phi},
\end{equation}
where $r$ is the received signal envelope, $ze^{j\phi _{0} }$ represents the LOS component, $we^{j\phi}$ represents other multipaths, the phases $\phi _{0}$ and $\phi$ are uniformly distributed between $0$ and $2\pi$, $z$ is log-normally distributed , and $w$ has a Rayleigh distribution. In recent years, meteorological effects have been gradually taken into consideration. In \cite{SAGIN_M_WZL}, a channel model taking into account weather impairments was proposed. In this case, the PDF of the received signal amplitude is
\begin{equation}
p(r)=p_{\beta}(r)p_{w}(r),
\end{equation}
where $p_{\beta}(r)$ and $p_{w}(r)$ are the PDFs of the mobile fading and weather impairments, respectively. Besides, \cite{SAGIN_SG_other2} proposes a tropical weather-aware land-mobile satellite channel model.
In this model, multipath fading of SG channels has been modeled based on the Rayleigh distribution, and meteorological effects such as rain, clouds, and tropospheric scintillation are modeled by adjusting the K-factor
\begin{equation}
K_{\rm total}=K_{\rm LOS}-K_{\rm Rain}-K_{\rm Cloud},
\end{equation}
where $K_{\rm LOS}$ is the K-factor under a clear LOS environment, $K_{\rm Rain}$ is the K-factor caused by the rain, and $K_{\rm Cloud}$ is the K-factor caused by the cloud and fog in the environment.

In addition to the above models, 3GPP TR 38.811 \cite{38811} provides channel models for non-terrestrial network (NTN) communication (NTN model). The framework of the NTN channel model is similar to \cite{38901} with a few modifications. First, the supported scenarios (dense urban, urban, suburban, and rural) and frequency bands (S-band and Ka-band) of the NTN model are modified. Second, the Earth-centered, Earth-fixed (ECEF) coordinate system is used in the NTN model to define the positions of satellites and ground stations instead of the traditional Cartesian coordinate system. Third, the NTN model introduces a circular aperture antenna model as a supplementary antenna configuration. Fourth, the elevation angle of the ground station becomes an important parameter for calculating large-scale and small-scale parameters. Fifth, a series of models are introduced to the NTN model to support the simulation of NTN channel characteristics, such as the clutter loss model used to capture interference from nearby buildings and ground objects. Besides, meteorological effects like atmospheric, rain, and ionospheric fading described previously are supported \cite{SAGIN_ITU2, SAGIN_ITU3, SAGIN_ITU4} in the NTN model. 

\subsection{Summary and Prospects}
As discussed above, SAGIN is considered a resolution to achieve the ubiquitous connectivity of 6G, and much research on the SAGIN channel has been carried out. Channel measurements of SAGIN segments like SG and AG have been conducted in many scenarios and frequency bands. Channel characteristics, such as propagation loss, shadow fading, meteorological effects, delay spread, K-factor, Doppler spread, and others have been studied and modeled. However, to improve the capability of SAGIN, further research on the channel is needed.

\subsubsection{More SAGIN channel measurements}
As discussed above, SAGIN includes satellite systems, aerial networks, and terrestrial communications. A number of channel measurements have been conducted in some segments of SAGIN, such as SG and AG. However, measurements for other segments of SAGIN, such as space-to-space (SS) and space-to-air (SA), are scarce. In addition, SAGIN segments always have large spatial scales. Thus the development of channel measurement platforms that have the ability to measure SAGIN channels is increasingly important. Besides, to obtain and analyze the channel characteristics of the integrated network, channel measurements in scenarios where different segments of SAGIN work together are needed.

\subsubsection{Futher exploring for SAGIN channel characteristics}
Channel characteristics, such as propagation loss and shadow fading, delay spread, Doppler spread, K-factor, and meteorological effects have been studied in recent years. However, research on other channel characteristics, such as angular spread, is scarce. In addition, because of the lack of measurements, the channel characteristics of some segments, such as SA, SS, and others, have not been deeply studied. Besides, the interaction between different segments of SAGIN should be considered and studied in future research.

\subsubsection{SAGIN channel modeling}
Channel models for SAGIN segments like SG and AG have been proposed in recent years. For large-scale channel characteristics like propagation loss, channel models of SG and AG can be obtained by modifying the existing propagation loss model. For small-scale channel characteristics, some statistical channel models were proposed in the early years, and 3GPP proposed channel models for simulating fast fading \cite{38811}. Besides, ITU provided channel models to capture meteorological effects like atmospheric attenuation, rain attenuation, and others \cite{SAGIN_ITU2, SAGIN_ITU3, SAGIN_ITU4}. However, channel models for other segments like SS and SA are still under research. Furthermore, there is a lack of a unified channel model for the integrated SAGIN segments, rather than channel models for a single segment.

\section{6G Channel Simulator}\label{SecSimu}
Channel simulator software is used to generate CIRs in different scenarios and configurations on the basis of the existing modeling framework and configuration parameters. A channel simulator can play an important role in designing, evaluating, and optimizing communication systems. In this section, channel simulators capable of supporting potential 6G technologies are reviewed, and BUPTCMCCCMG-IMT2030 is introduced as an example of a 6G channel simulator. A series of extensions for the 6G technologies and scenarios described in Sections III–VII were integrated. The simulation principles, structure, functions and applications are given below.

\subsection{Development of Channel Simulators for 6G}
In each wireless generation, a few channel simulators were developed and played important roles in performance evaluations of communication technologies. In the 4G era, WINNER II provided channel models and corresponding implementations for link- and system-level simulations \cite{WINNER2}, including generic and clustered delay line models defined for different propagation scenarios. In the 5G era, several organizations have developed simulators for link- and system-level channel simulations. In 2014, the Quasi Deterministic Radio Channel Generator (QuaDriGa) was released by the Fraunhofer Heinrich-Hertz Institute \cite{CMG_Ref2}. In 2017, New York University (NYU) released a channel simulator named NYUSIM \cite{CMG_Ref1}, which is based on the statistical spatial channel model proposed by NYU. In 2018, IMT-2020 CM-BUPT \cite{CMG_Ref3} was launched as an implementation of the widely recognized 5G standard GBSM channel model defined in ITU-R M.2412 \cite{2412} and 3GPP TR 38.901 \cite{38901}. In recent years, with the development of 6G research, a series of channel simulators have been released to capture the channel characteristics of 6G technologies. In 2020, SimRIS was released \cite{CMG_SimRIS}, providing a propagation channel for RIS-assisted communication research in various indoor and outdoor environments. In 2021, researchers from the Higher Institute for Applied Sciences and Technology (HIAST) published the TeraMIMO channel simulator for wideband MIMO THz communications \cite{CMG_TeraMIMO}. In addition, in 2023, the NYUSIM was updated to version 4.0, which supports simulations of THz frequency bands.
In response to potential 6G technologies, including ISAC, XL-MIMO, THz, and RIS, research institutions and universities worldwide have developed various simulation platforms. The team from New York University compiled a comprehensive global survey of simulators in 2023 \cite{NYUSIM}, encompassing channel simulators, link-level simulators, system-level simulators, and network simulators.
In this work, we specifically focus on conducting a systematic review of 6G channel simulators worldwide, as summarized in Table \ref{Sum of simulators}. The analysis reveals that the majority of channel simulators target THz band characterization, whereas a minority address specific domains. For example, NirvaWave specializes in near-field channel modeling, and SimRIS focuses on RIS channel emulation. Notably, BUPTCMCCCMG-IMT2030 stands out as one of the few simulators capable of supporting multiple emerging technologies simultaneously.

\begin{figure*}[t]
    \centering
    \includegraphics[width=14cm]{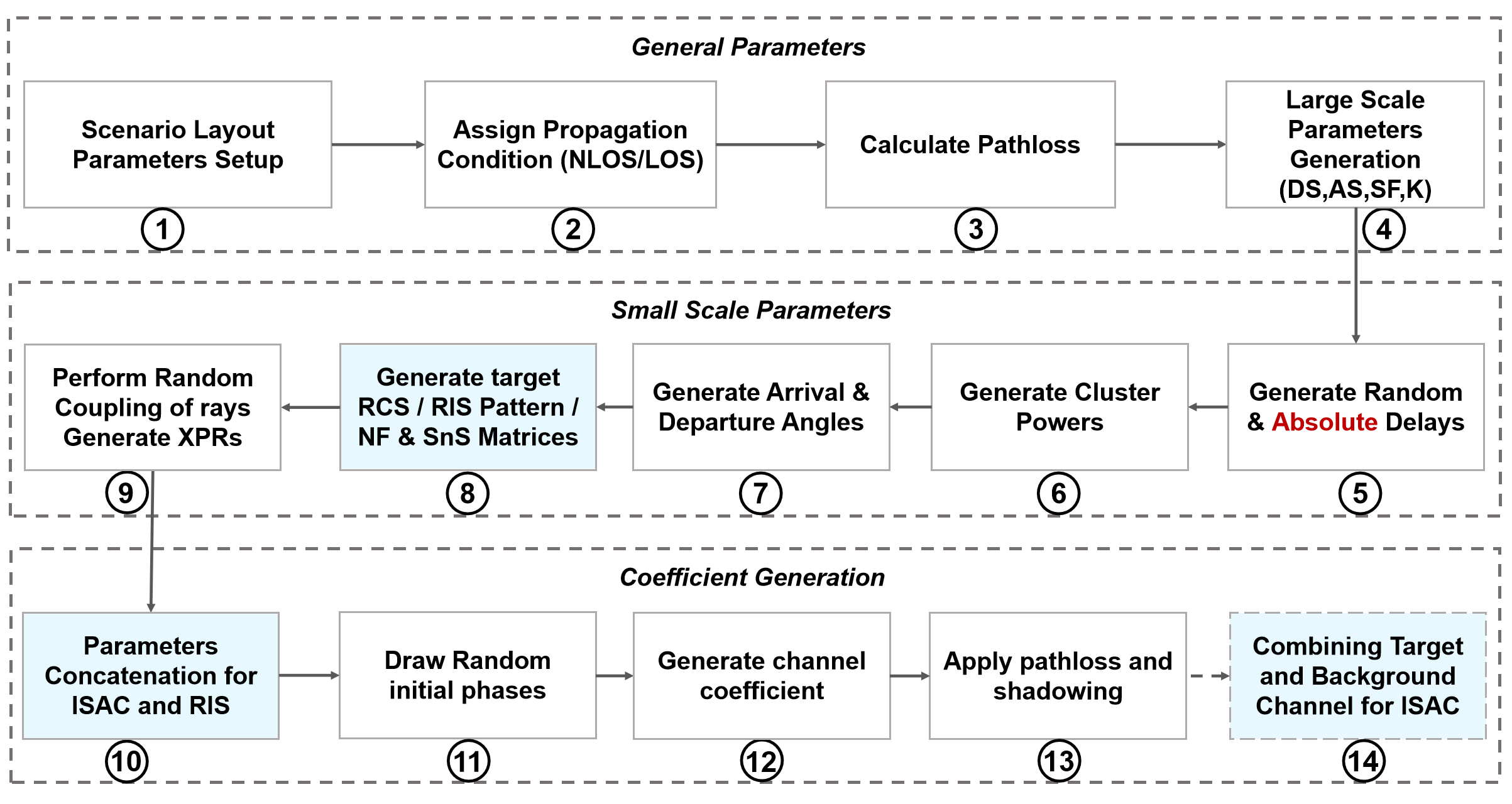}
\caption{Simulation flowchart for the BUPTCMCCCMG-IMT2030 based on the E-GBSM model.}
    \label{fig_CMG1_1}
\end{figure*}

\begin{figure*}
    \centering
    \includegraphics[width=18cm]{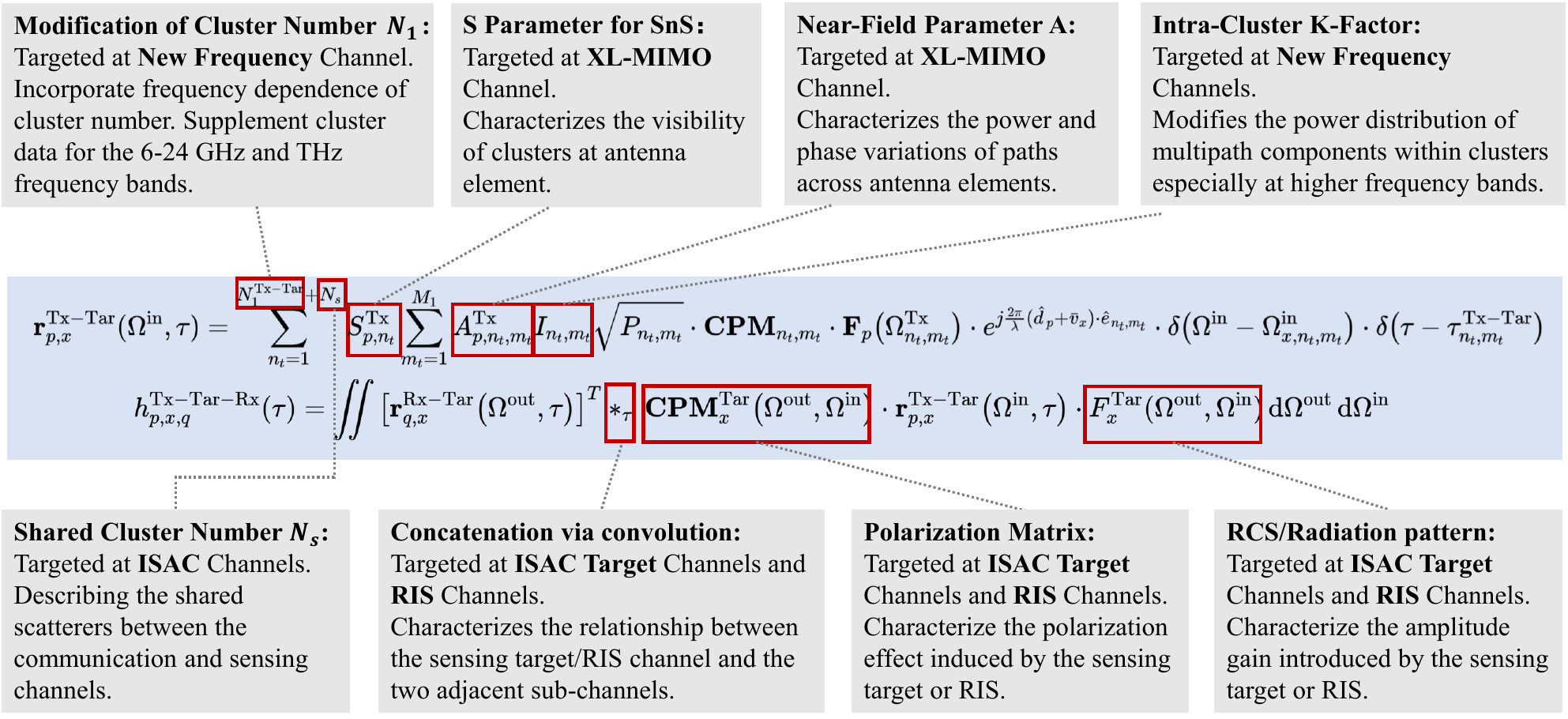}
    \caption{Principle of the BUPTCMCCCMG-IMT2030 simulator based on the E-GBSM model \cite{EGBSM}.}
    \label{fig:platform_H}
\end{figure*}

\begin{figure*}[htbp]
\centering
\includegraphics[width=14cm]{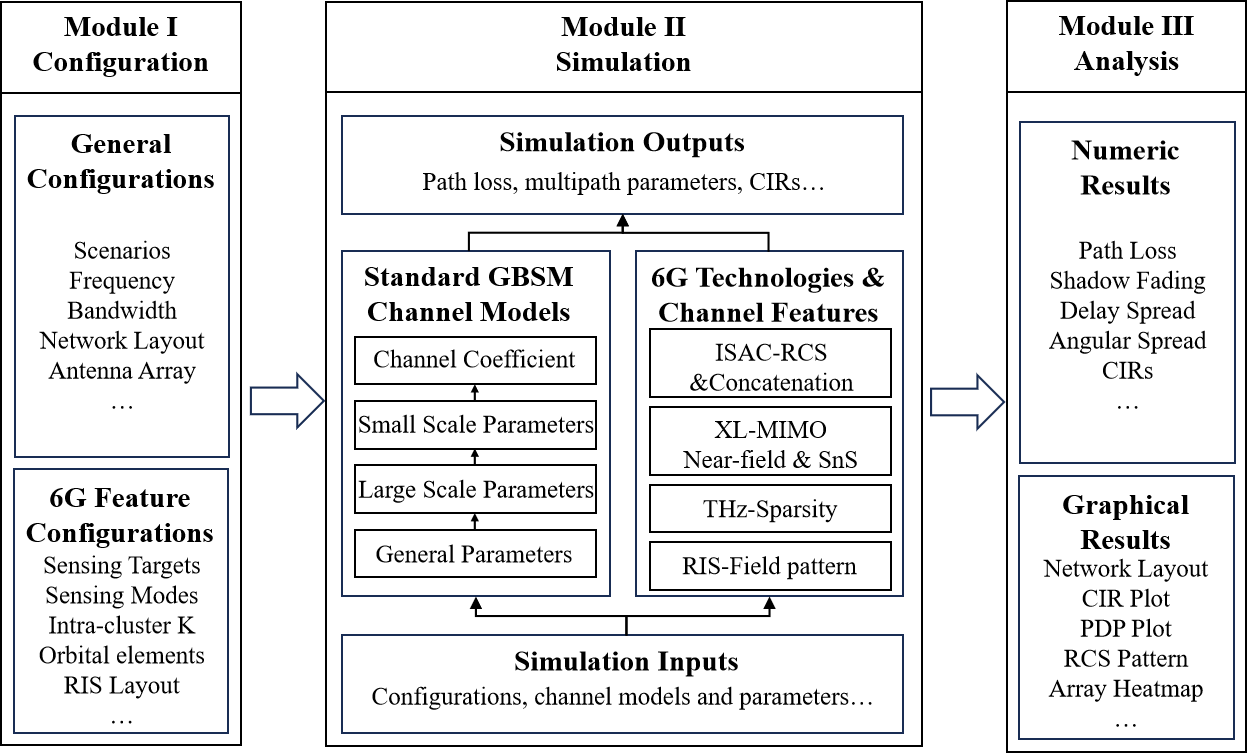}
\caption{Simulation framework of BUPTCMCCCMG-IMT2030.}
\label{fig_CMG1}
\end{figure*}

To meet the requirements of channel simulations supporting higher frequency bands, larger-scale antenna array, ISAC, RIS, and SAGIN, our team developed a 6G channel simulator called BUPTCMCCCMG-IMT2030 on the basis of the framework of the 5G standard GBSM channel model. It supports the generation of CIRs for various scenarios, including urban macrocells (UMa), UMi, rural macrocells (RMa), InH, and SAGIN.  New channel features and parameters are also integrated into the platform to support 6G key technologies, such as ISAC, XL-MIMO, THz, and RIS.

\subsection{Simulation principle of BUPTCMCCCMG-IMT2030}

BUPTCMCCCMG-IMT2030 is implemented on the basis of the procedure-oriented framework of the 5G standard GBSM channel model, with a series of extensions for 6G features. As shown in Fig.~\ref{fig_CMG1_1}, the framework of the basic simulation procedure is briefly summarized as follows \cite{38901}.
To enable these extensions in a systematic way, the platform relies on the extended GBSM (E-GBSM) principle as its theoretical foundation. This principle provides a unified mechanism to incorporate new channel characteristics beyond the 5G model.

The E-GBSM principle has been detailed in \cite{EGBSM}. In this paper, we take several representative 6G features—such as near-field effects, SnS, channel concatenation, and target RCS—as examples to illustrate how these characteristics can be incorporated into a unified expression under the E-GBSM framework. The basic principle is depicted in Fig.~\ref{fig:platform_H}.

The first equation, $\mathbf{r}_{p,x}$, represents the non-concatenated channel response in the angle and delay domains, such as for Tx–Target, Target–Rx, or standard Tx–Rx communication links. The components highlighted in red boxes indicate the extensions introduced for various 6G channel features. For example, $S_{p,n_t}^{\mathrm{Tx}}$ describes the SnS behavior at the Tx antenna array (i.e., visibility of clusters), as defined in Eq.~\ref{Eq_SnS}. $A_{p,n_t,m_t}^{\mathrm{Tx}}$ models the near-field effect through a second-order phase correction, with its expression provided in Eq.~\ref{equ:mod_mani}. When modeling high-frequency sparsity, the parameter $I_{n_t,m_t}$ (intra-cluster K-factor, ICK) can be introduced to redistribute power among paths within a cluster, concentrating energy around the LOS component \cite{THz_Sparsitylxm}. Additionally, if the sharing feature in the ISAC channel is considered, $N_s$ denotes the number of shared scatterers simultaneously contributing to both communication and sensing links, as described in earlier sections.

The second equation describes the time-domain representation of a concatenated channel, using the Tx–Target–Rx link as an example. Here, the amplitude gain whether representing a RIS radiation pattern or target RCS is denoted by $F^{\mathrm{Tar}}$, and the polarization interaction is modeled via the matrix $\mathrm{CPM}^{\mathrm{Tar}}$. The $*_{\tau}$ symbol denotes a delay-domain convolution operation at the small-scale level, indicating that the overall Tx–Target–Rx channel is constructed by convolving the two subchannels, Tx–Target and Target–Rx, in the delay domain.

\subsection{Software Structure and Functions}

The software platform structure of BUPTCMCCCMG-IMT2030 is illustrated in Fig.~\ref{fig_CMG1}. The software consists of three modules: configuration, simulation, and analysis. In Module I, the configuration module and system parameters, such as the scenario, center frequency, and bandwidth, are defined. In Module II, the simulation module, BUPTCMCCCMG-IMT2030, performs a simulation and generates a series of channel parameters. In Module III, the analysis module, simulation results are analyzed and presented in numerical and graphical formats. A stepwise description of the simulation is given below.

\subsubsection{Module I: Configuration}
In the configuration module, there are two types of configurations: general configurations and 6G feature configurations. The general configurations include basic system parameters for a complete simulation, including the scenario, center frequency, bandwidth, positions of the BS and UE, antenna type and layout, and mobility of the UE. 6G feature configurations are a series of parameters for new channel features for 6G. Various 6G channel features are supported, including an extension of fast-fading parameters and sparsity characteristics for the THz channel; near-field and SnS characteristics for the XL-MIMO channel; and other features for channels, i.e., SAGIN, ISAC, and RIS.

\begin{table*}[htbp]
\centering  
\caption{A comparative analysis of the channel simulators for 6G}
\begin{tabular}{p{3cm}|p{2cm}|p{8cm}|c}
\hline  \hline
 \centering Channel Simulator & \centering Developer & \centering Channel Simulation Capability  & Release Time

\\ \hline BUPTCMCCCMG-IMT2030 &  ARTT Lab of BUPT  & A channel model based on measurements ranging from 0.5 to 330 GHz \cite{THz_Ref_3}, supporting 6G new features and technologies such as ISAC, near field, SnS, RIS, NTN, etc. The simulator is fully compatible with standard channel models such as ITU-R M.2412, 3GPP TR 38.900/901, TR 36.777, TR 36.873 and TR 37.885, and has been calibrated based on TR 38.901 as the baseline. The developer was fully involved in the 3GPP R19 ISAC and 7--24 GHz channel modeling agenda. The simulator possesses the latest R19 channel simulation capabilities. & June 2023 \\ \hline  

QuaDRiGa v2.8.1 & Fraunhofer Heinrich Hertz Institute (HHI) \cite{CMG_Ref2} & Compatible with 3GPP TR 36.873, TR 37.885, TR 38.901 TDL, CDL channel models; compatible with mmMAGIC channel model, satellite channel modeling, and supports spatial consistency channel modeling.  & December 2023 \\ \hline

NYUSIM v4.0 & NYU WIRELESS \cite{NYUSIM} & Based on measured statistical channel models, supporting UMi, UMa, RMa, InH, InF scenarios simulation from 0.5 to 150 GHz. & June 2023  \\ \hline

Thor Simulator & Horizon 2020 Joint EU-Japan project \cite{thor} & A simulator for 300 GHz backhaul links. & June 2022  \\ \hline

KUCG v1.0.1 & Kyoto University \cite{kucg} & Used to generate channel impulse responses for the 60 GHz (millimeter wave), 95 GHz and 105 GHz (sub-THz) frequency bands. These responses are based on actual channel measurements conducted by the Orihara Laboratory of Kyoto University. & May 2024 \\ \hline 

SimRIS v2.0.1 & Koç University \cite{CMG_Ref2} & Can be used for channel modeling in RIS-assisted MIMO systems, with adjustable working frequency, terminal position, number of RIS components, and environment. & August 2023  \\ \hline 

NirvaWave & Princeton University \cite{nirvawave}&  A novel near-field channel simulator, built on scalar diffraction theory and Fourier principles, that precisely captures the propagation evolution of arbitrary user defined transmit electromagnetic (EM) signals in complex user defined wireless mediums.  & September 2024 \\ \hline

DeepMIMO v2 & Wireless Intelligence Lab REMCOM \cite{deepmimo} & Based on Remcom 3D ray tracing and parameter set generation to create large-scale MIMO datasets. & 2021 \\ \hline 

BUPTCMCC-6G-DataAI+ & ARTT Lab of BUPT \cite{buptcmcc} & The BUPTCMCC-6G-DataAI+ employs a generative data construction method, generating up to 300 million sets of channel data. It further expands into new scenarios, such as maritime scenario, ultra-high-speed mobility, multiple-perspective sensing, industrial internet, and indoor RIS, and introduces a new intermediate frequency band of 7--24 GHz. This supports the requirements of air interface AI algorithms, including CSI compressed feedback, occlusion and beam prediction, sensing positioning, air interface resource management, and network planning optimization. & September 2024 \\ \hline  \hline

\end{tabular}
\label{Sum of simulators}
\end{table*}

\begin{table*}[htbp]
\caption{Modifications to the standard GBSM model in BUPTCMCCCMG-IMT2030}
\begin{center}
\begin{tabular}{c|c|p{12.7cm}}
\toprule
    \hline
       \begin{tabular}{c}6G Features\end{tabular} & {\begin{tabular}{c} Steps \end{tabular}} & \multicolumn{1}{c}{\begin{tabular}{c}Descriptions\end{tabular}} \\ \hline
     {\begin{tabular}{c}ISAC\\(Shared cluster,\\target scattering)\end{tabular}}& {\begin{tabular}{c}Step 1,\\ Step 3-5,\\Step 7-12 \end{tabular}} &  {\begin{tabular}{l} Step 1: Set the target locations and geometric parameters;\\
        Step 3: Add the path loss calculation of the sensing channels;\\
        Step 4: Add the correlation between the sensing and communication channels;\\
        Step 5: Add the calculation of delays for the sensing channels;\\
        Steps 7-8: Add the calculation of ray angles, RCS, and clutter for the sensing channels;\\
        Steps 9-12: Add the calculation of the sensing CIR.\\
 \end{tabular}}   \\ \hline
      {\begin{tabular}{c} XL-MIMO\\(Near-field\\effect, SnS) \end{tabular}} & \begin{tabular}{c}Step 1,\\Step 11\end{tabular} &  {\begin{tabular}{l}Step 1: The size of the station region for the antenna array is configured;\\
Step 11: Apply the Near-field effect and SnS to channel coefficients.
 \end{tabular}}   \\ \hline
  
        \begin{tabular}{c}Multi-Frequency\\(Sparsity)\end{tabular} & \begin{tabular}{c}Step 3,\\Step 4,\\Step 11\end{tabular} &\begin{tabular}{l}Step 3: Extend the path loss model for the multi-frequency band;\\Step 4: Extend parameters for the multi-frequency band;\\Step 11: Apply selectively the sparsity to channel coefficients.\end{tabular}
   \\ \hline
       {\begin{tabular}{c}RIS\\(Concatenated channel)\end{tabular}} & Steps 1-11 &  {\begin{tabular}{l} Step 1: Set the configuration of RIS;\\

Steps 2-10: Propagation condition, path loss, LSP, and SSP are generated for both\\\hspace{4.75em}Tx-RIS and RIS-Rx links;\\
Step 11: The concatenated channel coefficient of Tx-RIS-Rx is calculated considering\\\hspace{4em}the radiation pattern of RIS.
 \end{tabular}}   \\ \hline

{\begin{tabular}{c}SAGIN\\(SG channel)\end{tabular}}& Step 1-4 &  {\begin{tabular}{l} Step 1: Set the configuration of BS and UE in the ECEF coordinate system;\\
Steps 2-4: Models and parameters for the SAGIN scenarios are used.\\
 \end{tabular}}   \\ \hline
       
       \bottomrule
    \end{tabular}
\end{center}
\label{modificationtable}
\end{table*}

\subsubsection{Module II: Simulation}
The simulation module is divided into 2 segments: the basic simulation and modifications for 6G. The basic simulation framework follows that of the 5G standard GBSM channel model, as shown in Fig.~\ref{fig_CMG1_1}. The realization of 6G features includes the modifications and extensions of the simulation steps.

The basic simulation segment divides the simulation framework in Fig.~\ref{fig_CMG1_1} into four parts. Part 1 includes steps 1 and 2, which generate general parameters, such as the network layout and link state. Part 2 includes steps 3 and 4, which generate the path loss and large-scale distribution parameters. Part 3 includes steps 5-9, which are used to generate small-scale parameters, such as cluster delays, powers, and angles. Part 4 includes steps 10-12, which generate channel coefficients.

The 6G extension segment includes five modifications for the new 6G channel features to the basic simulation segment. A summary of the modified steps and descriptions is shown in Table~\ref{modificationtable}.

\begin{itemize}
\item Modifications for the ISAC channel simulation.

For the ISAC channel simulation, a series of steps are modified. The original steps are used to generate the communication channels, and the modifications are used to generate sensing channels. In step 1, in addition to the layout of the BS and UE, the location and geometric parameters of the sensing targets are configured. In step 3, the path loss of the sensing channels is calculated according to the configurations of the sensing targets. In step 4, the correlation between the sensing and communication channels is applied. In step 5, the delays of the sensing channels are calculated. In steps 7-8, a series of SSPs of the sensing channels, such as ray angles, RCS, and clutter, are calculated. In steps 9-12, the channel coefficients of both the sensing channels and the communication channels are calculated. The detailed modifications and implementation procedures for ISAC channel simulation can be found in \cite{liu2024extend,ISAC-ZCS}.
\end{itemize}

\begin{itemize}
\item Modifications for the XL-MIMO channel simulation.

For the XL-MIMO channel simulation, steps 1 and 11 are modified. In step 1, the size of the station region, which is the number of antenna elements within the station region, is configured. In step 11, the spherical wave and SnS characteristics are applied to calculate the channel coefficient. The phase evolution of the spherical wave is calculated on the basis of the distance between the antenna elements and scatterers, which is estimated according to the angles and delays of the clusters. The SnS characteristic is simulated by applying the visibility variation of clusters on the basis of a two-state Markov chain. The simulation implementation of the XL-MIMO channel is shown in reference \cite{XLMIMO-GTY}.
\end{itemize}

\begin{itemize}
\item Modifications for the THz channel simulation.

For the THz channel simulation, steps 3, 4, and 11 are modified. In step 3, the path loss model is extended to the THz frequency band. In step 4, the parameters for generating multipaths are extended to the THz frequency band. In step 11, the model described in \cite{THz_Sparsitylxm} is applied to introduce sparsity into the THz channel. The detailed modifications and implementation process for THz channel simulation can be found in \cite{THz_Sparsityczw}.

\end{itemize}
\begin{itemize}
\item Modifications for the RIS-assisted channel simulation.

For RIS-assisted channel simulation, BUPTCMCCCMG-IMT2030 has modified a series of steps on the basis of the 5G standard channel model. In the RIS-assisted channel, the channel is divided into two links, Tx–RIS and RIS–Rx, and parameters are generated for them and calculated together. In step 1, the RIS is configured apart from the layout of the BS and UE. In steps 2–10, the propagation condition (LOS/NLOS), path loss, LSPs, and SSPs are generated for both Tx–RIS and RIS–Rx links. In step 11, the concatenated channel coefficients of the RIS-assisted channel are calculated by combining the channel coefficients of the Tx–RIS and RIS–Rx links and applying the radiation pattern of the RIS. For detailed modifications and implementation procedures related to RIS-assisted channel simulation, please refer to \cite{ghw}.

\end{itemize}

\begin{itemize}
\item Modifications for the SAGIN channel simulation.

For the SAGIN channel simulation, the channel model defined in 3GPP TR 38.811 \cite{38811} is implemented and integrated into BUPTCMCCCMG-IMT2030, including modifications to the coordinate system, models, and parameters. In step 1, the ECEF coordinate system is used to locate the BS and UE. In step 2, the propagation conditions (LOS/NLOS) are assigned via a different method. In step 3, the path loss model is modified as described in VII-C. In step 4, the fast-fading parameters for the SAGIN are used, which are related to both the frequency and elevation angle.
\end{itemize}

\begin{table*}[t]

\caption{Configurations for simulation for each 6G case\label{tab_CMG3}}
\begin{center}
\resizebox{\linewidth}{!}{
\begin{tabular}{c|c|c}
\toprule
\hline
6G Features & System Configurations                                                                                                                                   & Channel Feature Configurations                                                                                                                                                                          \\ \hline
ISAC     & \begin{tabular}[c]{@{}c@{}}Scenario: Indoor-office; LOS status: LOS\\ Center frequency: 28 GHz; Bandwidth: 200 MHz\\Position (x,y,z): BS:{[}0,0,1.5{]} (m); UE:{[}10,0,1.5{]} (m)\\Antenna:  BS: Single antenna; UE: Single antenna\end{tabular}    & \begin{tabular}[c]{@{}c@{}}Number of sensing targets: 3\\ Number of sensing clusters: 1\\ Echo path loss: Free space radar path loss\\ RCS parameters: Fixed\\ Method of ISAC: Shared cluster\end{tabular} \\ \hline

XL-MIMO      & \begin{tabular}[c]{@{}c@{}}Scenario: UMi-street canyon; LOS status: LOS\\ Center frequency: 28 GHz; Bandwidth: 200 MHz\\Position (x,y,z): BS:{[}0,0,3{]} (m); UE:{[}10,10,3{]} (m)\\Antenna:  BS: 256 elements ULA; UE: Single antenna\end{tabular}   & \begin{tabular}[c]{@{}c@{}}Stationary region size: 16 elements\end{tabular}                                                                                                       \\ \hline

THz      & \begin{tabular}[c]{@{}c@{}}Scenario: UMi-street canyon; LOS status: LOS\\ Center frequency: 132 GHz; Bandwidth: 1.2 GHz \\Position (x,y,z): BS:{[}0,0,3{]} (m); UE:{[}10,10,3{]} (m)\\Antenna: BS: Single antenna; UE: Single antenna\end{tabular} & Intracluster K-factor: 17.98 dB                          \\ \hline

RIS      & \begin{tabular}[c]{@{}c@{}}Scenario: UMi-street canyon; LOS status: LOS\\ Center frequency: 28 GHz; Bandwidth: 200 MHz\\Position (x,y,z): BS:{[}0,0,3{]} (m); UE:{[}10,10,3{]} (m)\\Antenna:  BS: Single antenna; UE: Single antenna\end{tabular}   & \begin{tabular}[c]{@{}c@{}} RIS size: 32$\times $32 elements\\ Codebook: Continuous anomalous reflection\\ASA: [$10^{\circ}$, $5^{\circ}$, and $1^{\circ}$]\end{tabular}                                                                 \\ \hline

SAGIN      & \begin{tabular}[c]{@{}c@{}}Scenario: Dense urban; LOS status: LOS\\ Center frequency: 2 GHz (S-band), 28 GHz (Ka-band)\\Antenna:  BS: Single antenna; UE: Single antenna\end{tabular}   & \begin{tabular}[c]{@{}c@{}} Latitude of BS: 0 \\ Longitude of BS: 0\\ Height of BS: 600-1500 km\\Elevation angle: $30^{\circ}$, $60^{\circ}$ \\Position of UE: Varies with above configurations\\ \end{tabular}                                                                 \\ \hline
\bottomrule
\end{tabular}
}
\end{center}
\end{table*}

\subsubsection{Module III: Analysis}
In Module III, a simple analysis of numerical and graphical results based on the generated channel coefficients is provided, including the delay spread, angular spread, reference signal receive power (RSRP), Gini index (used for sparsity analysis), network layout, and CIR of a single link.

\subsection{Applications}
\begin{figure*}[!h]
  \centering
    \begin{minipage}[b]{\linewidth}
      \subfigure[]{
        \includegraphics[width=0.5\linewidth]{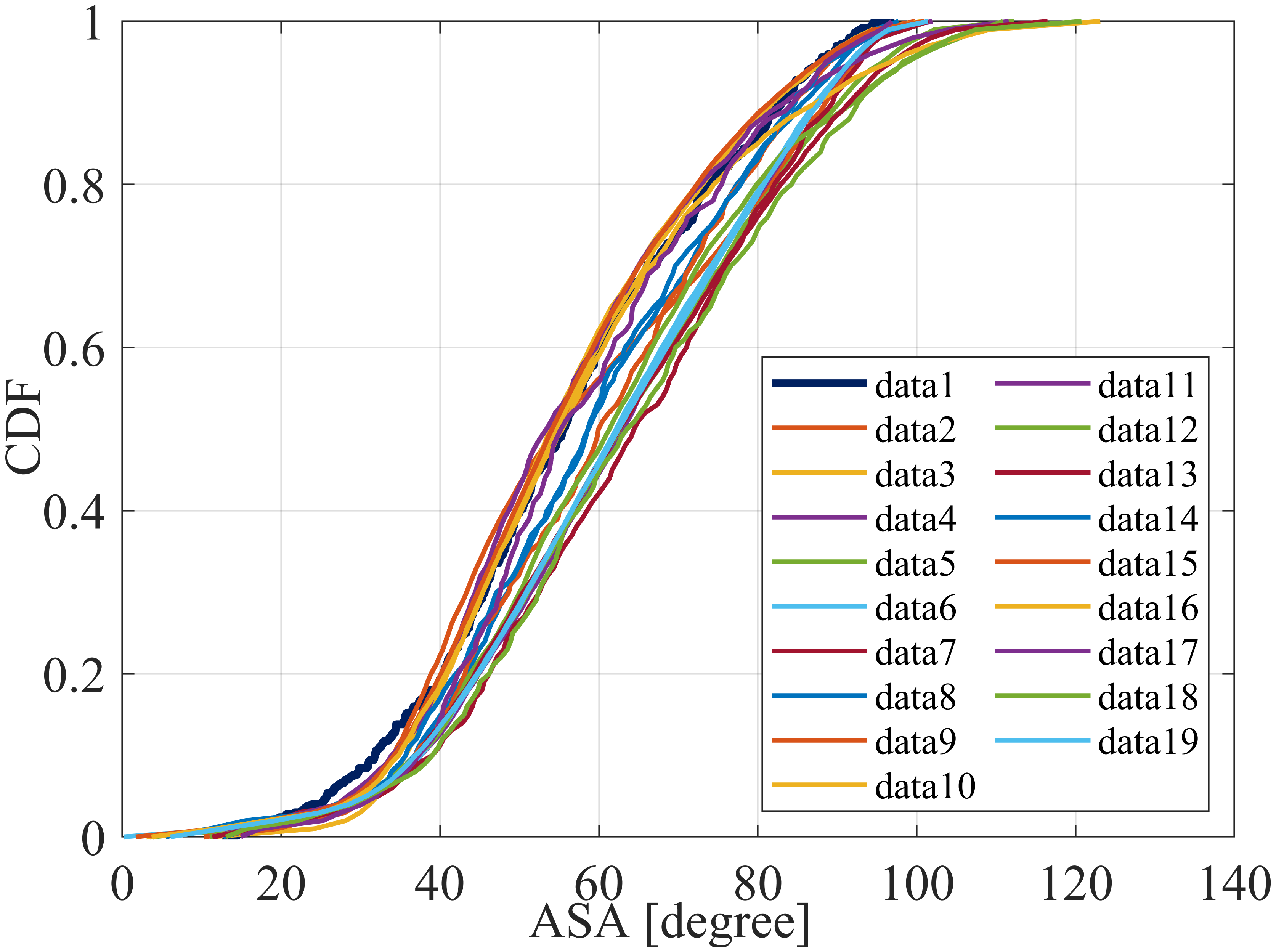}
      }
      \subfigure[]{
        \includegraphics[width=0.5\linewidth]{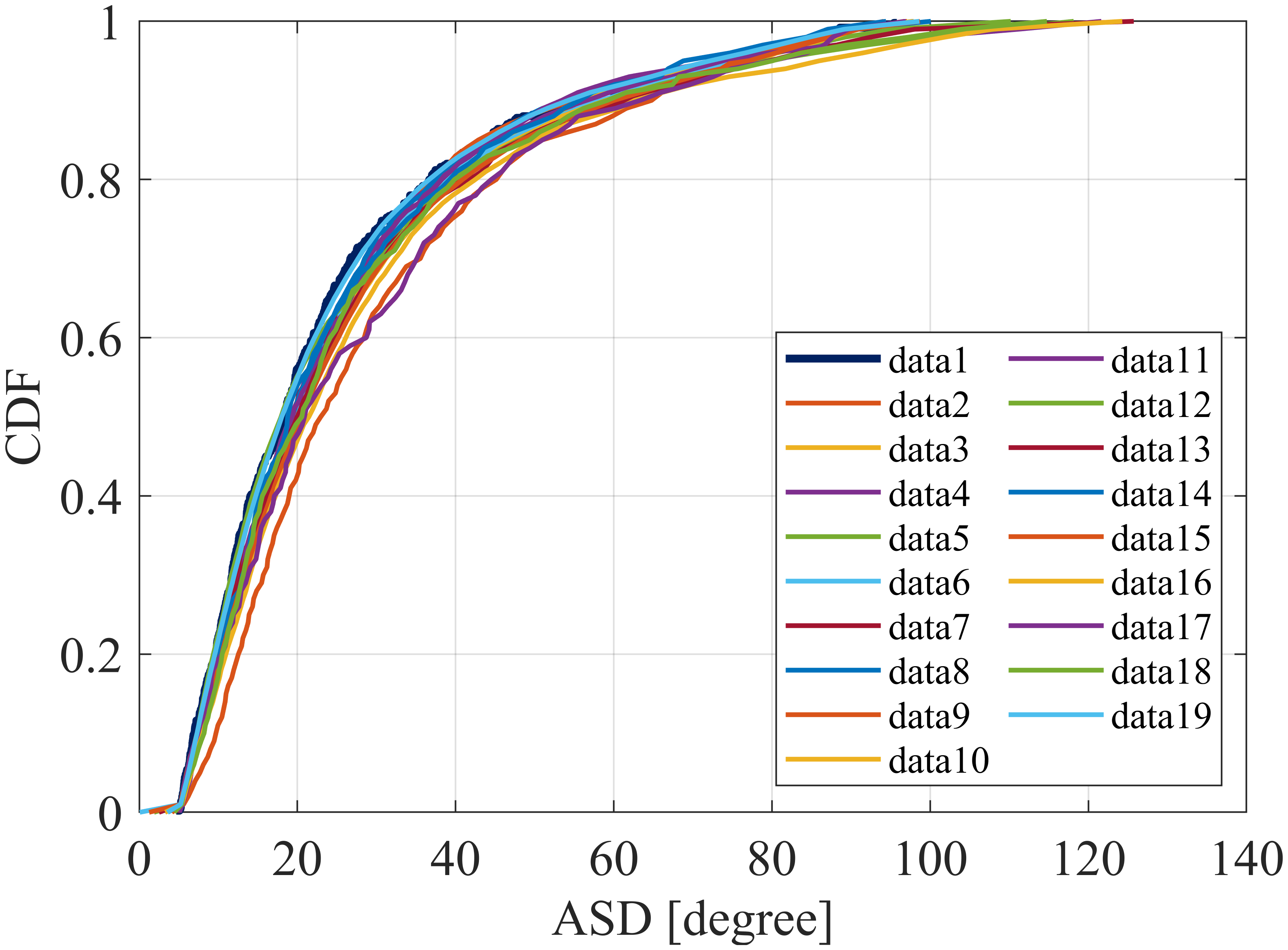}
      }

    \end{minipage}

    \begin{minipage}[b]{\linewidth}

        \subfigure[]{
        \includegraphics[width=0.5\linewidth]{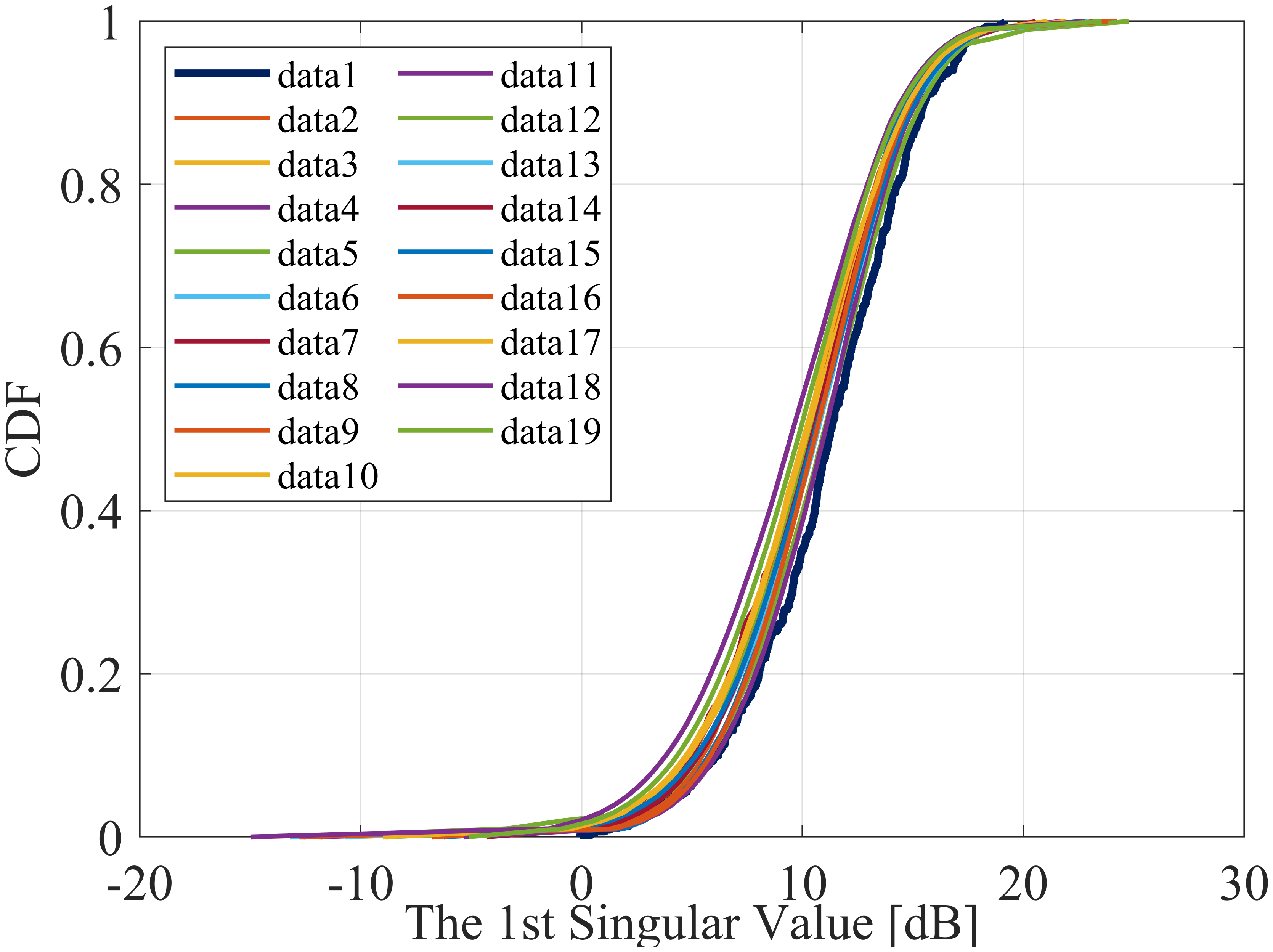}
      }
       \subfigure[]{
        \includegraphics[width=0.5\linewidth]{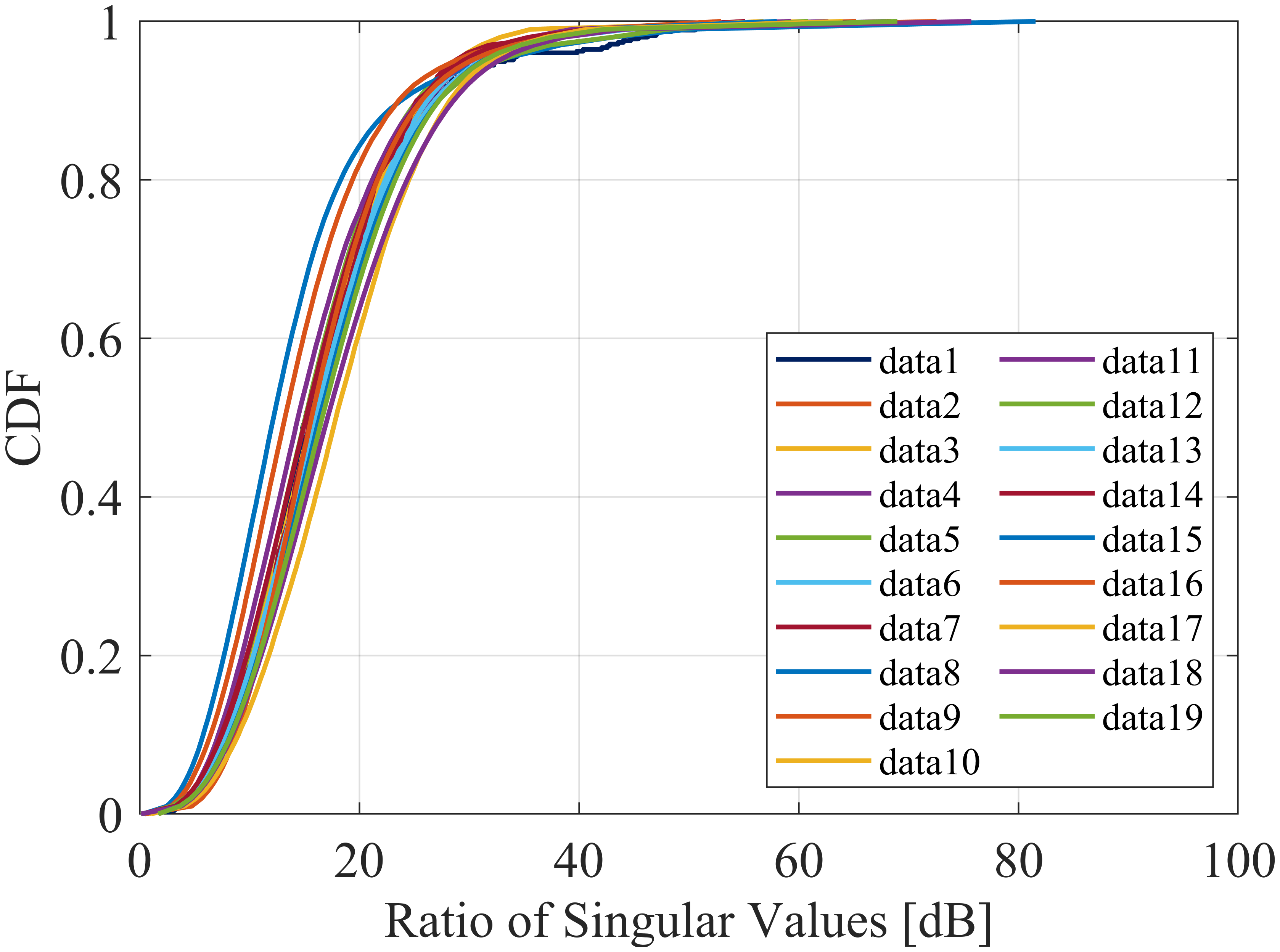}
      }

     \end{minipage}
          
    \begin{minipage}[b]{\linewidth}

      \subfigure[]{
        \includegraphics[width=0.5\linewidth]{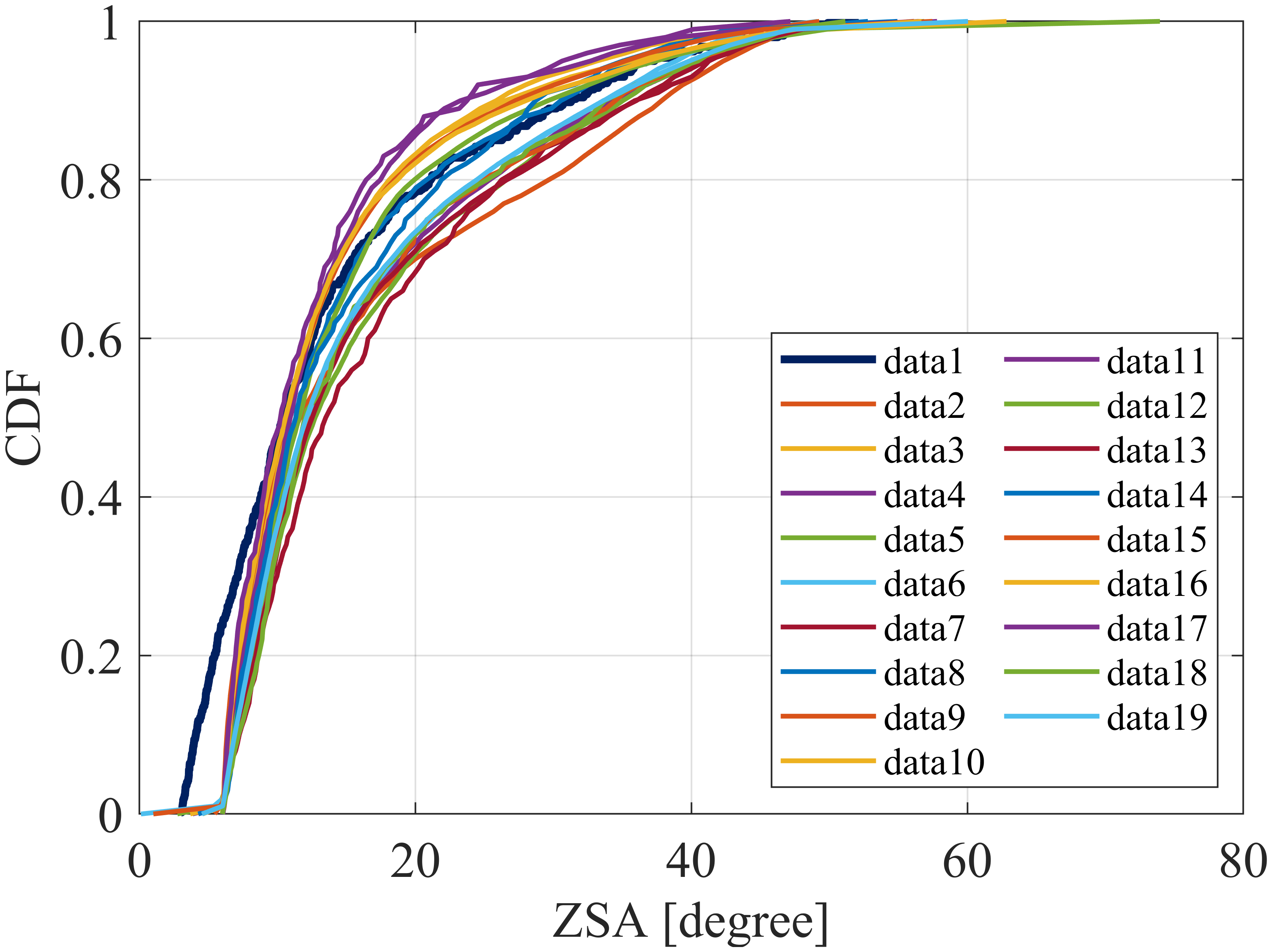}
      }
      \subfigure[]{
        \includegraphics[width=0.5\linewidth]{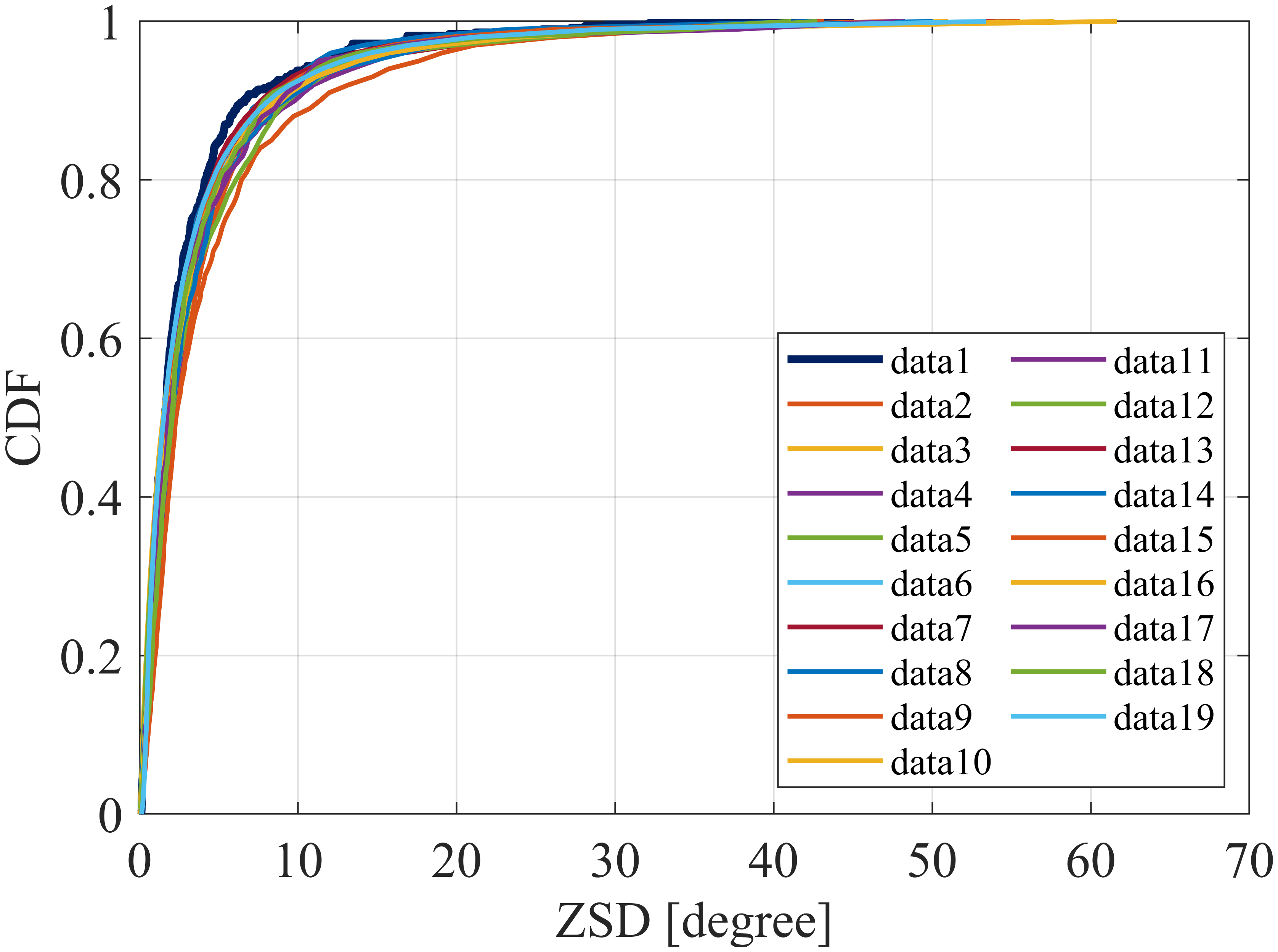}
      }

    \end{minipage}

    \caption{Comparison results with the calibration of 3GPP TR 38.901.}
    \label{fig:901}
  \end{figure*}

The E-GBSM model is based on modeling using measured data, and further utilizes this model, combined with parameters obtained through fitting test data, to construct a simulator. For detailed comparative analysis, please refer to references \cite{liu2024extend,ISAC-ZCS,XLMIMO-GTY,THz_Sparsityczw,ghw}.  Owing to space constraints, this paper will not elaborate further.
The BUPTCMCCCMG-IMT2030 channel model has undergone rigorous calibration and validation against 3GPP TR 38.901, including key channel parameters such as ASA (azimuth spread of arrival), ASD (azimuth spread of departure), ZSA (zenith spread of arrival), ZSD (zenith spread of departure), and singular values. As demonstrated in Fig. \ref{fig:901}, the calibration results (data 1) are in excellent agreement with the measurement data from 18 industrial partners.
The 6G technology extensions are developed on the basis of the TR 38.901 framework as the baseline, ensuring fundamental correctness. Through comprehensive analysis of real-world measurement data, we have incorporated newly observed channel characteristics and enhanced modeling methodologies to extend the framework for 6G applications. This approach maintains backward compatibility with existing standards while enabling accurate characterization of emerging propagation phenomena.

In this section, some application examples of the typical 6G technologies (ISAC, XL-MIMO, THz, RIS, and SAGIN) are provided. The main simulation configurations for different 6G cases are listed in Table~\ref{tab_CMG3}. 

\subsubsection{6G Feature 1--ISAC}

\begin{figure}
    \centering
    \includegraphics[width=3.1in]{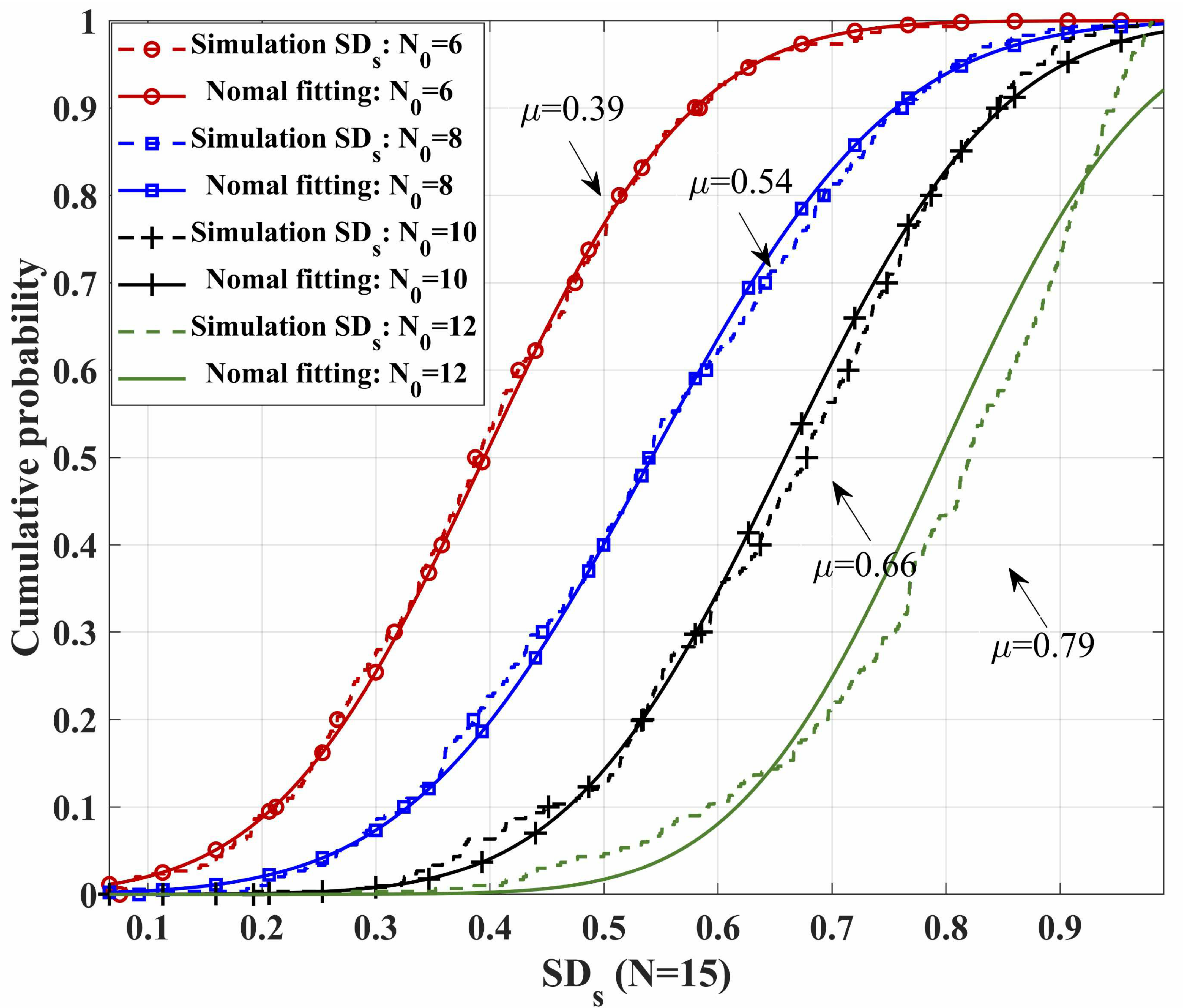}
\caption{Simulation results of the SD ($\text{SD}_\text{s}$) in the sensing channels under different numbers of shared clusters ($N_0$) when the total number of clusters ($N$) is 15.}
    \label{fig_CMG10_3}
\end{figure}

Using the configurations in Table~\ref{tab_CMG3}, we can generate the shared and nonshared clusters of the proposed ISAC channel model in (\ref{eqn_1}). The CDF of the SD metric in the sensing channels, which is the PR of the shared clusters ($N_0$) to the total clusters ($N$) in the ISAC channel, is shown in Fig.~\ref{fig_CMG10_3} under different numbers of shared clusters. A higher SD indicates greater potential to realize the mutual auxiliary functions of communication and sensing. As $N_{0}$ increases, $\text{SD}_\text{s}$ also increases. By controlling the appropriate $N_{0}$ of the proposed channel model, the desired SD value can be achieved, facilitating flexible simulation implementation of ISAC channels.

\subsubsection{6G Feature 2--XL-MIMO}

\begin{figure}
    \centering
    \includegraphics[width=3.1in]{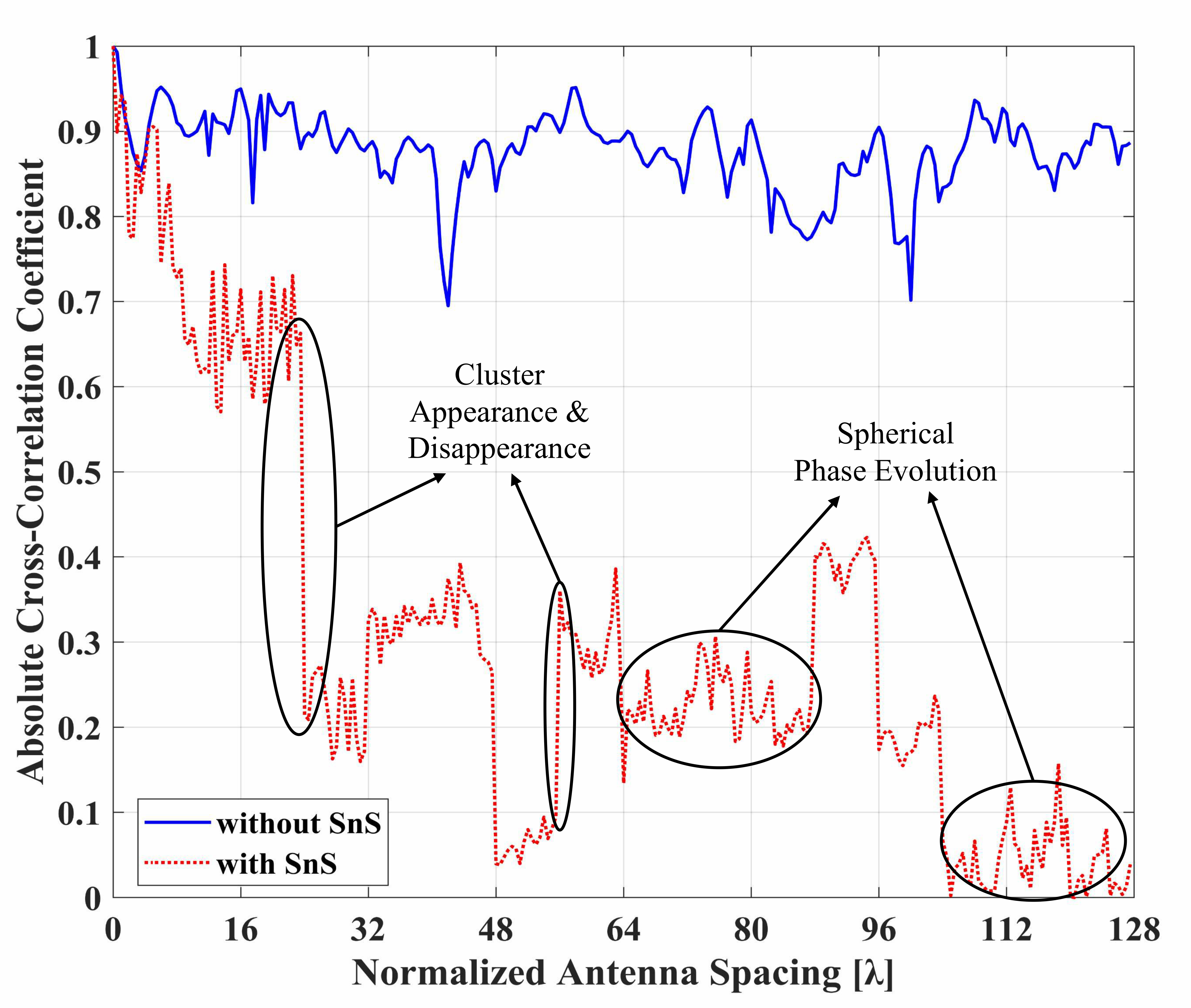}
\caption{Simulation results of the absolute cross-correlation coefficient between the reference element and other elements along the antenna array.}
    \label{fig_CMG10_2}
\end{figure}

On the basis of the configurations in Table~\ref{tab_CMG3}, the absolute cross-correlation coefficient of CIRs between the reference element and other elements in comparison with a channel generated via the standard channel model is shown in Fig.~\ref{fig_CMG10_2}. With increasing spacing between two elements, the cross-correlation coefficient of the standard channel model remains stable, but that of the XL-MIMO channel continues to decrease, indicating that the correlation of the two elements becomes weaker. In addition, the appearance or disappearance of clusters and the evolution of the phase also occur. The appearance or disappearance of clusters results in a sudden change in the cross-correlation coefficients, including an increase in the appearance and a decrease in the disappearance. When no cluster appears or disappears, the evolution of the phase results in a change in the cross-correlation coefficients.

\subsubsection{6G Feature 3--THz Frequency Band}

\begin{figure}
    \centering
    \includegraphics[width=3.1in]{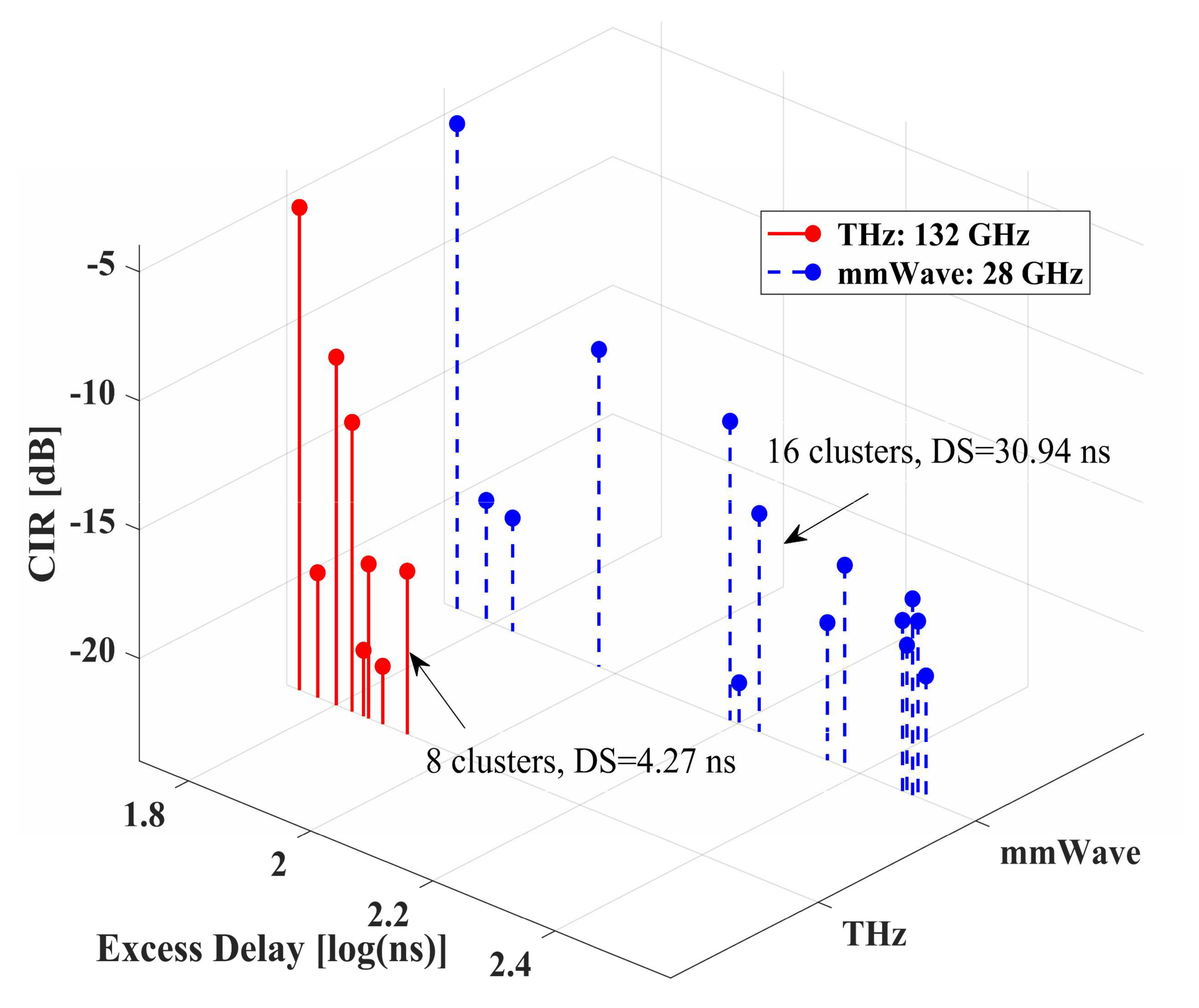}
\caption{Simulation results of channel CIR at THz frequency band (132 GHz) and mmWave frequency band (28 GHz). }
    \label{fig_CMG10_1}
\end{figure}

Using the configurations in Table~\ref{tab_CMG3}, the CIR results of the generated channel at 132 GHz in the THz band were compared with those at 28 GHz in the mmWave band, as shown in Fig.~\ref{fig_CMG10_1}. The THz channel has 8 clusters, and the mmWave channel has 16 clusters. Fewer multipaths are present in the THz band than in the mmWave band, which indicates that the THz channel has a sparser structure than the mmWave channel does. In addition, the delay spread in the THz channel (4.27 ns) is much lower than that in the mmWave channel (30.94 ns), according to Fig.~\ref{fig_CMG10_1}. The multipaths of the THz channel are concentrated in a narrower delay range, which means that the multipaths in the THz channel are concentrated within a small time period. 

\subsubsection{6G Feature 4--RIS}

\begin{figure}
    \centering
    \includegraphics[width=3.1in]{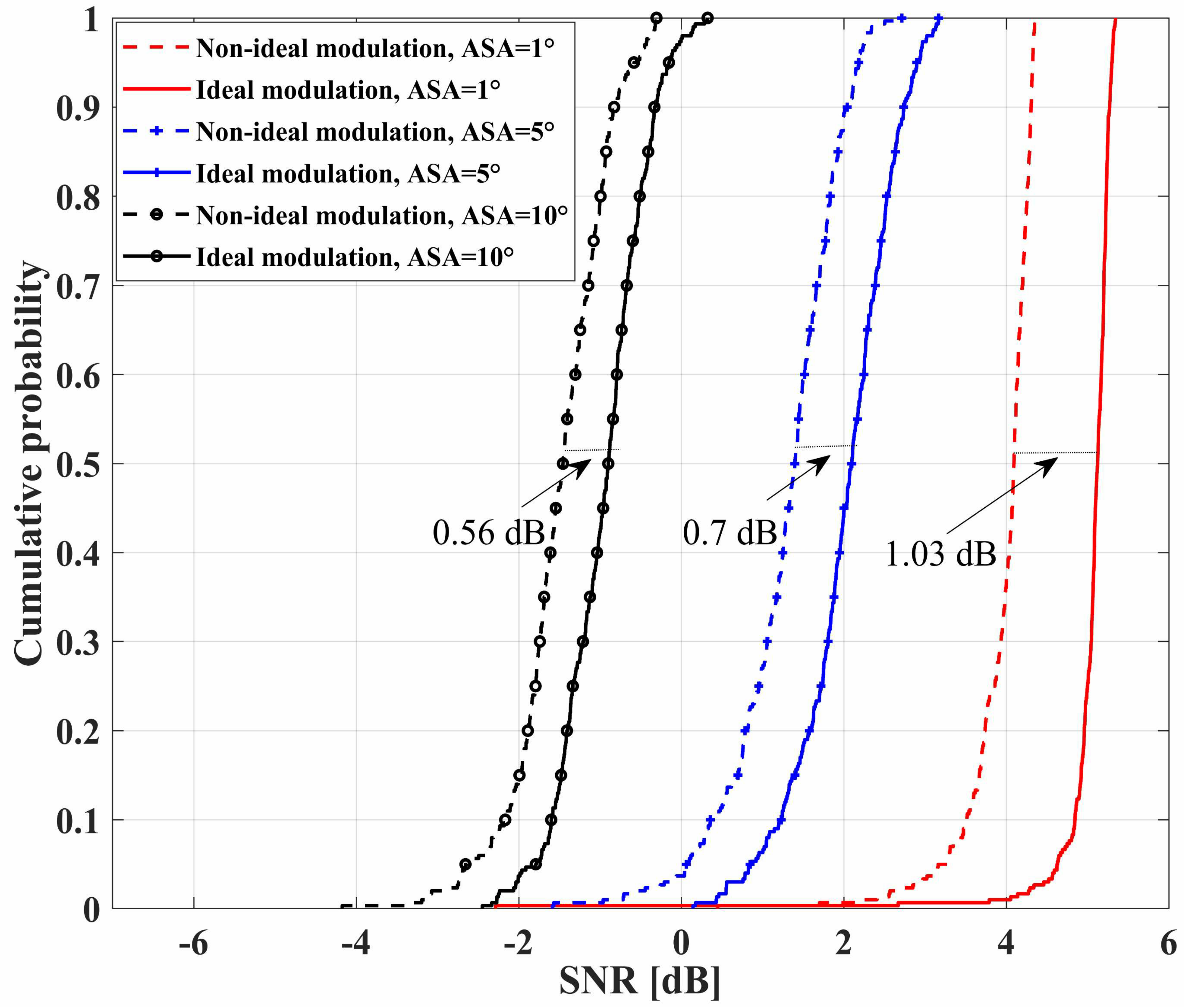}
\caption{Simulation results of instantaneous SNR at the receiver under different channel angular spread for both ideal and non-ideal modulation models of RIS.}
    \label{fig_CMG10_4}
\end{figure}

Using the configuration in Table~\ref{tab_CMG3}, the impact of channel angular spread on the performance of RIS-assisted communication is illustrated in Fig.~\ref{fig_CMG10_4}. With the increase in ASA in the Tx-RIS channel, the performance of the RIS concatenated communication link will worsen. This is because the greater the angular spread, the lower the energy in the specified incident direction.
Additionally, the impact of the angle-dependent characteristics of RIS reflection coefficients on RIS-assisted communication, as mentioned in Section \ref{reflection_coefficient}, has been considered in the simulation. In the legends, "Non-ideal modulation" represents the simulation results when considering this angle-dependent characteristic. It can be observed that the non-ideal modulation results in approximately 0.5, 0.7, and 1 dB of attenuation under the three angle spreads.

\subsubsection{6G Feature 5--SAGIN}

\begin{figure}
    \centering
    \includegraphics[width=3.1in]{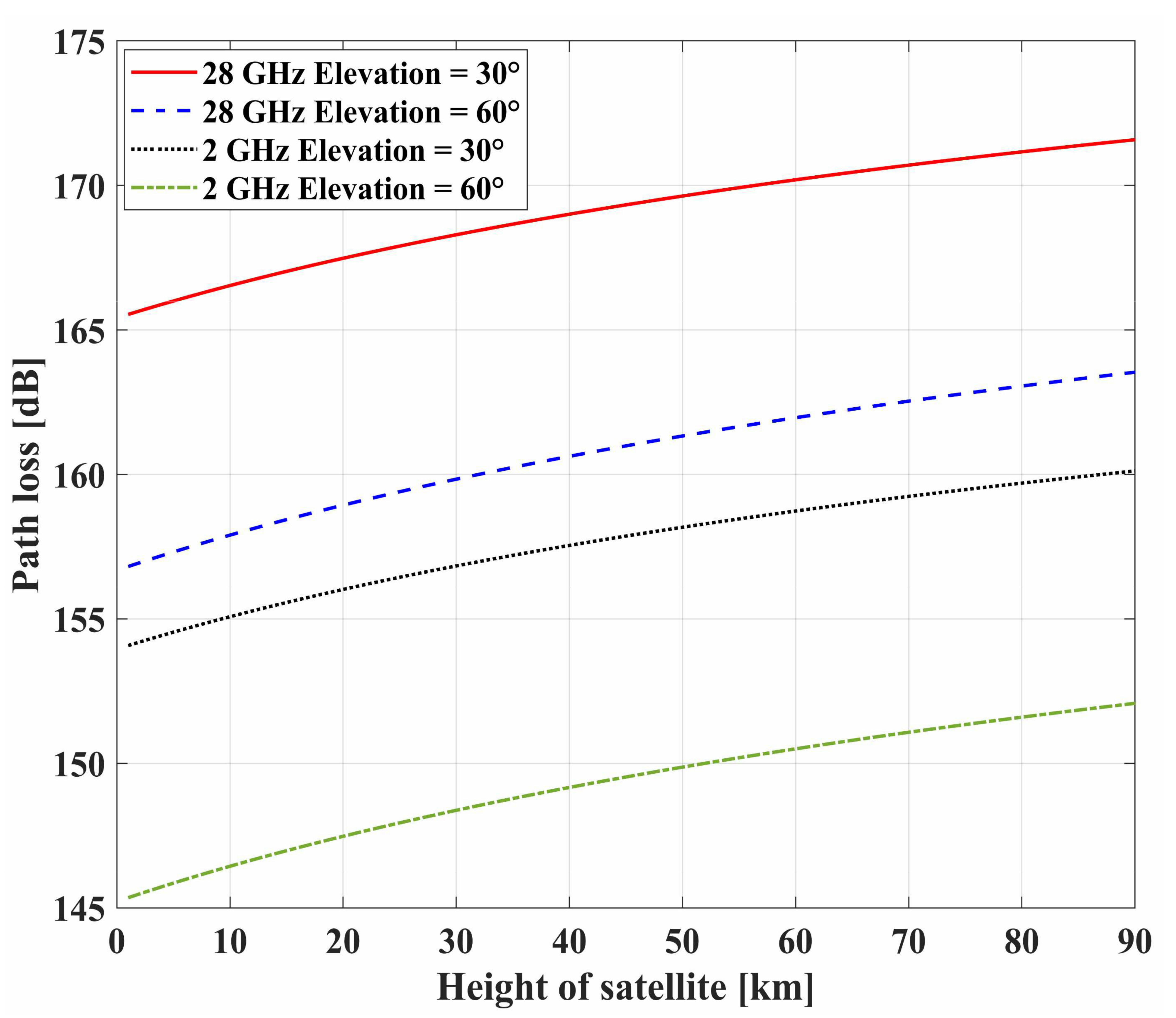}
\caption{Simulation results of the path loss of the SG channel under different elevation angle configurations.}
    \label{fig_CMG10_5}
\end{figure}

Using the configurations in Table~\ref{tab_CMG3}, the path loss of SG channels with different channel configurations is illustrated in Fig.~\ref{fig_CMG10_5}. The path loss of SG channels is affected by the height of the satellites, the elevation angle of the link, and the center frequencies. An increase in satellite height results in an increase in propagation distance, which increases path loss. Furthermore, the path loss increases when a higher frequency band is used. Additionally, the elevation angle of the link can affect path loss in SG scenarios; when the elevation angle increases, path loss increases.

\subsection{Summary and Prospects}
In this section, a summary of 6G channel simulators that support potential enabling technologies is provided. The BUPTCMCCCMG-IMT2030 channel simulator was subsequently introduced, including the design framework, user guidelines, and application examples for potential 6G technologies. To summarize, BUPTCMCCCMG-IMT2030 is a channel model simulator developed for 6G wireless communication by the authors. In the current version, BUPTCMCCCMG-IMT2030 supports simulations for the standardized channel model defined in ITU-R M.2412 \cite{2412} and adds a series of extensions for 6G features, such as ISAC, XL-MIMO, THz, RIS, and SAGIN. Specifically, BUPTCMCCCMG-IMT2030 supports channel simulation scenarios, including UMa, UMi, RMa, InH, and SG in SAGIN, and supports frequency bands, including centimeter waves, mmWaves, and THz waves at 132 GHz. In addition, new channel characteristics, such as the SnS of the multiantenna channel and the sparsity of the THz channel, are also included. Simulation frameworks for the RIS-assisted and ISAC channels are also implemented. At present, the 2.0 version of BUPTCMCCCMG-IMT2030 has been released at \cite{CMG_website}. 

The ISAC and RIS channel modeling methods employ a concatenated approach. For instance, consider a single-segment channel comprising 20 clusters, each containing 20 sub-paths. This concatenated structure results in a total of 20×20=400 clusters and 400×400=160000 paths. Each multipath incorporates parameters such as delay, power, angle, and Doppler, which introduces significant challenges for channel impulse response computation and data storage. Consequently, the simulation complexity escalates from $O(N)$ to $O(N^2)$. Concurrently, in XL-MIMO technology, the number of antenna elements significantly increases. Characterizing both near-field and spatially non-stationary characteristics further elevates the demand for simulation resources. To achieve an optimal trade-off between computational complexity and simulation fidelity, future research should focus on in-depth investigations into cluster pruning techniques for concatenated channels and the development of simplified concatenated methodologies. Additionally, exploring simplified algorithms for capturing the near-field and spatial non-stationary properties of XL-MIMO, as well as AI-assisted complexity reduction methods, represents promising directions for future investigations. In future versions, more scenarios (especially more space and air scenarios for SAGINs), frequency bands (more frequency points for THz systems), and channel characteristics of potential 6G technologies will be added on the basis of channel measurement and modeling studies to make BUPTCMCCCMG-IMT2030 a powerful assistance tool for channel model simulation.

\section{Open Issues on 6G Channel}\label{SecDis}
\subsection{Channel Model Standardization for 6G}
International standardization organizations have launched focused initiatives to advance channel model development in support of emerging 6G requirements. The 3GPP technical specification group (TSG) for radio access network (RAN) approved two new study items at the RAN \#102 meeting: ``channel modeling for ISAC'' \cite{3GPP_FR3_SID}, and ``channel modeling enhancements for 7--24 GHz'' \cite{3GPP_ISAC_SID}. Technical discussions commenced at the RAN1 \#116 meeting and concluded at the RAN1 \#121 meeting, spanning a total of nine meetings.

For the \textit{ISAC channel model} study item, typical sensing targets of UAVs, humans indoors, humans outdoors, automotive vehicles, automated guided vehicles, and objects creating hazards on roads/railways are approved as a starting point \cite{3gppRan116FL}. The common framework of the ISAC channel model is composed of a path component of the target channel and a component of the background channel. Several key new features of the ISAC channel have been discussed during the 3GPP standardization process \cite{zhang2024latest,3gppRan121FL}: (1) the RCS, which characterizes the target scattering effect and should be modeled for typical sensing targets; (2) the target channel, which is modeled through the concatenated sub-channels of Tx-target and target-Rx; (3) the background channel, where the bistatic mode can reuse existing communication TRs, while the modeling of the monostatic background channel is approved through multi-reference points; and (4) additional features such as spatial consistency, EO, Doppler, and blockage.

For the \textit{7--24 GHz channel model} study item, several key enhancements have been introduced \cite{3GPP_FR3}. Four additional modeling components have been incorporated \cite{3GPP_CR}: (1) near-field propagation, which accounts for spherical wavefront effects in array-based simulations; (2) SnS, which enables the modeling of power variations across antenna elements at both the BS and UE sides; (3) cluster number variability, which reflects the sparsity of high-frequency channels and captures the observation that the number of clusters can vary with factors such as environment, frequency, bandwidth, and spatial resolution; and (4) polarization power variability. In addition, large-scale and small-scale parameters have been updated based on newly collected FR3 measurement data, and a new suburban macro (SMa) scenario has been defined with a complete set of parameters \cite{TP_3GPP}.

Moreover, several standardization challenges remain. Additional technologies such as RIS and SAGIN require continued investigation to support their integration into standardized channel modeling frameworks. For example, further efforts are needed to address micro-Doppler in ISAC channel modeling, which is a key parameter to be used for detecting and tracking the target. Modeling RCS for different materials, distances, and frequencies is also essential for building a standardized RCS database to support identification.

\subsection{AI-enabled Methods in Channel Research}
AI has rapidly advanced in recent years and plays a vital role in wireless channel research. Early studies applied machine learning (ML) methods such as support vector machines (SVMs), relevance vector machines (RVMs), and principal component analysis (PCA) to tasks such as channel parameter extraction \cite{mxc_pca}, multipath clustering \cite{lyp_cluster}, and scenario classification \cite{ha_nlos}, primarily to improve the automation and robustness of channel modeling. With the rise of deep learning (DL) techniques, models can now extract high-dimensional, multiscale features from large-scale datasets and support accurate predictions of channel parameters, including channel state information, path loss, beam indices, and blockage conditions, by leveraging generative adversarial networks (GANs) \cite{zz_gan}, deep reinforcement learning \cite{robert_drl,zz_drl}, and other carefully designed DL networks \cite{zz_AI}. Furthermore, with the increasing use of environmental sensing technologies, AI models are now being developed to learn direct mappings between physical environments and wireless propagation behavior. In this context, we propose the concept of a digital twin channel (DTC) \cite{wh_dtc,wjl_rekp}, which leverages sensing technologies to obtain deterministic wireless environmental information and integrates AI models to generate real-time channel fading states. This approach lays the foundation for proactive and reliable communication decision-making and establishes a new online paradigm that actively addresses channel uncertainty.

The DTC framework holds promise for enabling real-time, environment-intelligent channel modeling and adaptive communication system operation. However, further improvements are needed in terms of accuracy, complexity, and generalizability. First, improving modeling accuracy will require the integration of fine-grained environmental representations and the use of high-resolution sensing data to capture subtle variations in environmental dynamics. Second, addressing computational complexity involves exploring lightweight model architectures, knowledge distillation techniques, and efficient inference strategies that support real-time performance. Third, to enhance generalizability, future models should be designed to adapt to a wide range of environments, mobility patterns, and previously unseen deployment scenarios. This may be achieved through the use of large-scale foundation models that are pretrained on diverse environments \cite{channelgpt}. These large channel models are capable of capturing common propagation patterns and can transfer knowledge across scenarios.

\subsection{Practical Channel Model-enabled System Performance Analysis}
The practical channel model is a key factor in system performance analysis. The existing channel models can effectively support the performance analysis of 5G and previous communication systems. However, for 6G, the technology is developing towards directions such as XL-MIMO and ISAC, which brings new characteristics to the channel. Therefore, the channel model needs to be updated. The channel model proposed in this paper has these features and can support various system performance analyses.

Unlike the existing base station antenna arrays, the extremely large-scale antenna array used in XL-MIMO has near-field and spatial non-stationary characteristics, and has been widely observed \cite{MIMO1-18,MIMO1-13,MIMO1-13-1,MIMO1-34-2,MIMO1-14,MIMO1-15,MIMO1-34,THz_resubmit_14,THz_resubmit_13,MIMO1-25,THz_resubmit_15,MIMO1-16-0,MIMO1-16,MIMO1-16-1,MIMO1-1-0,THz_resubmit_16,THz_resubmit_17}. Therefore, it is necessary to analyze various performance changes caused by these new characteristics, such as beam training \cite{2zhang2022,9738442,3wu2023,4zhang,10994282,ding2025farfieldvsnearfieldpropagation}, achievable rate \cite{8638522,9257469,10110335,10103817,miao2025theoreticalanalysisnearfieldmimo} and degrees of freedom \cite{9139337,10945401,9650519,10262267}. However, traditional channel models cannot characterize the new channel characteristics, making it difficult to support the performance analysis of XL-MIMO. \cite{MIMO1-25} based on the measured results, near-field parameters and spatial non-stationary parameters are introduced to represent the channel, and a near-field spatial non-stationary channel model is established. \cite{ding2025farfieldvsnearfieldpropagation} provided examples on how spherical wavefront and spatially non-stationarity affect the XL-MIMO system performance in terms of beamforming gain and achievable rate. \cite{miao2025theoreticalanalysisnearfieldmimo} focused on the theoretical analysis and empirical study of near-field MIMO capacity. Based on this foundation, more research directions, including how to consider the actual multipath channel, the spatial non-stationarity of the hybrid field channel in narrowband and wideband systems for near-field technology design, and how to design new beamforming in wideband hybrid field channels to overcome the beam splitting effect, can be further explored in future research.

ISAC, as a novel technology for 6G, relies heavily on accurate channel modeling for system performance and technical evaluation.
First, 3GPP and numerous academic studies have investigated the RCS characteristics of typical targets such as UAVs, humans, and vehicles. Precise RCS modeling is critical for target classification, recognition, and localization, especially in complex environments \cite{cao2022automatic}.
While traditional positioning depends on LoS paths, NLoS signals—leveraging single reflections with AoA and ToA—can also be exploited in cluttered scenarios \cite{al2002ml,chen2024and}. Modeling EOs helps enhance positioning via NLoS paths.
Micro-Doppler modeling can assist in recognizing the dynamic behavior and posture of sensing targets, such as small movements of human limb motion \cite{wei2019classification} or UAV rotor rotation \cite{wang2021lightweight}.
Moreover, accurate modeling of environmental effects improves clutter suppression and the extraction of target features (e.g., delay, and angle), thus enhancing localization robustness under multipath interference \cite{wang2020small,xu2023spatial}.
Finally, modeling the sharing feature of communication and sensing supports mutual optimization and performance evaluation, fostering efficient low-power ISAC system design \cite{nguyen2022access,zhang2022integrated}.

Similar to XL-MIMO, RIS deployments often operate in near-field conditions due to large physical apertures, making near-field propagation modeling essential for accurate system performance evaluation. Prior analyses under far-field assumptions suggest that the end-to-end RIS gain scales quadratically with the number of elements ($N^2$) \cite{basar_access,wqq_N2}, potentially outperforming MIMO systems with the same array size. However, recent studies incorporating spherical wave models show that near-field effects fundamentally limit such scaling, and passive RIS cannot surpass traditional MIMO under practical conditions \cite{bjornson_NorN2}.
In addition to near-field effects, existing RIS performance analyses often rely on simplified assumptions, such as Rayleigh/Rician fading, rich scattering, continuous phase shifts, and ideal reflection. These oversimplifications can significantly overestimate system performance, as practical environments may exhibit limited angular spread, discrete phase control, and non-ideal electromagnetic responses.
Therefore, future RIS-related performance evaluations should consider: (i) near-field propagation, (ii) realistic angular and fading models, (iii) hardware constraints such as discrete phase control, and (iv) electromagnetic characteristics of RIS elements. These aspects are critical for bridging the gap between theoretical modeling and real-world deployment.

\section{Conclusions}\label{SecCon}
The ITU's current work on the framework and overall objectives represents the official launch of 6G standardization. 
With the progression of standardization efforts, research on channel modeling has become increasingly critical. This paper presents a comprehensive survey of 6G channel requirements, field measurements, modeling methodologies and a released simulator. First, the representative requirements and challenges of 6G channel research are summarized. We discuss channel measurements and modeling of potential key technologies, including ISAC, XL-MIMO, mid-band, THz, RIS, and SAGIN. For channel measurement, a measurement platform with high accuracy that can support multiple scenarios is urgently needed. For channel modeling, new channel characteristics must be properly considered. Geometry-based stochastic modeling is one of the main modeling methods of 6G, but the model needs to be adjusted under different technologies and scenarios. Future research prospects are discussed for each technology.
A summary of the released 6G channel simulators is presented below. The BUPTCMCCCMG-IMT2030, developed on the basis of the ITU/3GPP GBSM framework, is introduced. It provides practical support for simulating a wide range of 6G technologies, ensures backward compatibility and is being extended to accommodate additional frequency bands, deployment scenarios, and channel characteristics. Finally, we address emerging perspectives in 6G channel research, including recent 3GPP activities on modeling for 7–24 GHz and ISAC, AI-enabled methods in channel research, and system performance analysis.

\section*{Acknowledgments}
This research is supported in part by National Natural Science Foundation of China under Grant (62525101), National Natural Science Foundation of China (92167202, 62101069, 62201086, 62201087), and Beijing University of Posts and Telecommunications-China Mobile Research Institute Joint Innovation Center.
\bibliographystyle{IEEEtran}
\bibliography{ref}

\begin{IEEEbiography}[{\includegraphics[width=1in,height=1.25in,clip,keepaspectratio]{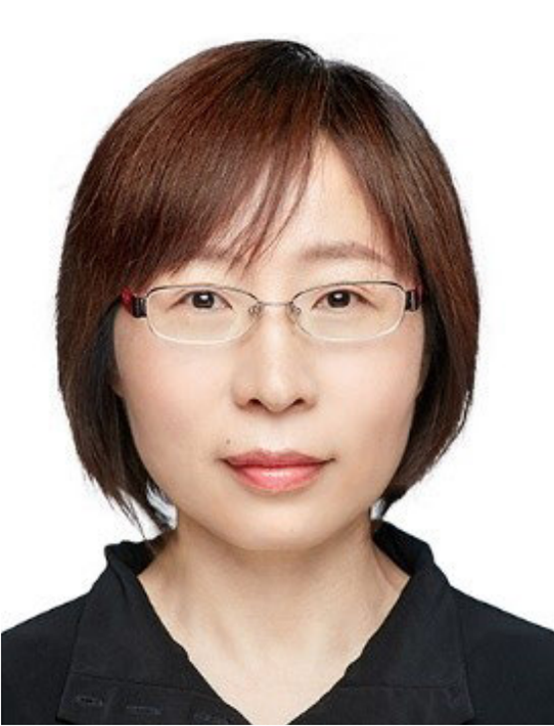}}]
{Jianhua Zhang} (Senior Member, IEEE) received the Ph.D. degree from the Beijing University of Posts and Telecommunications (BUPT) in 2003. She is currently a Professor with BUPT, the China Institute of Communications Fellow, and the Director of the BUPT--CMCC Joint Research Center. She has published more than 200 papers and authorized 40 patents. She received several paper awards, including 2019 SCIENCE China Information Hot Paper, 2016 China Comms Best Paper, and 2008 JCN Best Paper. She received several prizes for her contribution to ITU--R 4G channel model (ITU--R M.2135), 3GPP relay channel model (3GPP 36.814), and 3GPP 3D channel model (3GPP 36.873). She was also a member of 3GPP ``5G channel model for bands up to 100 GHz''. From 2016 to 2017, she was the Drafting Group (DG) Chairperson of ITU--R IMT--2020 Channel Model and led the drafting of the ITU--R M. 2412 Channel Model Section. She is also the Chairwomen of the China IMT--2030 Tech Group--Channel Measurement and Modeling Subgroup and works on 6G channel model. Her current research interests include beyond 5G and 6G, artificial intelligence, data mining, channel modeling for integrated sensing and communication, massive MIMO, millimeter wave, THz, visible light channel modeling, channel emulator, and OTA test.
\end{IEEEbiography}


\begin{IEEEbiography}[{\includegraphics[width=1in,height=1.25in,clip,keepaspectratio]{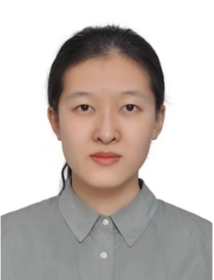}}]
{Jiaxin Lin} (Graduate Student Member, IEEE) received her B.S. degree in communication engineering from the Beijing University of Posts and Telecommunications (BUPT) in 2017. She is currently pursuing an M.S. degree at the School of Information and Communication Engineering, BUPT. Her current research interest is THz channel modeling.
\end{IEEEbiography}

\begin{IEEEbiography}[{\includegraphics[width=1in,height=1.25in,clip,keepaspectratio]{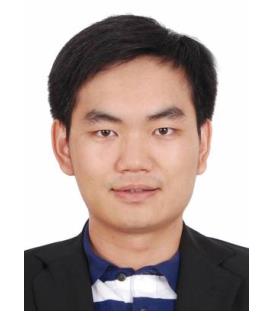}}]
{Pan Tang} (Member, IEEE) received the B.S. degree in electrical information engineering from the South China University of Technology, Guangzhou, China, in 2013, and the Ph.D. degree in information and communication engineering from the Beijing University of Posts and Telecommunications (BUPT), Beijing, China, in 2019. In 2017, he was a Visiting Scholar with the University of Southern California. From 2019 to 2021, he was a Postdoctoral Research Associate at BUPT, China. He is currently a associate researcher at BUPT. His current research interests include millimeter wave, THz, and visible light channel measurements and modeling.
\end{IEEEbiography}

\begin{IEEEbiography}[{\includegraphics[width=1in,height=1.25in,clip,keepaspectratio]{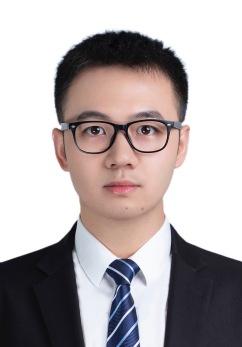}}]
{Yuxiang Zhang} (Member, IEEE) received his B.S. degree in electronic information engineering from Dalian University of Technology in 2014 and his Ph.D. degree from the Beijing University of Posts and Telecommunications (BUPT) in 2020. From 2018 to 2019, he was a visiting scholar with the University of Waterloo. He is now a associate researcher at BUPT, China. His current research interests include channel modeling, massive multiple-input multiple-output (MIMO), beamforming, and over-the-air testing.
\end{IEEEbiography}

\begin{IEEEbiography}[{\includegraphics[width=1in,height=1.25in,clip,keepaspectratio]{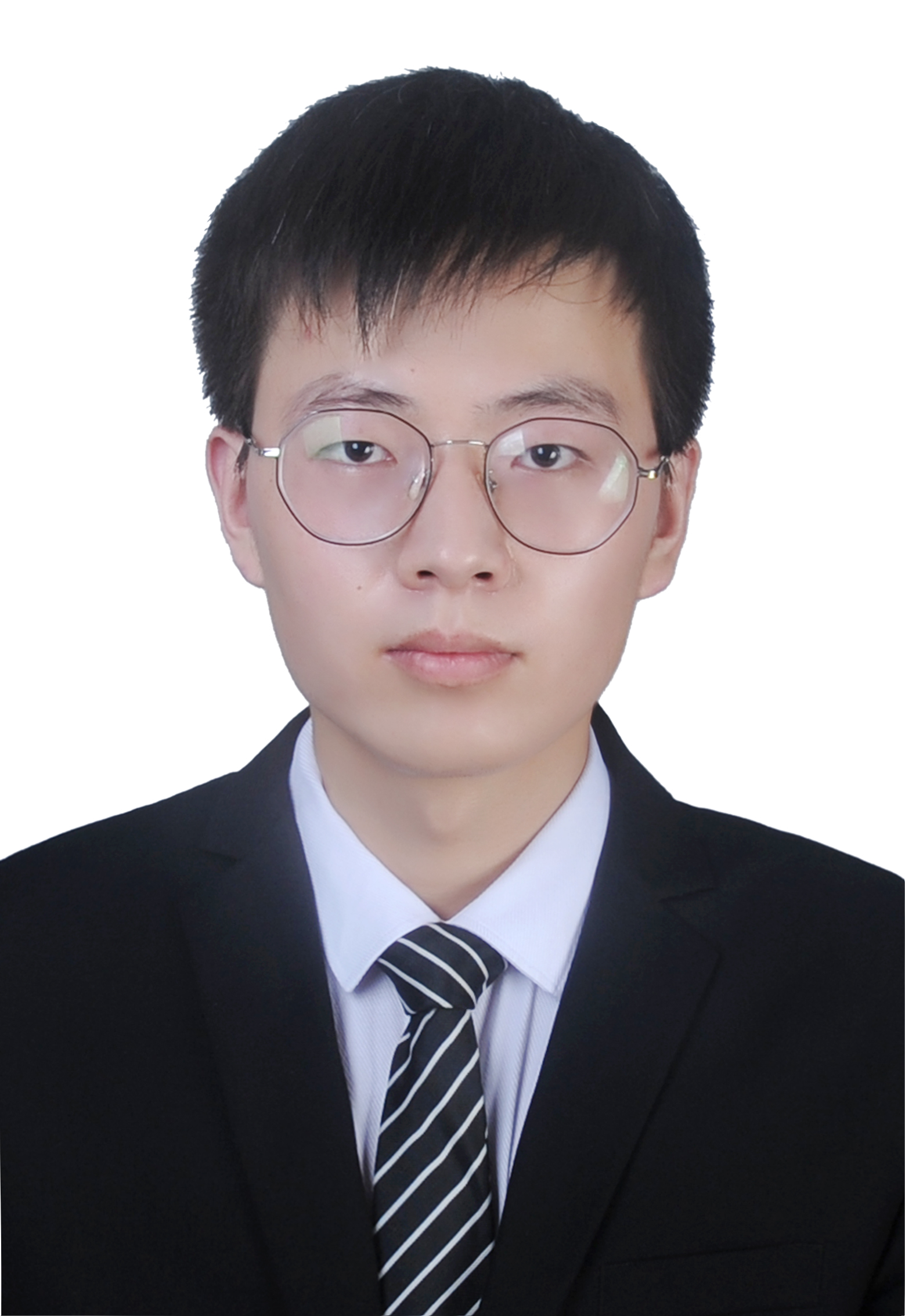}}]
{Huixin Xu} (Graduate Student Member, IEEE) received his B.S. degree in communication engineering from Chongqing University of Posts and Telecommunications in 2021. He is currently pursuing a Ph.D degree in the School of Information and Communication Engineering, Beijing University of Posts and Telecommunications. His research interests include THz channel modeling and channel reciprocity.
\end{IEEEbiography}

\begin{IEEEbiography}[{\includegraphics[width=1in,height=1.25in,clip,keepaspectratio]{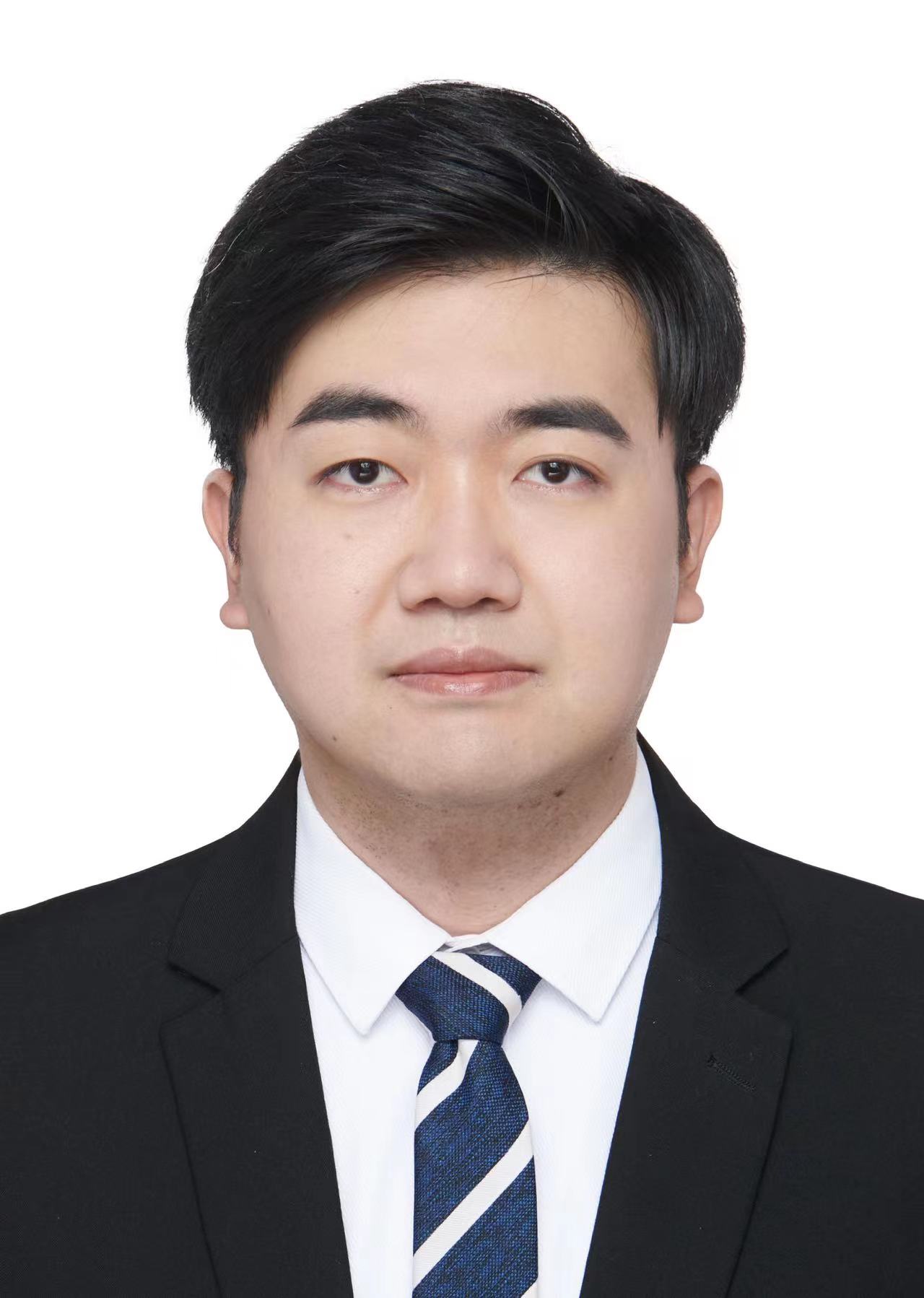}}]
{Tianyang Gao} received the B.S. degree from the Beijing University of Posts and Telecommunications (BUPT), Beijing, China, in 2020, where he is pursuing the M.S. degree. His research interest contains channel model simulation.
\end{IEEEbiography}

\begin{IEEEbiography}[{\includegraphics[width=1in,height=1.25in,clip,keepaspectratio]{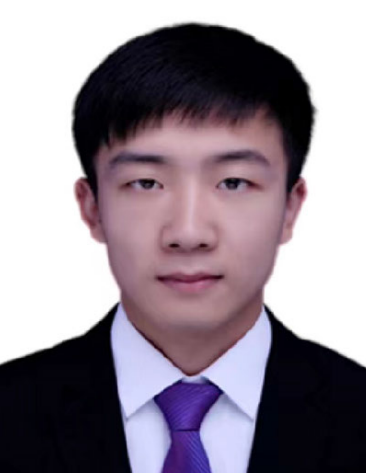}}]
{Haiyang Miao} (Graduate Student Member, IEEE) received the master's degree in communication and information systems from Beijing Jiaotong University (BJTU), Beijing, China, in 2021. He is currently pursuing the Ph.D. degree with the State Key Laboratory of Networking and Switching Technology, Beijing University of Posts and Telecommunications (BUPT). His current research interests include sub--6 GHz, mid--band, millimeter wave, and massive MIMO channel measurement and modeling.
\end{IEEEbiography}




\begin{IEEEbiography}[{\includegraphics[width=1in,height=1.25in,clip,keepaspectratio]{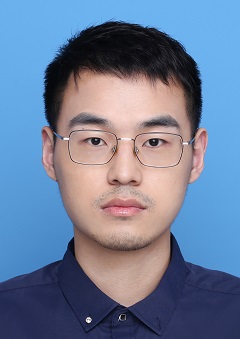}}]
{Huiwen Gong} received the B.S. degree from the Beijing University of Posts and Telecommunications (BUPT), Beijing, China, in 2021, where he is pursuing the M.S. degree. His research interest contains RIS channel modeling.
\end{IEEEbiography}

\begin{IEEEbiography}[{\includegraphics[width=1in,height=1.25in,clip,keepaspectratio]{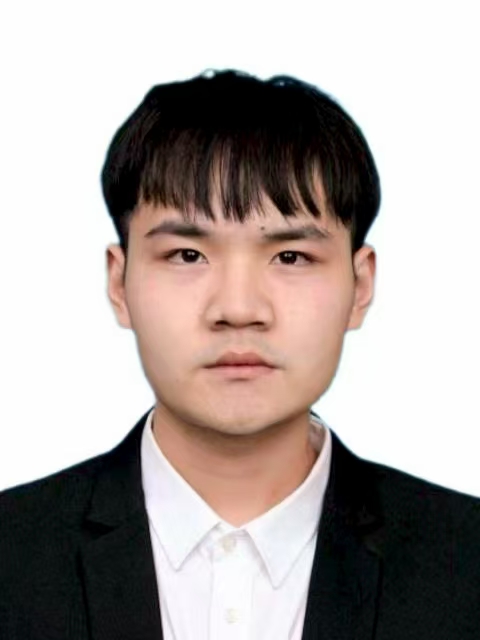}}]
{Changsheng Zhao} is currently working toward the Ph.D. degree in information and communication engineering with the Beijing University of Posts and Telecommunications, Beijing, China. His research interests include ISAC and over-the-air testing.
\end{IEEEbiography}

\begin{IEEEbiography}[{\includegraphics[width=1in,height=1.25in,clip,keepaspectratio]{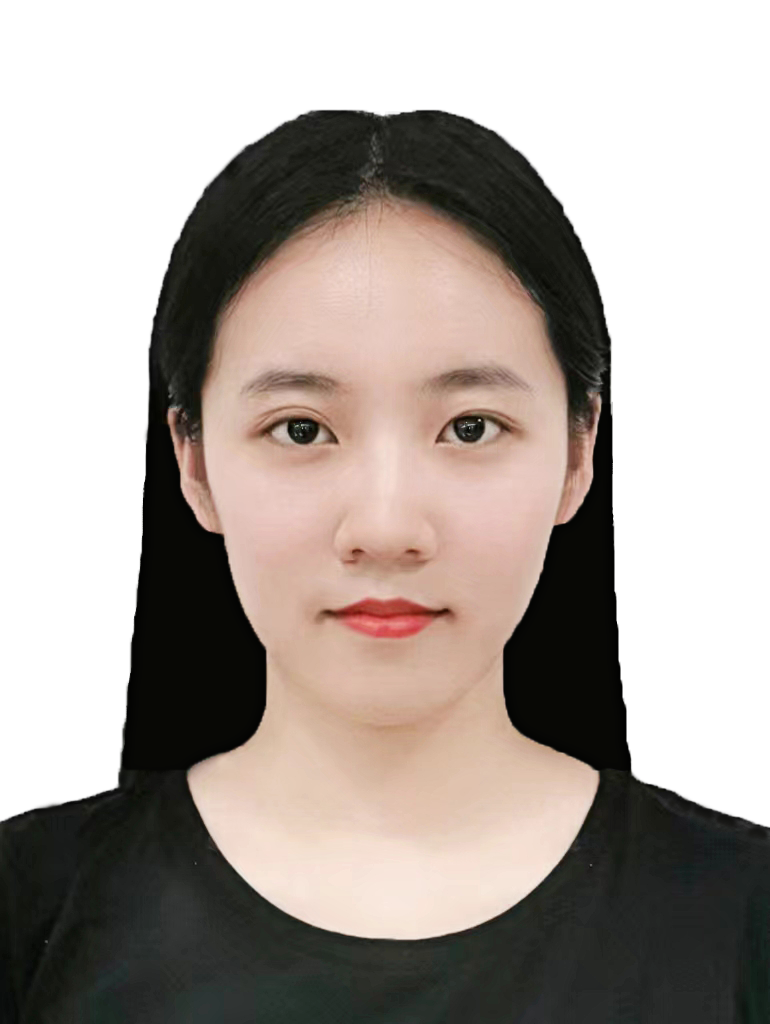}}]
{Yameng Liu} (Graduate Student Member, IEEE) received the B.S. degree in communication engineering from the Minzu University of China in 2021. She is currently pursuing the Ph.D. degree in information and communication engineering in the Beijing University of Posts and Telecommunications. Her current research interests include integrated sensing and communication, channel measurements and modeling.
\end{IEEEbiography}

\begin{IEEEbiography}[{\includegraphics[width=1in,height=1.25in,clip,keepaspectratio]{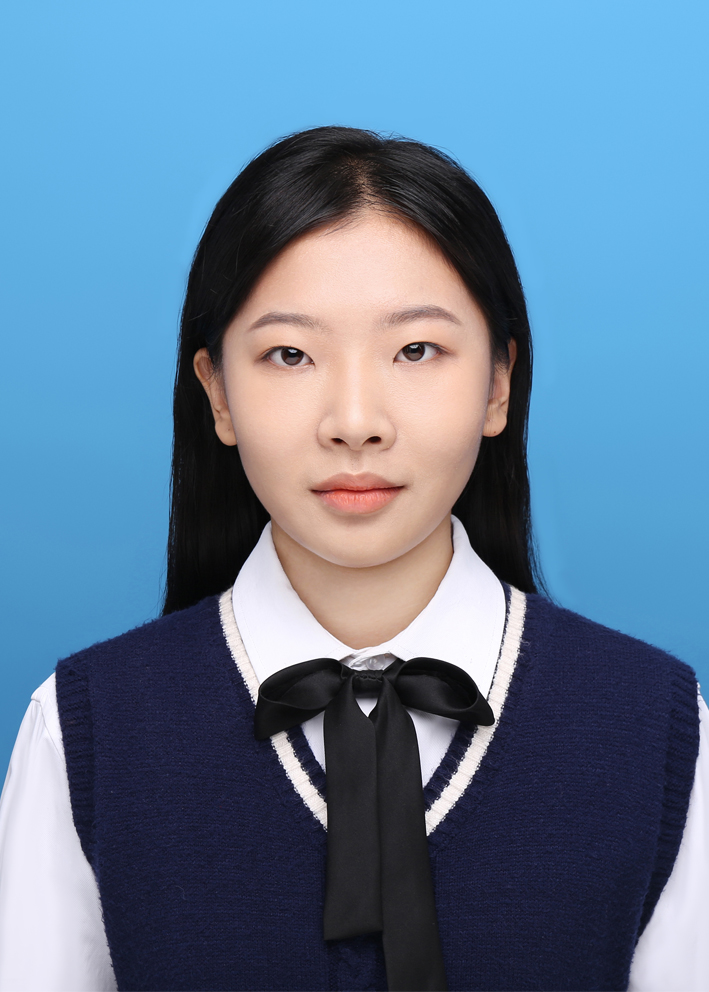}}]
{Yichen Cai} received the B.S. degree from the Beijing Jiaotong University (BJTU), Beijing, China, in 2019. She is currently pursuing the Ph.D. degree at the School of Information and Communication Engineering, Beijing University of Posts and Telecommunications (BUPT). Her current research interest is deep learning and channel prediction.
\end{IEEEbiography}

\begin{IEEEbiography}[{\includegraphics[width=1in,height=1.25in,clip,keepaspectratio]{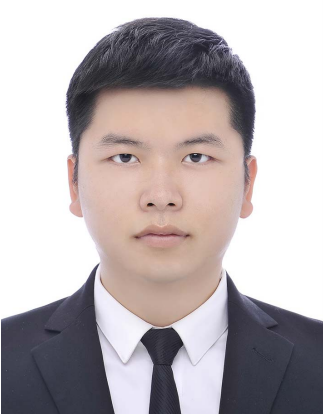}}]
{Zhiqiang Yuan} (Graduate Student Member, IEEE) received the B.S. degree from the Beijing University of Posts and Telecommunications (BUPT), Beijing, China, in 2018, where he is currently pursuing the Ph.D. degree. He had been a Visiting Ph.D. Student with the Antennas, Propagation, and Millimeter-Wave Systems (APMS) Section, Aalborg University, Aalborg, Denmark, from Jul. 2021 to Jul. 2023. He currently focuses on radio channel sounding and modeling for massive MIMO, millimeter-wave (mmWave), and THz systems, antenna array signal processing, and channel parameter estimation.
\end{IEEEbiography}

\begin{IEEEbiography}[{\includegraphics[width=1in,height=1.25in,clip,keepaspectratio]{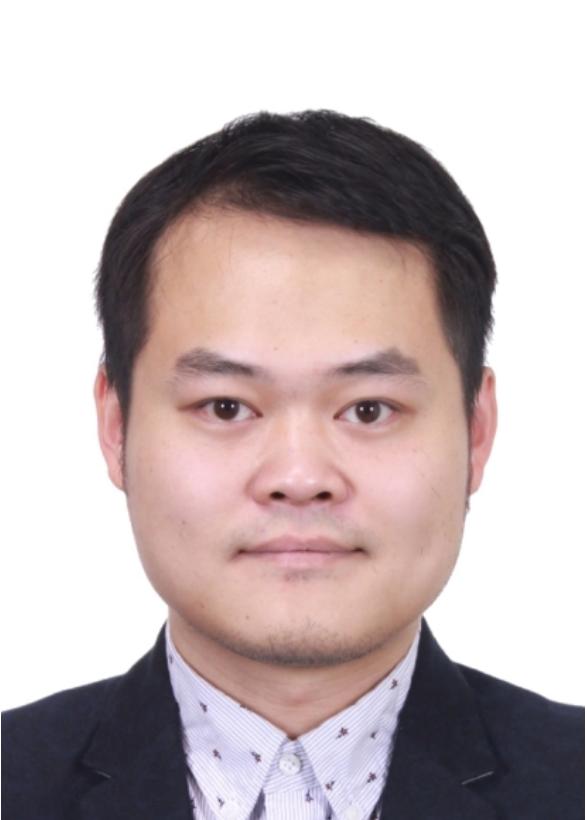}}]
{Lei Tian} (Member, IEEE) received the B.S. degree in communication engineering and the Ph.D. degree in information and communication systems from the Beijing University of Posts and Telecommunications in 2009 and 2015, respectively. His current research interests are wireless channel measurement, modeling, and simulation for B5G and 6G, including mmWave, THz, massive MIMO, and satellite-to-ground.
\end{IEEEbiography}

\begin{IEEEbiography}[{\includegraphics[width=1in,height=1.25in,clip,keepaspectratio]{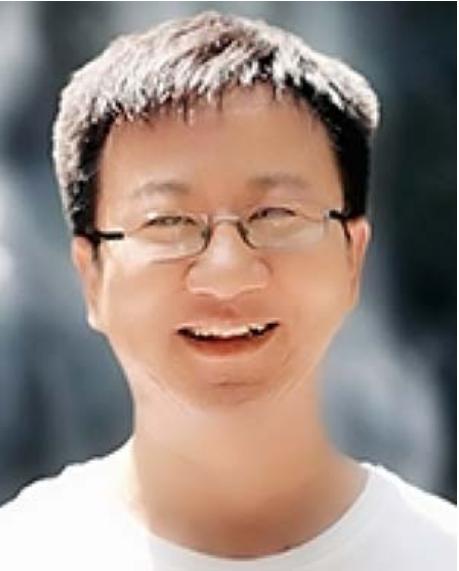}}]
{Shaoshi Yang} (Senior Member, IEEE) received the B.Eng. degree in information engineering from the Beijing University of Posts and Telecommunications (BUPT), Beijing, China, in 2006, and the Ph.D. degree in electronics and electrical engineering from the University of Southampton, U.K., in 2013. From 2008 to 2009, he was a Researcher of WiMAX standardization with Intel Labs China. From 2013 to 2016, he was a Research Fellow with the School of Electronics and Computer Science, University of Southampton, Southampton, U.K. From 2016 to 2018, he was a Principal Engineer with Huawei Technologies Co. Ltd. He is currently a Full Professor with BUPT. His research expertise includes 5G wireless networks, massive MIMO, iterative detection and decoding, mobile ad hoc networks, distributed artificial intelligence, and cloud gaming/VR.
\end{IEEEbiography}

\begin{IEEEbiography}[{\includegraphics[width=1in,height=1.25in,clip,keepaspectratio]{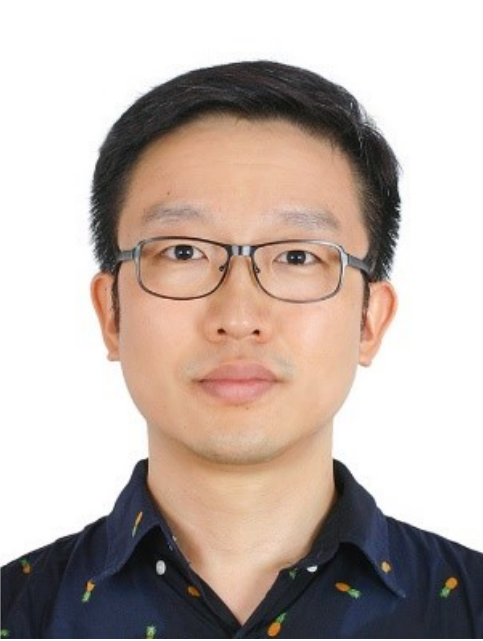}}]
{Liang Xia} received the B.S. degree in Electronic and Information Engineering in 2005 and the M.S. degree in Information and Communication Engineering in 2008 from Tsinghua University, Beijing, China. He was a researcher at LTE physical layer in Huawei from 2008 to 2014 and was a researcher at 5G physical layer in China Mobile from 2015 to 2019. His current research interests include millimetre-wave, THz, and visible light communication.
\end{IEEEbiography}

\begin{IEEEbiography}[{\includegraphics[width=1in,height=1.25in,clip,keepaspectratio]{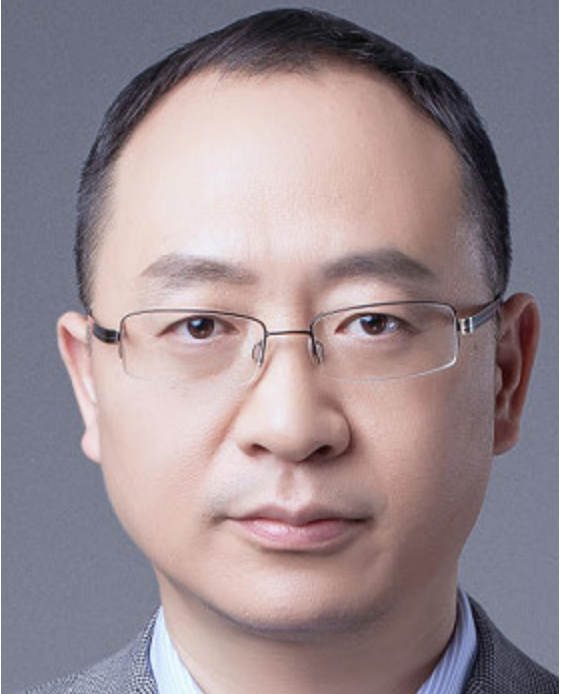}}]
{Guangyi Liu} (Member, IEEE) received the Ph.D. degree from the Beijing University of Posts and Telecommunications, Beijing, China, in 2006. He is currently a Chief Scientist with 6G in China Mobile Communication Corporation (CMCC), the Co--Chair of the 6G Alliance of Network AI (6GANA), the Vice Chair of THz Industry Alliance in China, and the Vice Chair of the Wireless Technology Working Group of IMT--2030 (6G) Promotion Group supported by the Ministry of Information and Industry Technology of China. He has been leading the 6G Research and Development of CMCC since 2018. He has acted as the Spectrum Working Group Chair and the Project Coordinator of LTE Evolution and 5G eMBB in Global TD--LTE Initiative (GTI) (2013--2020) and led the industrialization and globalization of TD--LTE evolution and 5G eMBB.
\end{IEEEbiography}


\begin{IEEEbiography}[{\includegraphics[width=1in,height=1.25in,clip,keepaspectratio]{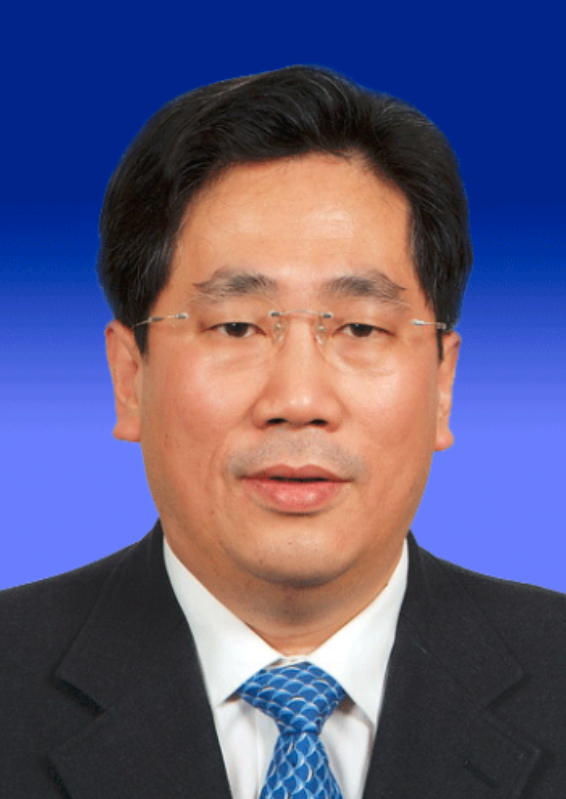}}]
{Ping Zhang} (Fellow, IEEE) received the M.S. degree in electrical engineering from Northwestern Polytechnical University, Xi'an, China, in 1986, and the Ph.D. degree in electric circuits and systems from the Beijing University of Posts and Telecommunications (BUPT), Beijing, China, in 1990. He is currently a Professor with BUPT. His research interests include cognitive wireless networks, fourth--generation mobile communication, fifth--generation mobile networks, communications factory test instruments, universal wireless signal detection instruments, and mobile Internet. He was the recipient of the First and Second Prizes from the National Technology Invention and Technological Progress awards and the First Prize Outstanding Achievement Award of Scientific Research in College. He is the Executive Associate Editor-in-Chief on information sciences of the Chinese Science Bulletin, a Guest Editor of IEEE Wireless Communications Magazine, and the Editor of China Communications.
\end{IEEEbiography}

\end{document}